\documentclass[
aps,
physrev,
reprint,
floatfix]{revtex4-2}
\usepackage{aas_macros}
\DeclareRobustCommand{\VAN}[3]{#2}
\let\VANthebibliography\thebibliography
\def\thebibliography{\DeclareRobustCommand{\VAN}[3]{##3}\VANthebibliography}
\usepackage{siunitx}	
\usepackage{bm}	        
\usepackage{amssymb}    
\usepackage{amsfonts}	
\usepackage{gensymb}    
\usepackage{mathtools}  
\usepackage{mathrsfs}   
\usepackage{graphicx}	
\usepackage{tikz}		
\usepackage[export]{adjustbox} 
\usepackage{overpic}    
\usepackage{dcolumn}    
\usepackage{multirow}   
\usepackage{makecell}   
\usepackage{xcolor}     
\definecolor{crimsonglory}{rgb}{0.75, 0.0, 0.2}
\definecolor{orangered}{rgb}{1.0, 0.27, 0.0}
\definecolor{forestgreen}{rgb}{0.13, 0.55, 0.13}
\usepackage{enumitem}   
\setlist[enumerate,1]{label=(\roman*)}
\usepackage{verbatim}
\usepackage{ifthen}
\usepackage[normalem]{ulem} 
\usepackage{hyperref}
\hypersetup{
  colorlinks,
  citecolor=blue,
  linkcolor=blue,
  urlcolor=blue}
\usepackage{cleveref}

\renewcommand{\autoref}[1]{\cref{#1}}
\newcommand{\Autoref}{\Cref}
\crefname{equation}{Eq.}{Eqs.}
\Crefname{equation}{Equation}{Equations}
\crefname{section}{\S}{\S\S}
\Crefname{section}{Section}{Sections}
\crefname{appendix}{Appendix}{Appendices}
\Crefname{appendix}{Appendix}{Appendices}
\crefname{figure}{Fig.}{Figs.}
\Crefname{figure}{Figure}{Figures}
\crefformat{section}{\S#2#1#3}
\crefrangeformat{section}{\S#3#1#4--#5#2#6}
\crefmultiformat{section}{\S#2#1#3}{, \S#2#1#3}{, \S#2#1#3}{, \S#2#1#3}
\newif\iffboxvisible
\fboxvisibletrue   
\newcommand{\myfbox}[1]{%
  \iffboxvisible
    \fbox{#1}%
  \else
    #1%
  \fi
}

\newcommand{\paran}[1]{\left(#1\right)}
\newcommand{\curly}[1]{\left\{#1\right\}}
\newcommand{\sqb}[1]{\left[#1\right]}
\newcommand{\Msun}{M_\odot}

\newcommand{\sinv}{\mathrm{s}^{-1}}
\newcommand{\cminv}{\mathrm{cm}^{-1}}
\newcommand{\ccinv}{\mathrm{cm}^{-3}}
\newcommand{\g}{\mathrm{g}}
\newcommand{\tage}{{\mathbb{T}}}
\newcommand{\tFB}{\mathbb{T}_\mathrm{FB}}
\newcommand{\tRB}{\mathbb{T}_\mathrm{RB}}

\newcommand{\half}{\frac{1}{2}}

\newcommand{\partialIlAt}[3]{\left(\partial #1/\partial #2\right)\vert_{#3}}

\newcommand{\nomBall}{\mathcal{B}_0}
\newcommand{\nomSlow}{\mathcal{S}_0}
\newcommand{\EJ}{E_{56}}
\newcommand{\vJ}{\beta_{-2}}
\newcommand{\thJ}{\theta_5}

\newcommand{\RB}{\mathrm{RB}}
\newcommand{\FB}{\mathrm{FB}}
\newcommand{\Mach}{\mathcal{M}}

\newcommand{\Tnorm}{T_{0.2}}

\newcommand{\srinv}{\mathrm{sr}^{-1}}

\newcommand\ion[2]{\ensuremath{\mathrm{#1\,\scriptstyle #2}}}
\newcommand{\tdel}[1][j]{\Delta\mathbb{T}_j}
\newcommand{\su}{sw} 
\newcommand{\trelax}{t_{\mathrm{relax}}}
\newcommand{\Max}{{\mathrm{max}}}
\newcommand{\Myeta}{{h}}

\newcommand{\ie}{\emph{i.e.,} }
\newcommand{\eg}{\emph{e.g.,} }

\newcommand{\be}{\begin{equation}}
\newcommand{\ee}{\end{equation}}
\newcommand{\bea}{\begin{equation*}}
\newcommand{\eea}{\end{equation*}}
\newcommand{\beqr}{\begin{eqnarray} \nonumber}
\newcommand{\eeqr}{\end{eqnarray}}
\newcommand{\beqrb}{\begin{eqnarray}}
\newcommand{\eeqrb}{\nonumber \end{eqnarray}}
\newcommand{\fin}{\mbox{ .}}
\newcommand{\coma}{\mbox{ ,}}

\newcommand{\cm}{\mbox{ cm}}

\newcommand{\se}{\mbox{ s}}

\newcommand{\Myr}{\mbox{ Myr}}

\newcommand{\erg}{\mbox{ erg}}

\newcommand{\km}{\mbox{ km}}
\newcommand{\pc}{\mbox{ pc}}
\newcommand{\kpc}{\mbox{ kpc}}

\newcommand{\keV}{\mbox{ keV}}

\newcommand{\ph}{\mbox{ ph}}

\newcommand{\gama}{$\gamma$}

\newcommand{\vect}[1]{\mathbf{#1}}

\newcommand{\dgrdot}{{\overset{^\circ}{.}}}
\defcitealias{Mondaletal2022}{M22}
\newcommand{\MK}{{\citetalias{Mondaletal2022}}}

\begin{document}
\title{Nested Fermi and eROSITA bubbles require very similar $\sim10^{56}$ erg collimated Galactic-center outbursts; their asymmetry indicates an eastern density gradient}

\author{Arka Ghosh}\thanks{E-mail: arka@post.bgu.ac.il}
\affiliation{Physics Department, Ben-Gurion University of the Negev, POB 653, Be'er-Sheva 84105, Israel}

\author{Uri Keshet}\thanks{E-mail: keshet.uri@gmail.com}
\affiliation{Physics Department, Ben-Gurion University of the Negev, POB 653, Be'er-Sheva 84105, Israel}

\author{Santanu Mondal}
\affiliation{Indian Institute of Astrophysics, II Block, Koramangala, Bangalore 560034, India}

\begin{abstract}
Observations indicate two nested pairs of extended bipolar bubbles emanating from the Milky-Way center --- the $|b|\sim80^\circ$ latitude \textit{eROSITA} bubbles (RBs), encompassing the smaller, $|b|\sim 50^{\circ}$ \emph{Fermi} bubbles (FBs) ---
and classify the edges of both bubble pairs as strong forward shocks. Identifying each bubble pair as driven by a distinct, collimated outburst, we evolve these bubbles and constrain their origin using a stratified 1D model verified by a suite of 2D and 3D hydrodynamic simulations which reproduce X-ray observations.
While the RBs are at the onset of slowdown, the FBs are still expanding ballistically into the RB-shocked medium.
Observational constraints indicate that both RB and FB outbursts had (up to factor $\sim2$--$4$ uncertainties) $\sim4^\circ$ half-opening angles and $\sim 2000$ km s$^{-1}$ velocities $100$ pc from their base, carrying $\sim10^{56}$ erg.
The FBs and RBs could thus arise from identical outbursts separated by $\sim10$ Myr; their longitudinal asymmetry favors an eastern ambient-density gradient over western wind suggestions.
\end{abstract}
\maketitle

\section{Introduction}
\label{sec:Intro}

Galactic outflows in the form of bipolar bubbles are ubiquitously observed in the local as well as distant Universe (see \citep{Bischetti+2017,LopezCoba2020} and references therein), in some cases (\eg \citep{JamrozyEtAl09}) arising from repeated outbursts of an active galactic nucleus (AGN). Broadband observations ranging from radio to $\gamma$-rays have shown that two pairs of bipolar, apparently\footnote{not only in projection, as we later show} nested bubbles emanate from the Galactic Center (GC) of the Milky Way.
The inner of these two pairs was first traced far from the Galactic plane in \textit{Fermi}-LAT data \citep{SuEtAl10, Dobler+2010} (dubbed the Fermi bubbles; FBs hereafter), reaching $\vert b\vert\simeq 50^\circ$ latitudes.
The outer bipolar bubbles, reaching $|b|\simeq 80^\circ$, were indicated in radio \citep{Sofue77} and X-ray \citep{Sofue94}, in particular using \textit{ROSAT} observations \citep{Sofue00}, and became later established by \textit{eROSITA} \citep{Predehletal2020}; this pair is designated the \textit{ROSAT/eROSITA} bubbles (RBs henceforth).
The northern RB coincides with the radio-to-\gama-ray Loop-I or northern polar spur (NPS); the southern RB was recently detected also in nonthermal radio and \gama-ray emission \citep{KeshetGhosh26}.

Multiple lines of evidence indicate that the edges of both FBs and RBs are forward shocks, implying that both pairs arise from energetic outbursts.
In both FBs \citep{Keshetgurwich18} and RBs \citep{Predehletal2020, KeshetGhosh26}, a high-latitude X-ray shell shows compressed, heated thermal electrons, while coincident \gama-rays \citep{SuEtAl10, KeshetGhosh26} indicate the fresh acceleration of relativistic electrons, consistent with diffusive shock acceleration. In the FBs, shock compression, particle acceleration, and magnetization are further indicated by inward \gama-ray hardening \citep{Keshetgurwich17}, a cooling break, and coincident synchrotron and thermal dust emission \citep{Keshet+2023}, which are both preferentially polarized perpendicular to the edge \citep{Keshet2024}, consistent with shock-amplified magnetic fields.

Despite contradictory claims in the literature \citep[][and references therein]{Predehletal2020,MouEtAl23,Sarkar2024}, there is substantial evidence identifying the edges of both FBs and RBs as strong forward shocks. In the FBs, the hard spectra of radio \citep{Keshet+2023} and \gama-ray \citep{Keshetgurwich17} emission from the edge, and of microwaves \citep{Dobler12, PlanckHaze13} and \gama-rays \citep{SuEtAl10} integrated over the bubbles, as well as the nearly uniform \gama-ray spectrum along the entire edge \citep{Keshetgurwich17}, imply a strong, $\Mach>5$ shock.
Lower Mach numbers inferred from the electron temperature jump, based on \ion{O}{VII} and \ion{O}{VIII} line emission ($\Mach=2.3_{-0.4}^{1.1}$; Refs.~\citep{MillerBregman2015, MillerBregman2016}) or \textit{ROSAT} R4-R7 band ratios ($\Mach\sim4$; Ref.~\citep{Keshetgurwich18}) suggest that shock-heated electrons and ions have not yet equilibrated, consistent with the young inferred age of the FBs \citep{Keshetgurwich18} and not \citep{Sarkar+2023} with weak shocks.

In the northern RB, a wide range of Mach numbers was inferred from radio \citep{ReichReich88, Borka07, Guzman11, Iwashita+2023, MouEtAl23} and \gama-rays \citep{MouEtAl23}, and a modest temperature jump was again invoked to argue for a weak, $\Mach\sim1.5$ shock (\eg \citep{Predehletal2020,MouEtAl23}).
However, the X-ray-traced edges of both RBs show a hard radio spectrum indicating a strong shock ($3\lesssim\Mach\lesssim 5$),
coincident with a large (factor $>3$) jump in X-ray brightness and \gama-ray emission similar to that of the FB edges \citep{KeshetGhosh26}.
The asymmetry between north and south bubbles is more pronounced in the RBs, which show a radio brightness nearly an order of magnitude lower in the southern hemisphere than in the north.
However, this difference was attributed to different ambient conditions \citep{Kataoka+2018}, in particular a southern density which could be modestly ($\sim20\%$) lower in some scenarios \citep{Sarkar19} but is measured to be $\sim2$ times lower than in the north \citep{KeshetGhosh26}.

The FB X-ray shells indicate a $\sim 10^{55\text{--}57} \erg$ outburst launched from the GC a few Myr ago \citep{Keshetgurwich18}. Modeling the NPS as an outburst from the GC suggests similar energy estimates, in the range $\sim 10^{55\text{--}57} \erg$, from an older, $\simeq 10\mbox{--}15\Myr$ outburst \citep{Akita+2018, LaRocca+2020}. Such consistent high-energy estimates favor two separate outbursts from the central super-massive black hole (SMBH), over alternative \citep[\eg][]{Sarkaretal15a, Zhang+2024, Scheffler+2025} scenarios.
Indeed, the FB morphology was shown to necessitate a collimated GC outburst, directed in projection approximately perpendicular to the Galactic plane \citep[][henceforth {\MK}]{Mondaletal2022}; a strong tilt is ruled out \citep{Sarkar+2023}. Note that most models \citep[\eg][]{Sarkar2017, Guo12, Guoetal2012, Zhang+2020, Yang+2022, Tseng+2024} fail to reproduce observations, obtaining over-inflated bubble morphologies or FB edges that are contact discontinuities, reverse shocks, or forward shocks of implausible Mach numbers or morphologies. Low-latitude features suggest more than one GC outburst \citep{SofueHanda1984, Baganoffetal03, Blandhawthorncohen03, Law2010, Foxetal2015, Lockman+2016, Sofue2017, Ponti+2019, Sofue+2021}, but it is difficult to link them directly to the high-latitude bubbles.

To our knowledge, no present study successfully produces both RB and FB edges as the forward shocks observed, consistent with observational constraints. Such a combined model is needed to characterize the outburst parameters because, as we show, the effect of the RBs on the CGM modifies the FB evolution. Here, we present a joint RB--FB model, using an analytic 1D approximation for stratified bubbles, corroborated by hydrodynamical 2D and 3D simulations. In \autoref{sec:burstModels}, we revisit key aspects of the FB stratified analytic model of {\MK}, generalize it to include the RBs, derive additional features such as analytic expressions for the projected bubbles, and use the results to constrain the outburst parameters.
Numerical simulations of two GC outbursts, producing both RBs and FBs, are described and analyzed in \autoref{sec:sims}.
We summarize our findings and conclude in \autoref{sec:discussion}.

Supplements to the text include descriptions of our Galactic Model (\autoref{appendix:galModel}), simulation convergence tests (\autoref{appendix:convTests}), initial conditions in viscous simulations (\autoref{appendix:transition}), a 3D demonstration (\autoref{appendix:3d}), numerical sampling of the simulation parameter phase space (\autoref{appendix:nonNominal}), and a critical discussion of concurrent \citep{Zhang+2025} numerical work (\autoref{appendix:critic}).

\section{Modeling Galactic bubbles}
\label{sec:burstModels}

Our Galactic model and jet setup, used in the analytical and numerical modelling of the bubbles, are outlined in \autoref{subsec:setup}, followed by an overview of the bubble evolutionary model in \autoref{subsec:modelOverview}.
The simple analytical model of {\MK} for stratified bubbles is revisited in \autoref{subsec:FBModel}, and generalized to include a late transition to isotropic expansion in \autoref{subsec:RBModel}.
We consider the effect of a circumgalactic medium (CGM) shocked by the RBs on the subsequent evolution of the FBs in \autoref{subsec:secondBurstModel}.
Both FB and RB outburst parameters are constrained using broadband observations in \autoref{subsec:estimates}, and by comparing the observed bubble morphologies to the analytic projection of the model along the line of sight in \autoref{subsec:bubbleProj}.

\subsection{Setup: Galactic model and jet injection}
\label{subsec:setup}

We adopt a cylindrically symmetric Galactic model, identical to that used in {\MK}, as outlined in \autoref{appendix:galModel}.
A rigid gravitational potential arises from baryonic and dark matter components. The density profile incorporates the Galactic disk, a central molecular zone, and the halo. We adopt spherical $\left(r, \theta,\phi\right)$ coordinates for numerical simulations and cylindrical coordinates $\left(R,\phi,z\right)$ for analytic modeling, with the GC at the origin and $\theta=0$ parallel to the $z$ axis. The azimuthal coordinate $\phi$ is frozen in both 2D simulations and analytic modeling. For simplicity, we focus on the $z>0$ hemisphere, assuming mirror symmetry about the Galactic plane unless otherwise stated.

Each GC jet is modeled as a radially-directed outflow of energy $E_j/2$ and a half-opening angle $\theta_j$, such that the total energy released in both hemispheres during a bipolar outburst is $E_j$. The jet is implemented numerically, as in {\MK}, by injecting momentum and energy at a radius $r=0.1$ kpc, with radial velocity $v_j$, over a time duration $\Delta t_j$. Successive outbursts are temporally separated by $\tdel$. The following injection parameter normalizations are introduced: $E_{56} \equiv E_j/10^{56}\erg$, $\Delta t_{-2} \equiv \Delta t_j/10^{-2}\Myr$, $\theta_5 \equiv \theta_j/5^\circ$, and $\beta_{-2} \equiv v_j/0.01 c$, where $c$ is the speed of light.

The relaxed CGM density profile may be approximated close to the $\theta=0$ axis and far from the disk as a $\rho(z) \simeq \rho_0 \left(z/10\kpc\right)^{-\alpha}$ planar power-law. We adopt $\rho_0 \simeq 4.4\times 10^{-4} m_p\text{ }\ccinv$ and $\alpha = 3/2$, where $m_p$ is the proton mass, consistent with the modified $\beta$-model inferred from \ion{O}{VIII} and \ion{O}{VII} line emissions \citep{MillerBregman2015}.
Such a quasi-planar CGM density profile motivates a stratified analytical description of the Galactic bubbles, presented for the FBs in {\MK}. Although this quasi-1D model is oversimplified, we demonstrate that it captures the basic properties of the simulated bubbles, including the high-latitude RBs.

\subsection{Three stages of bubble evolution}
\label{subsec:modelOverview}

A bubble inflated in the CGM by a collimated outburst initially evolves ballistically: its head moves at an approximately constant vertical (\ie along $z$) velocity, while its lateral (along $R$) expansion is much slower, such that $R_b\ll z_H$ (subscripts $b$ for bubble and $H$ for head).
Ballistic motion corresponds to a bubble shape $R_b(z)$ that reasonably matches in projection the observed FBs (\MK), and to a lesser extent also the RBs; see \autoref{subsec:bubbleProj}.
This ballistic stage is followed by a slowdown phase, where the head velocity decreases, approximately as a temporal power-law, and the laterally-inflated bubble becomes increasingly spherical. There is a range of parameters where slowing-down bubbles approximately reproduce FB and RB observations, although the viability of the former is quite limited even in the absence of the RBs (\MK) and practically ruled out in their presence (\autoref{subsec:secondBurstModel}).

The transition between the ballistic and slowdown phases occurs when the height $z_H$ of the bubble reaches a characteristic value $z_s$ (subscript $s$ for transition to slowdown), where the energy deposited in the mass
\begin{equation}
    M_\mathrm{\su} \simeq \pi\theta_j^2\int_0^{z_H}z^2\rho(z)dz
    \simeq
    \frac{\pi}{3-\alpha}C_\rho\theta_j^2 z_H^{3-\alpha}
    \label{eq:sweptMass}
\end{equation}
swept up by the head of the bubble exceeds the injected energy.
Here, we defined a constant $C_\rho \equiv (10 \kpc)^\alpha \rho_0\simeq 4.0\times 10^6 \mbox{ g cm}^{-3/2}$.
Namely, $M_\mathrm{\su} v_j^2/2 \simeq E_j/2$ at the transition height
\begin{equation}
z_s \simeq \left[ \frac{(3-\alpha) E_j}{\pi C_\rho v_j^2 \theta_j^2}\right]^{\frac{1}{3-\alpha}} \simeq 10.5 \, \xi^{2/3} \kpc \label{eq:Lc}
\end{equation}
and time
\begin{equation}
t_s = \frac{z_s}{v_j} \simeq 3.4~\vJ^{-1}\,\xi^{2/3} \Myr \label{eq:Tc}\,,
\end{equation}
where we introduced a dimensionless parameter
\begin{equation}
    \xi \equiv \frac{3\EJ}{\paran{\vJ\thJ}^2} \label{eq:xiDef}
\end{equation}
regulating the ballistic-to-slowdown transition.
The numerical factor in \autoref{eq:xiDef} is chosen such that for $\xi \simeq 1$, the transition $z_s\simeq 10\kpc$ approximately matches the FB height; then $\xi\ll1$ ($\xi\gg1$) FBs propagating into an undisturbed CGM are slowing down (still ballistic).

Beyond this transition, for $z_H \gtrsim z_s$, the bubble generally expands faster laterally than vertically. During this stage, the bubble height evolves approximately as a $z_H \sim z_s (t/t_s)^{\tau_z}$ power-law of index $\tau_z \ll 1$, as shown in {\MK}, and verified numerically therein for the FBs and below more generally. At time $t_\circ$ after the outburst, this lateral expansion renders the bubble quasi-spherical, in the sense that its half height $z_H/2$ equals its half-height width $R_b(z_H/2)$, such that $R_b(z_\circ/2) = z_\circ/2$ (subscript $\circ$ representing this stage, where each bubble is approximately circular in its hemisphere). Therefore, $z_\circ \equiv z_H(t_\circ) \sim z_s\left(t_\circ/t_s\right)^{\tau_z}$.
As numerically demonstrated in \autoref{sec:sims}, the maximal non-projected bubble width $R_\mathrm{max}\simeq R_b(z_H/2)$ throughout the bubble evolution, so we may use either quantity as convenient.

The bubble's non-projected aspect ratio, defined as
\begin{equation}
    \mathcal{A} \equiv \frac{2R_\mathrm {max}}{z_H}\,,\label{eq:aspDef}
\end{equation}
is therefore unity at $t\simeq t_\circ$.
A higher degree of spherical symmetry with respect to the GC entails larger $\mathcal{A}$, approaching 2 in the limit where the bipolar flow relaxes into a fully spherical distribution.
In this limit, a bubble bounded by a sufficiently strong shock approximately follows the spherical Sedov-Taylor-von Neumann similarity solution for a power-law medium, as the Galactic potential and implied density profile approach spherical symmetry
far ($\gtrsim 10\kpc$) from the disk. Therefore, in this phase, the evolution
of a sufficiently energetic bubble is dictated by the injected energy, and its head approximately follows $z_H \propto t^{2/\paran{5-\alpha}}=t^{4/7}$. We refer to $1\lesssim \mathcal{A} < 2$ bubbles as quasi-spherical, and derive estimates for $z_\circ$ and $t_\circ$ in \autoref{subsec:RBModel}.

As shown below, observations imply that the RBs are still approximately ballistic or just recently started slowing down, whereas the FBs can safely be approximated as ballistic. Indeed, the presence of the RBs implies a rarefied FB upstream, severely limiting the $5\kpc\lesssim z_s\lesssim 10\kpc$ slowing down {\MK} branch of FB solutions.

\subsection{Stratified ballistic and slowdown evolution}
\label{subsec:FBModel}

In the ballistic phase, the bubble head vertically rises through the CGM with a velocity that remains approximately constant and equal to the injection velocity, $v_H(t<t_s) \simeq v_j$, so the bubble height is given by
\begin{equation}
    z_H(t<t_s) \simeq v_jt \label{eq:ballCond}\,.
\end{equation}
Therefore, the age (defined as the time since injection begun) of a ballistic bubble observed at a height $z_H$ can be estimated as
\begin{equation}
    \tage(z_H<z_s) \simeq \frac{z_H}{v_j} \simeq 3.3z_{10}\beta_{-2}^{-1}\Myr\label{eq:ageBall}\,,
\end{equation}
where $z_{10} \equiv z_H/10\kpc$. The shock Mach number at the bubble head is given by
\begin{equation}
    \Mach_H(t<t_s) \simeq \frac{v_j}{c_s} \simeq 13~\beta_{-2}\Tnorm^{-1/2} \label{eq:machHeadBall}\,,
\end{equation}
where $c_s$ is the upstream speed of sound, and we defined $\Tnorm \equiv k_BT/0.2\keV$ upstream.

Approximately equating the lateral kinetic energy in each Eulerian $z$-slice to its thermal energy as deposited by the head shock, {\MK} showed that the maximal half-width of the bubble is given by
\begin{equation}
    R_\mathrm{max} (t<t_s) \simeq  R_b(z_H/2,t) \simeq \frac{v_jt}{2}(3\theta_j)^{1/2} \label{eq:RmaxBall}\,,
\end{equation}
so Eqs.~\eqref{eq:aspDef}, \eqref{eq:ballCond} and \eqref{eq:RmaxBall} imply an aspect ratio
\begin{equation}
    \mathcal{A}(t<t_s) \simeq (3\theta_j)^{1/2}\label{eq:aspRatioBall}\,,
\end{equation}
preserved during this ballistic phase.

The outburst energy can be estimated by adopting a Primakoff-like model \citep{Keshetgurwich18} for the bubble pressure
\begin{equation}\label{eq:P}
  P_b(\vect{r})\simeq (r/r_{sh})^3P_d(\vect{r}_{sh})\simeq (r/r_{sh})^3 P_d
\end{equation}
and mass density
\begin{equation}\label{eq:P}
  \rho_b(\vect{r})\simeq \frac{r}{r_{sh}}\rho_d(\vect{r}_{sh})
\end{equation}
profiles along a radial ray emanating from the GC and terminating at the radius $r_{sh}$ of the shock, assumed strong.
Here, the downstream (subscript $d$) pressure is approximately constant along the entire shock surface \citep{Keshetgurwich18}, and the kinetic-to-internal energy ratio is constant along a radial ray emanating from the GC,
\begin{equation}\label{eq:Frac}
  \frac{\rho_b v^2/2}{P_b/(\Gamma-1)} = \frac{(\Mach^2-1)^2}{(\Mach^2+3)(\Mach^2-1/5)}\simeq 1 \coma
\end{equation}
so the total outburst energy is
\begin{eqnarray}
\!\!\!  E_j & \simeq & 4\int_{z>0} \frac{P_b \,dV}{\Gamma-1} \simeq
  \frac{4\pi P_d}{3(\Gamma-1)}\int_0^{\pi/2} r_s(\theta)^3\sin{\theta}\,d\theta \nonumber \\
  & \simeq & \pi \left(1+\frac{2\theta_j}{3}\right) \theta_j z_H^3 \frac{P_d}{\Gamma-1}
  \simeq \frac{4\pi}{3}  \frac{R_{\Max}^2 z_HP_d}{\Gamma-1}\fin
  \label{eq:TotE}
\end{eqnarray}

The subsequent, slowdown phase is characterized by strong deceleration of the head due to the inertia of the swept-up mass. Conservation of $z$-momentum implies (\MK)
\begin{equation}
    z_H(t_s<t<t_\circ)\simeq\sqb{1 + \paran{-1+t/t_s}\tau_z^{-1}}^{\tau_z}z_s
    \label{eq:zHSlow}\,,
\end{equation}
where $0<\tau_z\ll1$, depending on the CGM density parameter $\alpha$ and the structure of the head. The bubble age at this phase is therefore
\begin{equation}
    \tage(z_s<z_H<z_\circ)
    \simeq
    \sqb{\tau_v + \tau_z\paran{z_H/z_s}^{1/\tau_z}}
    t_s
    \label{eq:slowAge}\,,
\end{equation}
where $\tau_v \equiv 1-\tau_z$ is near unity.
At late, $t_s\ll t \lesssim t_\circ$ slowdown times, $z_H \simeq
\sqb{t/\paran{\tau_zt_s}}^{\tau_z} z_s$ and so $v_H \simeq \sqb{t/\paran{\tau_zt_s}}^{-\tau_v} v_j$. Therefore, $\tau_z$ and $\tau_v$ are the
approximate late-time power-law indices of $z_H$ and $v_H$ (respectively; hence their $z$ and $v$ subscripts). At this stage, the head velocity can be written as
\begin{equation}
    v_H(t_s\ll t\lesssim t_\circ) \simeq v_j\paran{z_H/z_s}^{-\tau_v/\tau_z} \label{eq:slowVel}\,,
\end{equation}
which implies a head Mach number
\begin{equation}
    \Mach_H(t_s\ll t\lesssim t_\circ) = \frac{v_H}{c_s}  \simeq
    13~\beta_{-2}\Tnorm^{-1/2}\paran{\frac{z_H}{z_s}}^{-\tau_v/\tau_z}
    \, .
    \label{eq:machHeadSlow}
\end{equation}
In the {\MK} approximation, the maximal half-width of the bubble in the slowdown phase is found near $z_s$ and evolves as
\begin{equation}
    R_\mathrm{max}
    (t_s<t<t_\circ)
    \simeq \paran{3\theta_jv_jz_st}^{1/2}
    \label{eq:RMaxSlow}\,,
\end{equation}
corresponding to
\begin{equation}
    \mathcal{A}
    (t_s<t<t_\circ)
    \simeq 2\sqrt{3\theta_j\tau_z}\paran{z_H/z_s}^{-1+1/(2\tau_z)} \,,
    \label{eq:aspRatioEvol}
\end{equation}
as seen from Eqs. \eqref{eq:zHSlow} (for $t\gg t_s$) and \eqref{eq:RMaxSlow}.

\subsection{Quasi-spherical expansion}
\label{subsec:RBModel}

During the slowdown phase, the lateral $R_{\mathrm{max}} \propto t^{1/2}$ expansion of the bubble is faster than its $z_H \propto t^{\tau_z}$ vertical rise, as $\tau_z \ll 1$. Namely, the bubble widens significantly while its head velocity decelerates. At time $t_\circ$ after jet injection, the bubble attains $\mathcal{A}\simeq 1$ quasi-sphericity. Subsequently, the aspect ratio continues to grow and may asymptotically approach $\mathcal{A}=2$, which corresponds to one spherical bubble about the GC. As the Galactic potential and CGM density far ($\gtrsim 10 \kpc$) from the disk approach spherical symmetry, an $\mathcal{A}\simeq 2$ bubble with a sufficiently strong shock is governed by the spherical Sedov-Taylor-von Neumann (STvN) solution
\begin{equation}
    z_H(t> t_\circ)
    \sim
    \paran{E_j/C_\rho}^{2/7}t^{4/7} \label{eq:selfSimEvol}\,,
\end{equation}
up to a factor of order unity ignored below.

Sufficiently close to $\mathcal{A}=2$, an amply energetic quasi-spherical bubble thus expands with an approximately isotropic velocity
\begin{equation}
    v_H(t> t_\circ)
    \simeq
    \dot{z}_H
    \sim
    300~E_{ 56}^{2/7}t_{20}^{-3/7} \label{eq:vHSelfSim} \km~\sinv\,,
\end{equation}
where $t_{20} \equiv t/20\Myr$, corresponding to
\begin{equation}
    \Mach_H(t>t_\circ) = \frac{v_H}{c_s}
    \sim
    1.4~E_{56}^{2/7}\Tnorm^{-1/2}t_{20}^{-3/7} \label{eq:machHSelfSim}\,.
\end{equation}
The age of such an STvN bubble is of order
\begin{equation}
    \mathbb{T}(z_H>z_\circ)
    \simeq \frac{4}{7}\frac{z_H}{v_H} \sim 26\left(\frac{\Mach}{2}\right)^{-1}\left(\frac{z_{10}}{2}\right)\Tnorm^{-1/2} \Myr \label{eq:selfSimAge}\,.
\end{equation}
This approximation does not account for the initial ballistic and slowdown phases of a jetted explosion, as it applies asymptotically to a point-like, spherical explosion; a more careful estimate is provided below in \autoref{eq:ageSedov}.

\autoref{eq:aspRatioEvol} indicates that the bubble becomes quasi-spherical, \ie $\mathcal{A} = 1$, at a height
\begin{equation}
    z_\circ \simeq \left(12 \theta_j\tau_z\right)^{\tau_z/\left(2\tau_z-1\right)} z_s
    \simeq  1.3 \theta_5^{-1/8} z_s \,, \label{eq:approxZcirc}
\end{equation}
where $\tau_z = 0.1$ is adopted in the last approximation. The implied proximity of $z_\circ$ and $z_s$ demonstrates the modest change in $z_H$ as the bubble widens after $t_s$. The time $t_\circ$ of quasi-sphericity can be estimated by plugging $z_H=z_\circ$ in \autoref{eq:slowAge},
\begin{align}
    t_\circ & \simeq
    \sqb{\tau_v +
    \paran{12\theta_j\tau_z}^{1/\paran{2\tau_z-1}}
    \tau_z}
    t_s
    \nonumber\\
    & \simeq \paran{0.9 + 1.7\theta_5^{-5/4}}t_s\,,\label{eq:slowLifeTime}
\end{align}
where we again adopted $\tau_z=0.1$ in the last result, although such an approximation worsens as $\mathcal{A}$ approaches or exceeds unity.
Thus, the $\sim t_s$ duration of the $t_s<t<t_\circ$ slowdown phase is comparable to that of the $t<t_s$ ballistic stage, but $z_H$ increases only by $\sim30\%$; a non-circularized ($z_H<z_\circ$) bubble implies
$\xi>0.63z_{10}^{3/2}\theta_5^{3/16}$.

As shown numerically in \autoref{subsec:singleJets}, after the bubble reaches $\mathcal{A}=1$, it slowly evolves towards \autoref{eq:selfSimEvol}. The age of a sufficiently spherical bubble born from a jetted outburst is
\begin{align}
 \tage(z_H>z_\circ) &\simeq t_\circ + (C_\rho/E_j)^{1/2} (z_H^{7/4}-z_o^{7/4}) \label{eq:ageSedov} \\
 &\simeq 14\left( z_{10}^{7/4}-1.4\xi^{7/6} \right)E_{56}^{-1/2}\Myr\,, \nonumber
\end{align}
where the last expression adopts for simplicity $\theta_5=1$ and $\tau_z = 0.1$; here, the second term in the parenthesis becomes more negative with increasing $\xi$, but so does $z_H>z_s\propto \xi^{2/3}$ in the positive first term.

Finally, the downstream temperature $T_d$ at the head of a sufficiently spherical bubble is obtained from the Rankine-Hugoniot jump conditions as
\begin{eqnarray}
    k_BT_d & \simeq & m v_H^2\paran{\frac{3}{16} + \frac{21}{40}\Mach_H^{-2}-\frac{9}{80}\Mach_H^{-4}}
    \label{eq:tempIonDown}
    \\\nonumber
    & \simeq & 0.13 \paran{1 + 2.8\Mach_H^{-2} - 0.6\Mach_H^{-4}}\ E_{56}^{4/7}t_{20}^{-6/7}\keV\,,
\end{eqnarray}
where $m \simeq 0.6m_p$ is the mean particle mass, consistent with the cosmic value for low-metallicity, ionized gas.

\subsection{FBs in a pre-shocked CGM}
\label{subsec:secondBurstModel}

Consider two collimated GC outbursts, as motivated by \autoref{sec:Intro}: the first launched at $t=0$ to produce the RBs, and the second launched at $t=\tdel[01]$, resulting in the FBs. As the FBs were shown (\MK) to require a collimated outburst directed nearly perpendicular to the Galactic plane, and the RBs share a similar albeit more evolved morphology (see \autoref{subsec:bubbleProj}), the same applies to the RB outburst.
For simplicity, we assume a relaxed CGM before the first --- but not the second --- outburst.

The disturbed, shocked gas downstream of the RBs gradually relaxes, and in the absence of another outburst would asymptotically approach the initial, unperturbed CGM conditions dictated by the rigid potential.
The corresponding relaxation time at a height $z$,
\begin{equation}
    \trelax > \frac{z}{c_s} \simeq 42~ z_{10}\Tnorm^{-1/2}~\Myr\,, \label{eq:trelax}
\end{equation}
is longer than $\tdel[01]$ during much of the evolution of the FBs, as shown in \S\ref{sec:sims}.
Therefore, in order to model the FBs as a function of their upstream ambient medium, one must take into account how the RBs affect the CGM in their downstream.

The main modification is a lower density and a higher temperature upstream of the FBs, with respect to the relaxed CGM conditions assumed upstream of the RBs. The lower density implies a smaller swept-up bubble mass $M_\mathrm{\su}$ in \autoref{eq:sweptMass} at a given $z_H$. Consequently, the ballistic-to-slowdown transition height $z_s$ and age $t_s$ are increased for the FBs with respect to Eqs.~\eqref{eq:Lc} and \eqref{eq:Tc}, which assume a relaxed CGM upstream. These elevated $z_s$ and $t_s$ shrink the {\MK} phase space of possible slowdown FB solutions, as they require an even larger $\xi$, nearly ruling out such solutions.

For instance, if one assumes for simplicity that the power-law slope of the upstream $\rho\propto z^{-3/2}$ density profile is preserved ahead of the FBs, such that $\rho$ diminishes uniformly by a constant multiplicative factor $F$, of order $3\mbox{--}20$ according to our RB simulations, then the FB transition heights $z_s$ and $z_\circ$ and times $t_s$ and $t_\circ$ all increase by the same constant factor $F^{2/3}\sim 2\mbox{--}7$.
Then the FB slowdown condition $z_s<z_{\FB}\simeq 10\kpc$ implies a stricter, $\xi \lesssim 0.93F^{-1}$ condition, where we henceforth use FB and RB subscripts to distinguish between the respective parameters of the two outbursts.

The presence of the RBs renders ballistic solutions more plausible for the FBs, and their alternative slowdown solutions more restricted, also by elevating the temperature upstream of the FBs, by a factor of order $F$.
The ballistic FB constraint $\Mach_{\FB}\gtrsim 14(T_{0.2}F)^{-1/2}(\tage_{\FB}/3\Myr)^{-1}z_{10}$ thus loosens, where we combined \autoref{eq:machHeadBall} with $\beta_j\simeq 0.011(\tage_{\FB}/3\Myr)^{-1}z_{10}$, which increases for spectrally-young \citep{Keshet+2023} FBs according to \autoref{eq:ageBall}. Similarly, for slowdown FBs, $\Mach_{\FB}$ diminishes not only by a high power of $(z_{\FB}/z_s)$ in \autoref{eq:machHeadSlow}, but also by $F^{-1/2}$, implying that $z_s$ must be even closer to $z_\mathrm{FB}$ than constrained by {\MK}. We conclude that the FBs are most likely ballistic, otherwise they transitioned to slowdown just below $z_{\FB}$.
This conclusion is supported by the ballistic FB morphology shown in \S\ref{subsec:bubbleProj}.

\subsection{Observation-based constraints}
\label{subsec:estimates}

\subsubsection{FB inferences}
\label{subsubsec:ObsFBs}

The presence of the RBs strengthens the {\MK} constraints on the FB outburst parameters because, as discussed in \S\ref{subsec:secondBurstModel}, it renders it safer to approximate the FBs as presently still ballistic.
The observed FB edge morphology thus constrains $\theta_j\simeq 4^\circ$ for its outburst (\MK), corresponding by \autoref{eq:aspRatioBall} to $\mathcal{A}\simeq 0.5$ and hence $R_\mathrm{max}\simeq z_H\mathcal{A}/2\simeq 2.5z_{10}\kpc$, valid for different FB deprojections \citep{Keshetgurwich17} and models \cite[\MK]{Sarkaretal15b, Zhang+2020} as they recover a similar $z_{10}\simeq 1$ bubble height even when its edge is not modeled as a forward shock.

The injection velocity of the FBs can be inferred from their $\sim$few Myr spectral age \citep{Keshet+2023}, implying by \autoref{eq:ageBall} that $\beta_j\simeq 0.011(\tage/3\Myr)^{-1}z_{10}$, or constrained by their high, $\Mach_{\FB}\gtrsim 5$ Mach numbers \citep{Keshetgurwich17}, using \autoref{eq:machHeadBall} to yield $\beta_j\simeq 0.012(\Mach_{\FB}/5)(\Tnorm F_{10})^{1/2}$, where we defined $F_{10}\equiv F/10$. Combined, these results indicate that $\Mach_{\FB}\simeq 4.5(\tage/3\Myr)^{-1}(\Tnorm F_{10})^{-1/2}z_{10}$ is consistent with the high-Mach inferred from observations for $F\lesssim 10$, disfavoring small $z_{\RB}\lesssim 15\kpc$ deprojections of slowdown RBs in which $F\simeq 30$ is typically found in \S\ref{sec:sims} to be large.

Similarly, the $\xi F \gtrsim 0.93z_{10}^{3/2}$ ballistic FB condition can be written in terms of injected energy either as $E_{56}\gtrsim 0.023(\tage_{\FB}/3\Myr)^{-2}F_{10}^{-1}z_{10}^{7/2}$ based on age constraints or independently of $F$ as $E_{56}> 0.029 (\Mach_{\FB}/5)^2\Tnorm z_{10}^{3/2}$ based on Mach number constraints.
Overall, the elevated $F>1$ inferred from the presence of the RBs renders the FBs faster and younger (for a given $\Mach_{\FB}$), or less energetic (for a given $\tage_{\FB}$). Similar constraints can be obtained more robustly, even without assuming ballistic FBs, from the $\xi F \gtrsim 0.60 z_{10}^{3/2}$ condition, arising because their non-circularized morphology implies that $z_\circ\gtrsim z_H$.

Such modeling places only lower limits on the total FB outburst energy, as do energy estimates of individual bubble components.
The most direct estimate, $10^{55\mbox{--}56}\erg$ in thermal electrons, derives from the FB X-ray shells, but thermal ions could carry $\sim10$ times more energy \citep{Keshetgurwich18}.
Adopting a strong shock where $P_d\simeq (3/4)\rho(z_H) v_j^2$, the Primakoff-like approximation \eqref{eq:TotE} indicates that
\begin{equation}
E_{\FB}\simeq 5\times 10^{55} \beta_{-2}^2z_{10}^{3/2}\theta_4 F_{10}^{-1}\erg \coma
\end{equation}
and thus $\xi\simeq 2.4z_{10}^{3/2}(\theta_4F_{10})^{-1}$.
We conclude that the FB outburst carried $\sim 10^{56}\erg$, up to an uncertainty factor $\sim4$, and that at $z=100\pc$ the collimated flow had $\theta_j\simeq 4\degree$ and $\beta_j\simeq 10^{-2}$, up to an uncertainty factor $\sim2$. Even in the regime where $\xi\simeq 0.5$ drops below unity, $F\gtrsim3$ keeps the FBs ballistic.

\subsubsection{RB inferences}
\label{subsubsec:ObsRBs}

The RBs are more difficult to deproject than the FBs, with plausible projections of very different $z_{\RB}$ still acceptable, as shown in \autoref{subsec:singleJets}. Nevertheless, the width $R_{\mathrm{max}}$ of the RBs is robustly deprojected as $7\mbox{--}8\kpc$, as shown in \autoref{subsec:singleJets} and supported by simulations  \citep{Yang+2022, Zhang+2025} and even spherically symmetric RB modeling \citep{Predehletal2020}.
Indeed, as the RBs reach latitudes close to the Galactic poles, their maximal radius should be only slightly smaller than the $R_\odot \simeq 8.5$ kpc Galactocentric radius of the sun.
Another observational constraint on the RBs is their strong forward shock, nominally taken as $\Mach_{\RB}\simeq 4$, although $3<\Mach_{\RB}<5$ values are also possible \citep{KeshetGhosh26}. Their $\mathcal{A}=2R_{\Max,\RB}/z_{\RB}<2R_{\Max,\RB}/z_{\FB}\simeq 1.5$ aspect ratio indicates that the RBs are not yet sufficiently spherical to be considered deep in the STvN phase, so one should consider the earlier, \ie ballistic or slowdown, regimes.

The slowdown phase is sensitive to upstream conditions, as even small changes in $\tau_z$ lead to very different $z_H$ and $\Mach$ as the bubble evolves.
The similar projected heights of the north vs. south RBs, despite their different upstream conditions as reflected in their different radio and \gama-ray signatures, thus indicate that they are not deep in the slowdown phase.
So, like the FBs, the RBs too are either ballistic, requiring $\xi>1.7 (z_H/15\kpc)^{3/2}$, or just recently started slowing down ($\xi\simeq 1$). In any case, combining the slowdown Mach \autoref{eq:machHeadSlow} with $z_{\RB}>z_{\FB}$ implies that for the RBs
\begin{equation}\label{eq:RBsXi}
  \xi\gtrsim 0.76 \left(\frac{\Mach_H}{4}\beta_{-2}^{-1}T_{0.2}^{1/2}\right)^{1/6}_{\RB} z_{10,\FB}^{3/2} \coma
\end{equation}
where we again took $\tau_z=0.1$, depending only weakly on CGM conditions and the unknown $\beta_j$. Thus, $\xi$ cannot be small for the RBs.

In the ballistic case, the RB width constrains
\begin{equation}\label{eq:RBsConstA}
  \beta_{-2}   \simeq 0.68\left(\frac{t}{15\Myr}\right)^{-1}
  \left(\frac{R_{\Max}}{8\kpc}\right)\theta_5^{-1/2}
\end{equation}
to be near unity according to \autoref{eq:RmaxBall}, so the energy implied by \autoref{eq:xiDef},
\begin{equation}\label{eq:RBsEA}
\!\!\!  E_{\RB} \simeq 3.1\times 10^{55}\theta_5
  \left(\frac{R_{\Max}}{8\kpc}\right)^2
  \left(\frac{t}{15\Myr}\right)^{-2}
  \left(\frac{\xi}{2}\right) \erg \, ,
\end{equation}
is linear in $\xi\gtrsim 2$.
The projection of the RBs, derived for the stratified model in \autoref{subsec:bubbleProj} and verified by simulations in \autoref{subsec:singleJets}, indicates that ballistic bubbles are viable only for $\{\theta_\RB, z_\RB\} \simeq \{4\degree, 30\kpc\}$ within a factor of $\sim 2$.
Therefore, we may use the Mach number constraint in \autoref{eq:machHeadBall} to obtain robust ballistic estimates
\begin{equation}\label{eq:betaRB}
  \beta_{\RB}\simeq 3.1\times 10^{-3} \left(\frac{\Mach_{\RB}}{4}\right)T_{0.2}^{1/2}\coma
\end{equation}
\begin{equation}\label{eq:tRB}
  \tRB\simeq 32\left(\frac{z_\RB}{30\kpc}\right)\left(\frac{\Mach_{\RB}}{4}\right)^{-1}T_{0.2}^{-1/2}\Myr\coma
\end{equation}
and
\begin{equation}\label{eq:RBsEB}
  E_{\RB} \simeq 0.63\times 10^{55} \theta_5^2 \left(\frac{\Mach_{\RB}}{4}\right)^2 \left(\frac{\xi}{2}\right) T_{0.2} \erg \coma
\end{equation}
again for $\xi\gtrsim 2$.

In the slowdown case, the RB width constrains, via \autoref{eq:RMaxSlow},
\begin{equation}\label{eq:RBsConst}
  \beta_{-2} \theta_5  \simeq 0.51\left(\frac{t}{15\Myr}\right)^{-1}
  \left(\frac{R_{\Max}}{8\kpc}\right)^2 \xi^{-2/3}
\end{equation}
to be near unity, so the RB energy implied by \autoref{eq:xiDef},
\begin{equation}\label{eq:RBsE}
  E_{\RB} \simeq 0.86\times 10^{55}
  \left(\frac{R_{\Max}}{8\kpc}\right)^4
  \left(\frac{\tRB}{15\Myr}\right)^{-2}
  \xi^{-1/3} \erg \, ,
\end{equation}
depends weakly on $0.8\lesssim\xi<1$, which is tightly constrained anyway in this regime.

The morphology of the RBs, unlike that of the FBs, shows deviations from ballistic evolution and evidence for some slowdown, as discussed below in \S\ref{subsec:bubbleProj}, so their $\xi$ cannot be much larger than unity. On the other hand, the RBs are inconsistent with substantial slowdown, in terms of morphology (\S\ref{subsec:bubbleProj}), approximate north-south symmetry, the large $\xi$ implied by their high Mach in \autoref{eq:RBsXi}, and the limited $F\lesssim 10$ rarefaction behind them implied by the high FB Mach number (\autoref{subsubsec:ObsFBs}). We conclude that the RBs are at the onset of slowdown, such that the above ballistic and slowdown estimates both approximately hold with $\xi\sim 1$.
Therefore, the RBs have $E_j\simeq 10^{56}\erg$, $\tage\simeq 15\Myr$, $\beta_{-2}\simeq 0.5$, and $\theta_j\simeq 5\degree$, up to factor $\sim 2$ uncertainties.
Indeed, our simulations reproduce RB observations only if they are $\sim10$--$20\Myr$ old.
Note the inferred similarity of the RB and FB outburst parameters $E_j$, $\beta_j$, and $\theta_j$.
Hence, the ballistic Primakoff-like \eqref{eq:TotE} estimate
\begin{equation}
E_{56}\simeq 3.3\left(\frac{\Mach_{\RB}}{4}\right)^2\left(\frac{R_{\Max}}{8\kpc}\right)^2 T_{0.2} \left(\frac{z_\RB}{30\kpc}\right)^{-\frac{1}{2}}\erg
\end{equation}
imposes an upper limit for the RBs, corresponding to a high $\xi\simeq 85(z/30\kpc)^{7/2}(R_{\max}/8\kpc)^{-2}$.

\subsection{Stratified bubble projection}
\label{subsec:bubbleProj}

The edge of the stratified ballistic bubble follows (\MK)
\begin{equation}\label{eq:BallRb}
  x^2+y^2 = R_b^2(z)\simeq z^2\left[ 3\left(\frac{z_H}{z}-1\right)\theta_j+\theta_j^2 \right] \coma
\end{equation}
which can be projected with respect to the sun by $\{x,\tilde{y},z\}\cdot \bm{\nabla}[x^2+y^2-R_b(z)^2]=0$, to yield
\begin{equation}\label{eq:BallProj_l}
  \tan l(z) = \frac{x}{\tilde{y}}
  = \frac{ \sqrt{ [ 12 \Myeta - 4\zeta (3 - \theta_j) -
   9 \zeta \Myeta^2 \theta_j ] \zeta \theta_j} } {2 - 3 \zeta \Myeta \theta_j}
\end{equation}
and
\begin{equation}\label{eq:BallProj_b}
  \frac{1}{\tan^2 b(z)} = \frac{x^2+\tilde{y}^2}{z^2}
  = \frac{1}{\zeta^2} -  (3-\theta_j)\theta_j
\end{equation}
for a range of $0<z<z_H$.
Here, we placed the sun at Cartesian coordinates $\{x,y,z\}=\{0,-R_\odot,0\}$ such that $\tilde{y}\equiv y+R_\odot$, and defined normalized  $\zeta\equiv z/R_\odot$ and $\Myeta\equiv z_H/R_\odot$ lengthscales.
Conversely, we can estimate parameters such as $\theta_j$ and $z_H$ from an observed bubble projection, for example from its maximal (in absolute value) longitude $l_{\Max}>0$ and latitude $b_{\Max}>0$,
\begin{equation}\label{eq:BallProj_thj}
  \frac{2}{3}\theta_j = 1 -  \sqrt{1 - \left[\frac{4 \sin l_{\Max}}{3 \cos^2 \left(l_{\Max}\right) \tan b_{\Max}}\right]^2 }
\end{equation}
and
\begin{equation}\label{eq:BallProj_zH}
  3z_H = 2 R_\odot \sin \left(l_{\Max}\right) \sqrt{3\theta_j^{-1}-1} \fin
\end{equation}

One may similarly project the bubble in the slowdown regime, where (\MK)
\begin{equation}\label{eq:SlowRb}
  \frac{R_b^2(z,t)}{3z^2\theta_j}-\frac{\theta_j}{3} \simeq \begin{cases}
    -1+v_j t/z  & \mbox{ for } z<z_s \\
    \left(z_H/z\right)^{1/\tau_z}\tau_z & \mbox{ for } z>z_s
   \end{cases}
\end{equation}
is continuous (albeit not smooth) at $z_s$ for $z_H\simeq z_s^{\tau_v} (v_j t/\tau_z)^{\tau_z}$.
Due to its strong $R_b(z)$ dependence, the $z>z_s$ region is confined in projection within the $z\leq z_s$ edge, unless the jet is very narrow, with
$12\tau_z\theta_j^{-1}>9\zeta^2(z_H/z)^{1/\tau_z}-4(z_H/z)^{-1/\tau_z}$
typically corresponding to $\theta_j\lesssim 2\degree$.
The high-latitude part of the edge is thus given by $z=z_s$, and its projection by
\begin{equation}\label{eq:SlowProj_l}
  \tan l = \frac{\sqrt{\sin^2 l_{\Max}-\eta^2}}{1+\eta}
\end{equation}
and
\begin{equation}\label{eq:SlowProj_b}
  \tan b = \frac{(1-\sin l_{\Max})\tan b_{\Max}}{\sqrt{1+2\eta+\sin^2 l_{\Max}}}\,
\end{equation}
with varying $\eta\equiv y/R_\odot$.
Maximal $|b|$ at the minimal $\eta=-\sin l_{\Max}$ and maximal $|l|$ at $l'(\eta)=0$ yield the relations
\begin{equation}\label{eq:zsEdge}
 z_s = R_\odot (1-\sin l_{\Max})\tan b_{\Max}
\end{equation}
and
\begin{equation}\label{eq:QEdge}
 3\theta_j \tau_z \left(\frac{z_H}{z_s}\right)^{1/\tau_z} = \left( \frac{1/\tan b_{\Max}}{1-\csc l_{\Max}}\right)^{2} \coma
\end{equation}
where here we safely neglect the second term on the LHS of Eq.~\eqref{eq:SlowRb} far from the head.
This outer part of the edge connects continuously (albeit not smoothly) to the $z<z_s$ projection,
formally given by Eqs.~\eqref{eq:BallProj_l} and \eqref{eq:BallProj_b} but with $\Myeta$ replaced by $v_j t/R_\odot$.

\begin{figure}[h!]  \includegraphics[width=0.95\columnwidth, trim=0cm 0cm 0cm 0cm, clip=False]{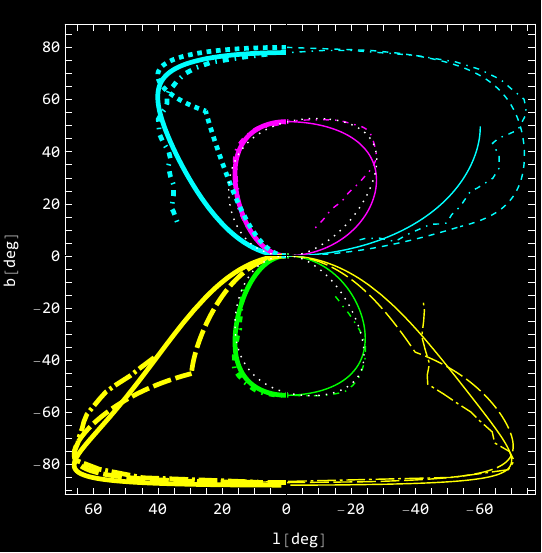}
    \caption{Coarse-grained edge-detector edges (dot-dashed) of FBs \citep{Keshetgurwich18} and RBs \citep{KeshetGhosh26}, and their analytically projected ballistic (solid) and slowdown (dashed) models, which are based on maximal $|l|$ and $|b|$ values in each sector: north (short dashing) vs. south (long dashing), east (thick curves) vs. west (thin).
    Ballistic bubbles with a westward tilt are also shown (dotted white curves) for the FB.
    \label{fig:edges1}
    }
\end{figure}

Figure \ref{fig:edges1} shows (as dot-dashed curves) the edges of the \emph{Fermi}-LAT FBs \citep{Keshetgurwich18} and the \emph{eROSITA} RBs \citep{KeshetGhosh26}, as inferred from coarse-grained edge detectors.
Ballistic (solid) and slowdown (dashed) models projected along the line of sight are shown in each sector for each edge, according to its observed $\{l_{\Max},b_{\max}\}$ values.
One does not expect an agreement between east and west sectors, as the bubbles are significantly stretched westward.
Some differences between north and south sectors are also expected, as the southern bubbles are fainter; in particular, the RB upstream is significantly more dilute in the south \citep{KeshetGhosh26}.

\subsubsection{FB edges}

The FB edges are fairly well reproduced by the projected ballistic model in the eastern sectors, where Eqs.~\eqref{eq:BallProj_thj} and \eqref{eq:BallProj_zH} indicate consistent $\theta_j\simeq 4\dgrdot4$ ($\theta_j\simeq 3\dgrdot8$) and $z_H\simeq 9.6\kpc$ ($z_H\simeq 10.4\kpc$) in the north (south), as anticipated above; these results align with numerical simulations (\MK).
The observed bubbles bend westwards at higher $|b|$, so a ballistic profile matching a western edge corresponds to different parameters than its eastern counterpart, and provides a weaker match to the observed edge.
Here, the inferred $l_{\Max}$ and $b_{\Max}$ correspond to $\theta_j\simeq 19\dgrdot8$ ($\theta_j\simeq 11\dgrdot2$) and $z_H\simeq 7.4\kpc$ ($z_H\simeq 8.9\kpc$) in the north (south).

The similar westward stretching of the RBs, the FBs, and the polarized lobes associated with the RBs (not with the FBs \citep{Keshet2024}) was previously attributed to a westward or southwestward Galactic wind of $\sim 200\km\se^{-1}$ velocity, disfavoring the possibility of a sufficiently strong asymmetry in the density profile \citep{Sofue94, Mou+2018, MouEtAl23Wind}.
Naively, the east--west asymmetry in the above $\theta_j$ values may be interpreted as a $\Delta \psi\sim 8\degree$ ($4\degree$) westward offset in the northern (southern) FB outflow, roughly corresponding to a horizontal velocity $v_j \tan(\Delta\psi)\sim 400\km\se^{-1}$ ($200\km\se^{-1}$).
Quantitatively, the figure illustrates (white dotted curves) the projection of FBs pushed westward at a velocity increasing linearly with $z$, from zero at $0.5\kpc$ to $v_{\Max}$ at $10\kpc$; such projections can be derived analytically starting from Eq.~\eqref{eq:BallRb} with $x$ shifted by a cumulative arbitrary offset $x_w(z)$.
The resulting westward elongation provides a reasonable match to the observed edges, with the same model parameters in respective east and west sectors.
This and similar wind models indicate that a horizontal wind driving the asymmetry would need to be fast, $v_{\Max}\gtrsim 400\km\se^{-1}$ ($300\km\se^{-1}$) across the northern (southern) FB.

However, a western $\gtrsim 200\km~\sinv$ wind is unlikely to surround the FBs, because it would (i) be too fast and possibly supersonic; and (ii) not remain fast after colliding with the strong outward shocks of the enclosing RBs.
A more natural explanation for the westward elongation is an eastward density gradient, allowing both RBs and FBs to expand faster towards their west; see \autoref{sec:discussion} for a plausibility argument.
Indeed, such a density gradient is already suggested by the enhanced brightness of the FBs and the RBs in their eastern sectors, despite their uniformly (at least across the former \citep{Keshetgurwich17}) strong shocks.
Quantitatively, a $\sim 50\%$ ($\sim 25\%$) faster westward shock propagation suffices to recover the observed asymmetry in the northern (southern) FB, producing projected bubbles very similar to those in the fast wind; an analytic derivation follows from Eq.~\eqref{eq:BallRb} with a multiplicative factor $w(\phi)$ on $x$.
As the westward shocks propagate faster, the accumulated mass may transition the bubbles faster to their slowdown regime in their western sectors; it is interesting to check if the RBs are consistent with such an evolution.

\subsubsection{RB edges}

For the RBs, argued in \ref{subsubsec:ObsRBs} to have intermediate, $\xi\sim 1$ values, the observed edges lie in general between ballistic and slowdown projections, with  eastern (western) edges in better agreement with the ballistic (slowdown) profiles as anticipated above.
In particular, the western edges suggest a break, resembling that of the slowdown regime, and the northwestern RB extends too far west to admit any ballistic solution.

The ballistic models corresponding to the $\{l_{\Max},b_{\Max}\}$ values of the eastern RB edges yields $\theta_j\simeq 4\dgrdot2$ ($2\dgrdot9$) in the north (south), with $z_H\simeq 22.9\kpc$ ($z_H\simeq 39.6\kpc$), whereas the southwest sector corresponds to $\theta_j\simeq 6\dgrdot2$ and $z_H\simeq 27.4\kpc$.
While the ballistic morphological fit is imperfect, these results suggest that to within a factor of $\sim2$, the RBs have $\theta_j\simeq 4\degree$ and $z_H\simeq 30\kpc$.

For slowdown RB projections, we may estimate $z_s$ from Eq.~\eqref{eq:zsEdge} and, if we adopt for simplicity the same $\theta_j\simeq 4\degree$ inferred from nearly ballistic RBs and the $\tau\simeq 0.1$ typically found in \S\ref{sec:sims}, also $z_H$ from Eq.~\eqref{eq:QEdge}.
The southern RBs then give $\{z_s,z_H\}/\mbox{kpc}\simeq \{21,25\}$ ($\{14,18\}$) in the east (west), whereas the northern $\{l_{\Max},b_{\Max}\}$ values correspond to $\{17,20\}$ ($\{1.6,3.3\}$).
The small lengthscales in the northwest sector suggest an early transition to slowdown, but may be biased by a westward expansion faster than incorporated in the nominal model.
The dependence of $z_H$ on $\theta_j$ and $\tau_z$ is weak; it changes by $<30\%$ even if these two parameters are simultaneously modified by any factor $<2$. Conversely, it is difficult to determine $\theta_j$ and $\tau_z$ from the morphology of a slowed-down bubble.

We conclude that morphologically, the RBs are at the onset of slowdown, reaching $z_H\sim 30\kpc$ within a factor of $\sim 2$, with higher (lower) values for the more ballistic (slowed down) bubbles.
In particular, this implies that the FBs are physically nested within the RBs, not only in projection.
The westward bulging of the RBs, reminiscent of their nested FB counterparts, likely arises from an eastward density gradient, largely preserved downstream of the RB shocks, rather than from a westward wind.

\section{Simulated Galactic bubbles}
\label{sec:sims}

We present a suite of purely hydrodynamic, 2D simulations of axisymmetric collimated outbursts from the GC, using \verb|PLUTO| \citep[version 4.0]{Mignoneetal07}. Some of our results are verified qualitatively using 3D simulations, briefly discussed in \autoref{appendix:3d}. We present single-outburst simulations reproducing the RBs in \autoref{subsec:singleJets}, and double-outburst simulations producing both RBs and FBs in \autoref{subsec:doubleJets}. All simulations are (unless otherwise stated) converged with resolution, as demonstrated in \autoref{appendix:convTests}.

We consider both ballistic ($\xi\gg1$; denoted $\mathcal{B}$-simulations) and slowdown ($\xi\ll1$; $\mathcal{S}$-simulations) bubbles; see \autoref{table:twinParams} for a list of simulations and their parameters.
Low energy, high velocity ($\xi \ll 1$) simulations require viscosity to suppress instabilities along the injected flow, effectively modifying the injection parameters governing the evolution modelled in \autoref{sec:burstModels}; see {\MK} and \autoref{appendix:transition}.
The intermediate, $\xi \simeq 1$ regime is of particular interest (see \autoref{sec:burstModels}), but here the diffusion-limited numerical time-steps become prohibitively small, so convergence cannot be demonstrated in the present setup. Some marginally converged results are presented nonetheless, simulated both with and without viscosity to bracket the outcome, and at different resolutions for plausible Richardson extrapolations.

We present the thermal structure of the simulated bubbles, as well as their projected hard (to minimize foreground), $2$--$10\keV$ bremsstrahlung emission assuming $0.2Z_\odot$ metallicity; a modified version of \verb|PASS| \citep{Sarkaretal15a} is used to construct the projected maps. The Mach number $\Mach_H$ at the head of each simulated bubble is inferred from the Rankine-Hugoniot conditions by measuring the $\theta \leq \theta_j$ averaged pressure jump across the bubble head. The shock morphology is compared to the bubble edges obtained by applying coarse-grained gradient filters to \emph{ROSAT} FB \citep{Keshetgurwich17} and \emph{eROSITA} RB \citep{KeshetGhosh26} data.

\subsection{RB simulations}
\label{subsec:singleJets}

We simulate RB-like bubbles produced by outbursts that are either high-energy and low-velocity ($\xi \gg 1$) or low-energy and high-velocity ($\xi \ll 1$). Our nominal $\xi \gg 1$ setup (denoted $\nomBall$, as it yields ballistic RBs) has parameters $E_{56}=30$, $\theta_5=1$, and $\beta_{-2}=0.8$, which correspond to $\xi \simeq 141$, and $\Delta t_{-2}=4$. Our nominal $\xi \ll 1$ setup ($\nomSlow$, producing slowing-down RBs) has parameters $E_{56}=0.1$, $\theta_5=0.8$, and $\beta_{-2}=8$, which correspond to $\xi \simeq 0.007$, and $\Delta t_{-2}=4$.

\begin{figure}[h!]
    \centering{
        \hspace{0cm}
        \raisebox{0.3mm}[0pt][0pt]{\includegraphics[width=0.263\textwidth]{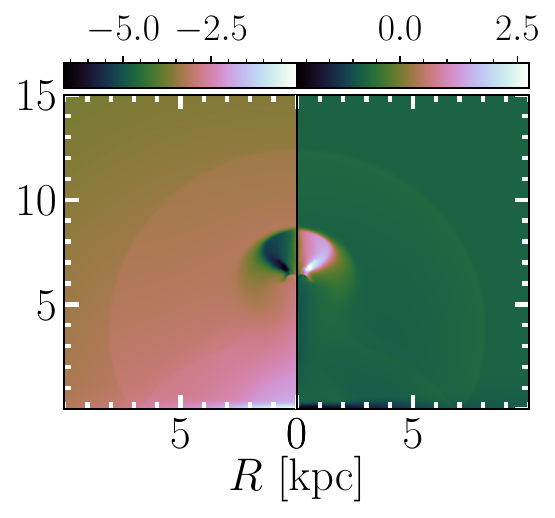}}
        \hspace{-9.25cm}\includegraphics[width=0.277\textwidth, trim=0cm 0cm 0cm 0cm]{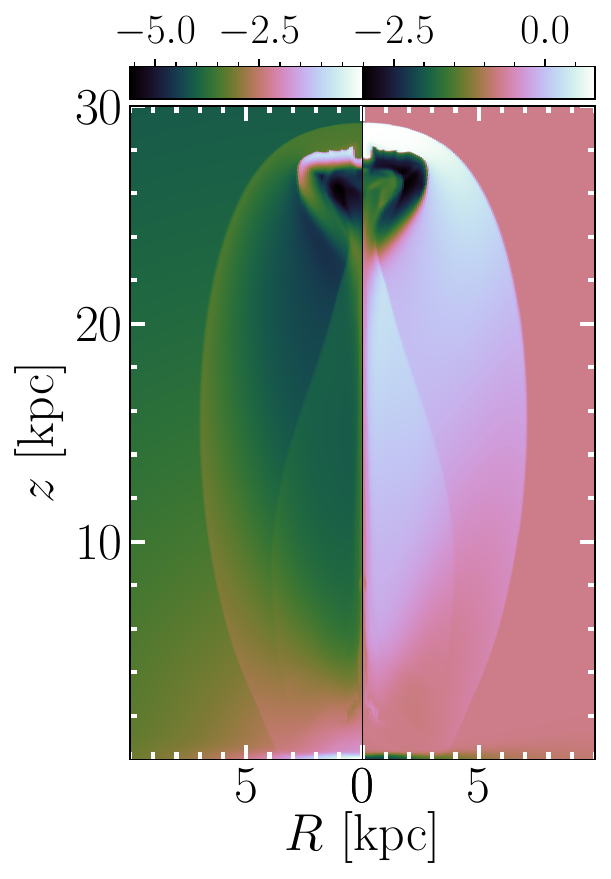}\\
        \hspace{0cm}\includegraphics[width=0.27\textwidth, trim=0cm 0cm 0cm 0cm, clip=False]{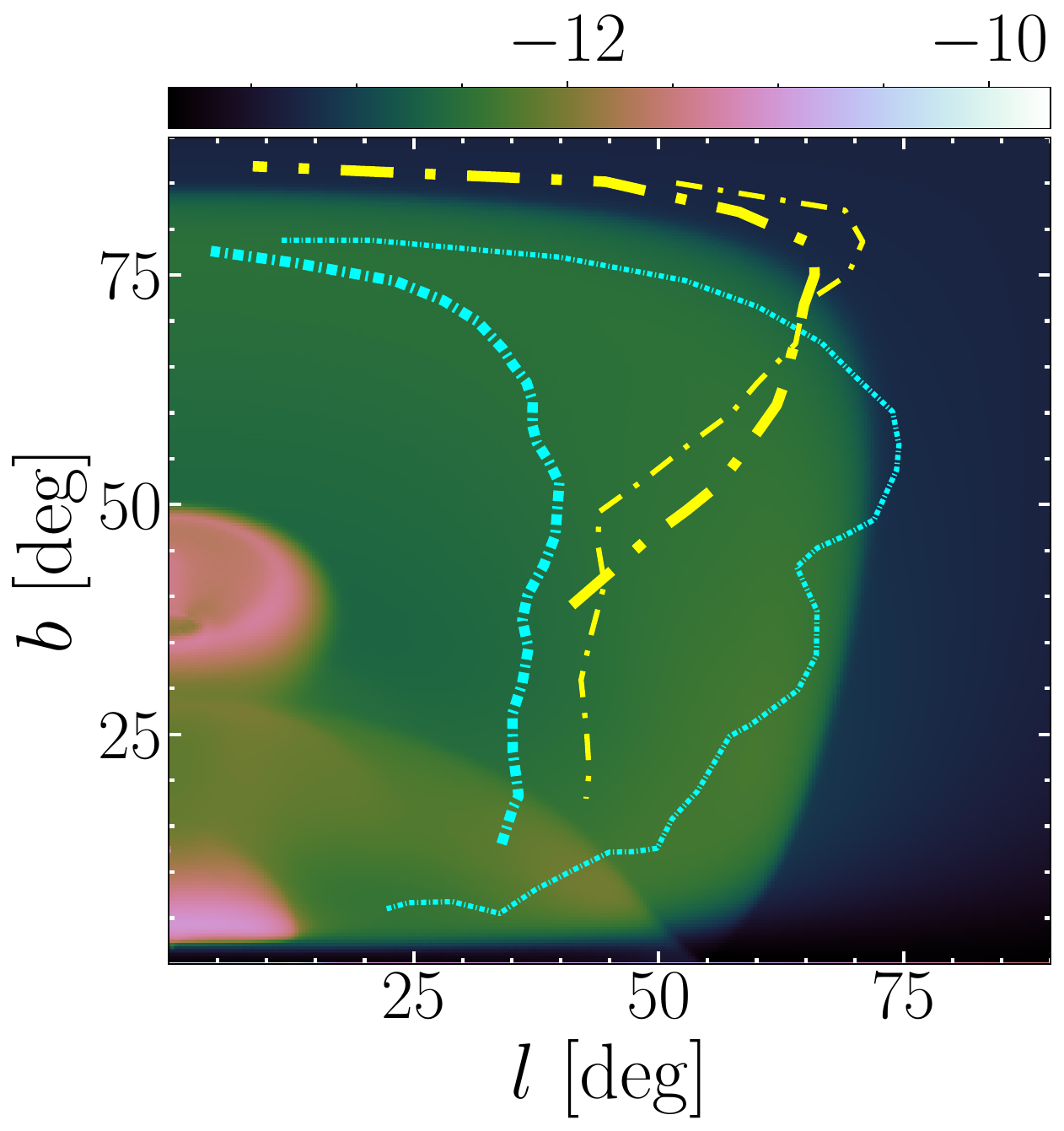}
        \hspace{-9.05cm}\includegraphics[width=0.27\textwidth, trim=0cm 0cm 0cm 0cm, clip=False]{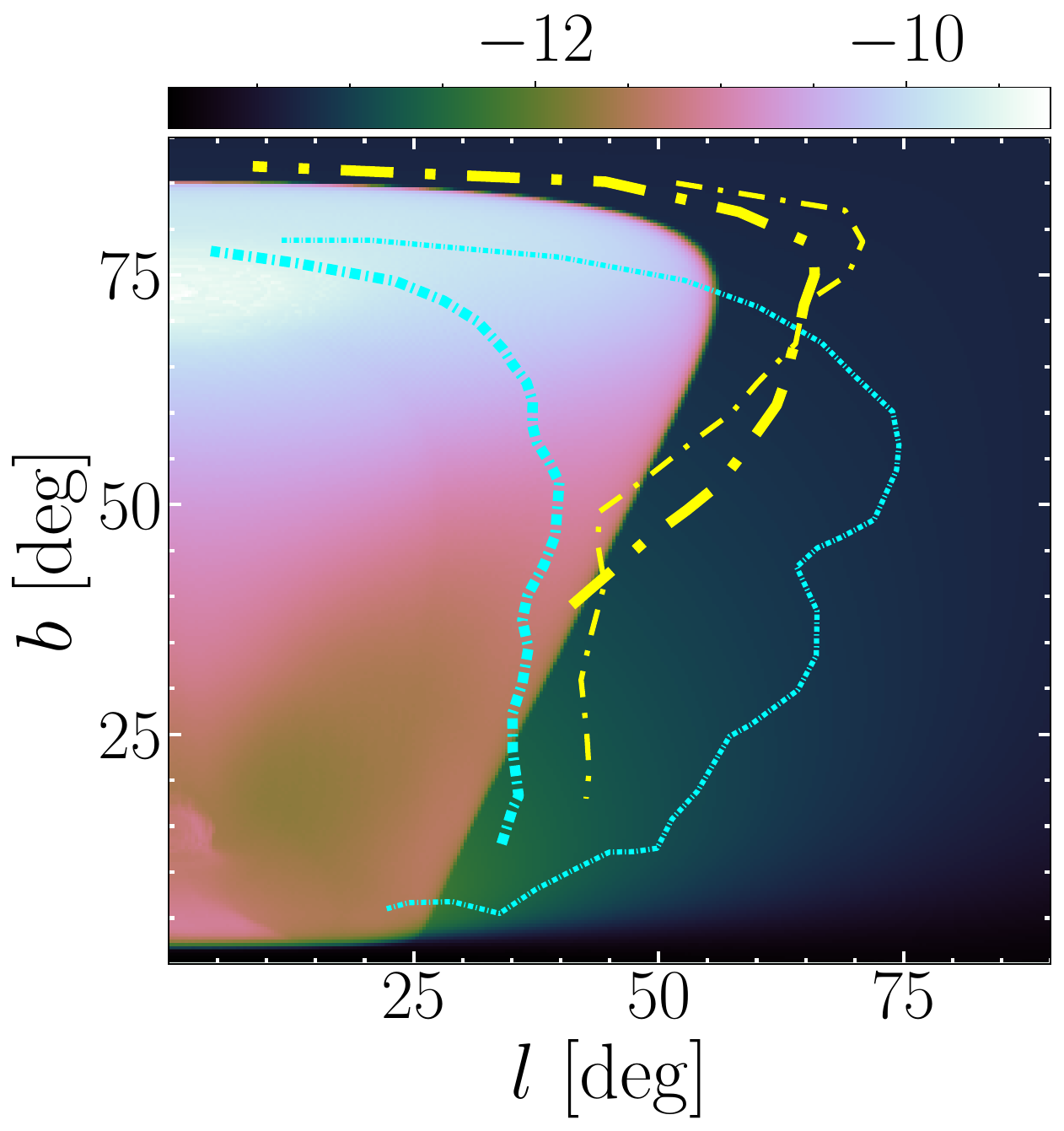}\\

    }
    \caption{Thermal structure (top row) and projected X-ray brightness (bottom) for RB-only, nominal simulations $\nomBall$ (left column; $t\simeq 11.2 \Myr$) and $\nomSlow$ (right column; $t\simeq 22.3\Myr$), shown with identical spatial scales. Thermal images combine density ($\ccinv$ units, left half) and temperature (keV, right half), whereas surface brightness is shown for $2\text{--}10\keV$ bremsstrahlung (erg$~\sinv\cm^{-2}\,\srinv$); color scales are base $10$ logarithmic. RB edges extracted from gradient filters \citep{KeshetGhosh26} are overlaid on the projections as in \autoref{fig:edges1}, for north (short dot-dashed cyan curves) vs. south (long dot-dashed yellow), east (thick curves) vs. west (thin) sectors.
    \label{fig:gen0Nominal}}
\end{figure}

Figure \ref{fig:gen0Nominal} presents these two nominal simulations as the bubbles reach the typical peak latitude of the RBs, chosen as $\vert b\vert \simeq 84^\circ$ \citep{KeshetGhosh26}. Thus, the left (right) column shows $\nomBall$ ($\nomSlow$) at $\tRB \simeq 11.2 \Myr$ ($\tRB \simeq 22.3 \Myr$).
For both simulations, the top row shows (on the same spatial scale for a fair comparison) the density and pressure distributions in a fixed $\phi$ plane, whereas the bottom row shows the projected X-ray brightness. Although the $\nomBall$ bubbles are significantly taller than their $\nomSlow$ counterparts, the projections of both simulations are plausible in the sense that neither one can be ruled out by the measured RB morphology. Indeed, the observed edges differ markedly between north and south hemispheres, and between east and west parts of the bubbles, as shown by the four different contours in the figure.

Morphological and thermal differences between RBs of such very different $\xi$ values are prominent. The forward shock of $\nomSlow$ appears quasi-spherical compared to $\nomBall$, as anticipated from our model. The ballistic $\nomBall$ shows a strong, $\Mach_H \simeq 12$ shock producing a $\sim\keV$-hot interior, whereas $\nomSlow$ drives an $\Mach_H \simeq 1.2$ shock for a $\sim 0.3\keV$ interior. The ballistic $\nomBall$ is X-ray brighter than $\nomSlow$ due to the significantly stronger shock compression. The $\nomSlow$ edges appear approximately uniformly bright, while the high, $b > 60^\circ$, latitude region of $\nomBall$ is brighter than its low-latitude region. The projected $\nomSlow$ and $\nomBall$ shocks roughly coincide with the northwest and southwest RB edges, respectively, so only the latter are presented in subsequent figures, to reduce clutter.

In $\nomBall$, a conical discontinuity trails the forward shock closely, followed by colder, denser gas compared to its immediate surroundings. In contrast, $\nomSlow$ shows a discontinuity structure lagging
the forward shock by a large, $\Delta z \simeq 5 \kpc$ distance, harboring hot and tenuous gas. Overall, these conclusions are similar to those inferred from analogous simulations of the FBs (\MK). One may expect that the low-latitude thermal structure downstream of an RB would affect a subsequent outburst, as discussed in \autoref{subsec:doubleJets}.

As expected, $\nomBall$ coasts ballistically, while $\nomSlow$ undergoes a prolonged slowdown phase. The $\nomBall$ bubble grows nearly linearly up to its designated projected height, both vertically, $z_H \propto t^{0.96 \pm 0.01}$, and horizontally, $R_\mathrm{max} \propto t^{0.98 \pm 0.09}$. The average head velocity is consistent with the injected $\beta_{-2} \simeq 0.8$, as can be inferred from the figure. The bubble aspect ratio remains constant at $\mathcal{A}\simeq 0.5$ throughout the evolution, consistent with the injected $\theta_5 \simeq 1$ by \autoref{eq:aspRatioBall}.
\begin{figure}
    \def\pw{0.8\linewidth}
    \def\ph{\linewidth}
    \fboxvisiblefalse
    \centering
    \begin{minipage}{\pw}
    \myfbox{\includegraphics[width=\linewidth, trim=0.25cm 0.25cm 0.5cm 0.25cm]{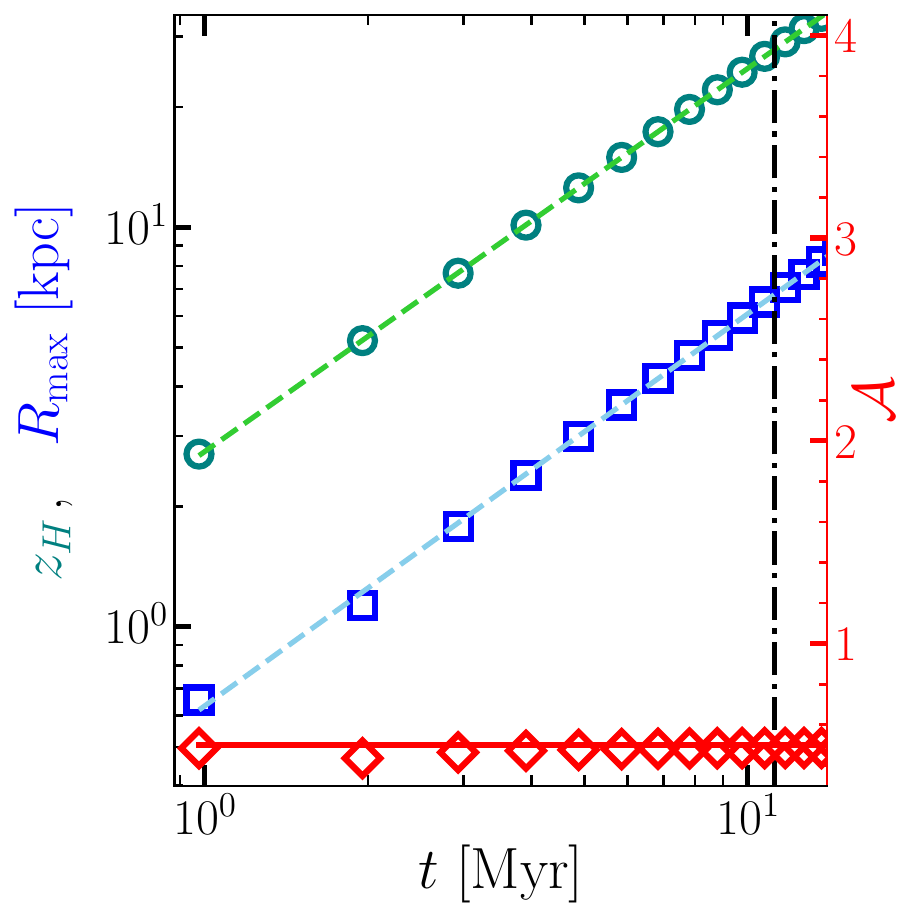}}
    \end{minipage}\\\vspace*{0.2cm}
    \begin{minipage}{\pw}
    \myfbox{\includegraphics[width=\linewidth, trim=0.25cm 0.25cm 0.5cm 0.25cm]{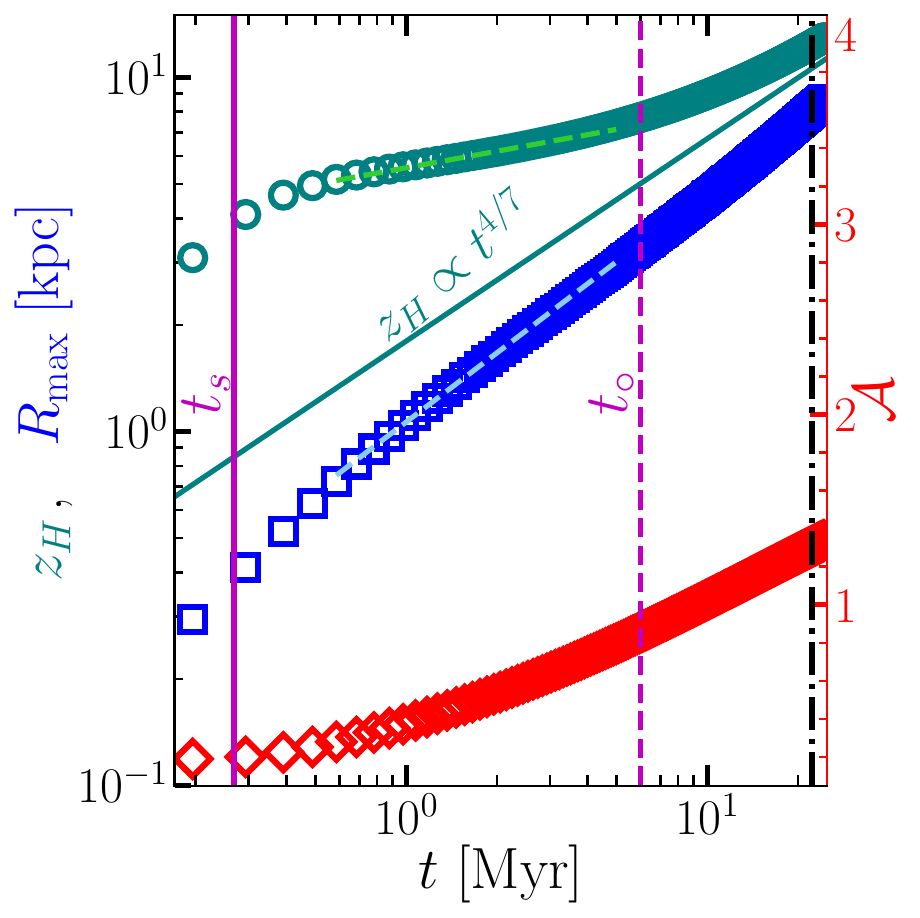}}
    \end{minipage}
    \caption{
    RB evolution in the nominal $\nomBall$ (top) and $\nomSlow$ (bottom) simulations: bubble head height (teal circles), half-width at half-height (blue squares), and aspect ratio (red diamonds with right axis); the time of crossing $|b|\simeq 84^\circ$ is highlighted (vertical dash-dotted line). Also shown are power-law slopes (where well-fitted; dashed lines), the ballistic-model aspect ratio (for $\nomBall$; solid red), and for $\nomSlow$, the estimated $t_s$ (labeled vertical solid line; \autoref{eq:Tc}) and $t_\circ$ (vertical dashed line; \autoref{eq:slowLifeTime}) transitions based on the viscosity-modified parameters (see text), and the isotropic $z_H$ expansion (solid teal; \autoref{eq:selfSimEvol}).
    \label{fig:G0RBEvol}}
\end{figure}

High-velocity simulations such as $\nomSlow$ require the aforementioned introduction of viscosity to suppress instabilities caused by the strong shear during injection. The viscosity effectively modifies the outburst parameters governing the subsequent dynamics, in a way that can be recovered (\MK) from the early evolution. We illustrate this for $\nomSlow$ by using a high temporal resolution to resolve the ballistic phase of the bubble, finding modified $\vJ' \simeq 5.2$ and $\thJ' \simeq 0.20$ parameters (see \autoref{appendix:transition}).
These modified parameters reflect initial flow collimation resulting in an effective $\xi'\simeq 0.26$, which is only mildly in the slowdown regime.
Using Eqs.~(\ref{eq:Lc}) and (\ref{eq:Tc}), the ballistic-to-slowdown transition is then expected at $z_s\simeq 4.3\kpc$ and $t_s\simeq 0.27\Myr$, which indeed agree with the $\nomSlow$-simulated transition (see $t_s$-labeled line in the bottom panel of \autoref{fig:G0RBEvol}).

The bottom panel of \autoref{fig:G0RBEvol} demonstrates that the $\nomSlow$ bubble head undergoes three phases of evolution: an initial quasi-linear, $z_H \propto t$ ballistic propagation, followed by $z_H \propto t^{\tau_z}$ slowdown, which at late time gradually approaches the $z_H \propto t^{4/7}$ STvN solution. We obtain $\tau_z\simeq 0.16 \pm 0.02$ by fitting this evolution for $2t_s \lesssim t \lesssim 5 \Myr$, where the upper (lower) limit is chosen to avoid the gradual, slowdown-to-isotropic (ballistic-to-slowdown) transition. Using this $\tau_z$ and the viscosity-modified outburst parameters, we obtain $\{z_\circ,t_\circ\} \simeq \{9.3 \kpc, 6.0\Myr\}$ from \autoref{eq:approxZcirc} and \autoref{eq:slowLifeTime}. Throughout $t_s \lesssim t\lesssim t_\circ$ Myr, the evolution of the maximal half-width follows $R_\mathrm{max} \propto t^{0.65 \pm 0.14}$, consistent with the expected behavior from \autoref{eq:RMaxSlow}. At $t_\circ$ we find an $\mathcal{A}\simeq 0.86$ bubble approaching quasi-sphericity, as anticipated from the model.

\begin{figure}
    \def\pw{0.49\linewidth}
    \def\ph{0.49\linewidth}
    \fboxvisiblefalse
    \centering
    \begin{minipage}{\pw}
    \myfbox{\includegraphics[width=\linewidth]{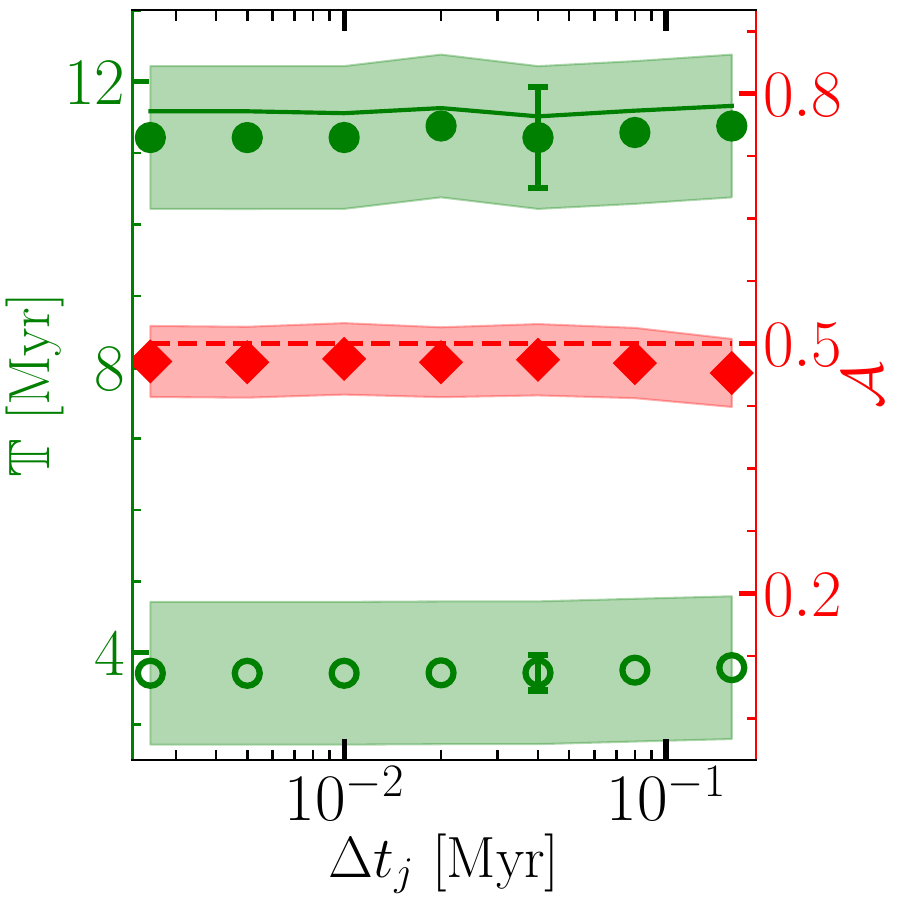}}
    \end{minipage}
    \begin{minipage}{\pw}
    \hspace*{0.2cm}
    \myfbox{\includegraphics[width=\linewidth]{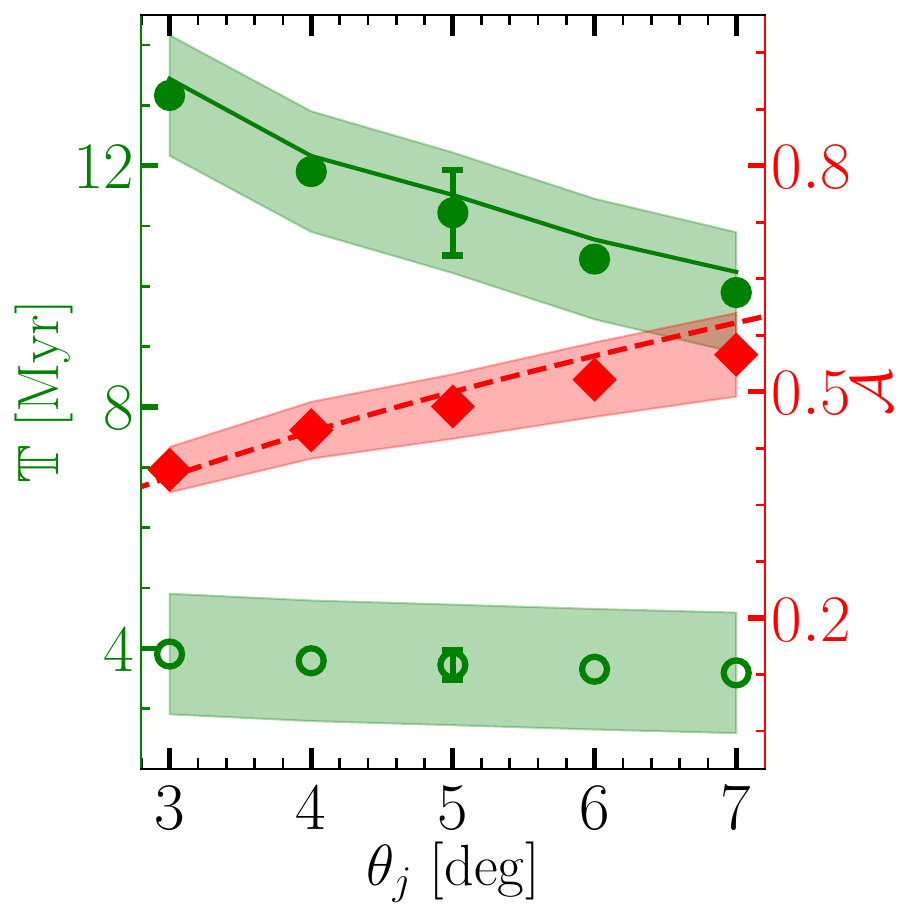}}
    \end{minipage}

    \begin{minipage}{\pw}
    \myfbox{\includegraphics[width=\linewidth]{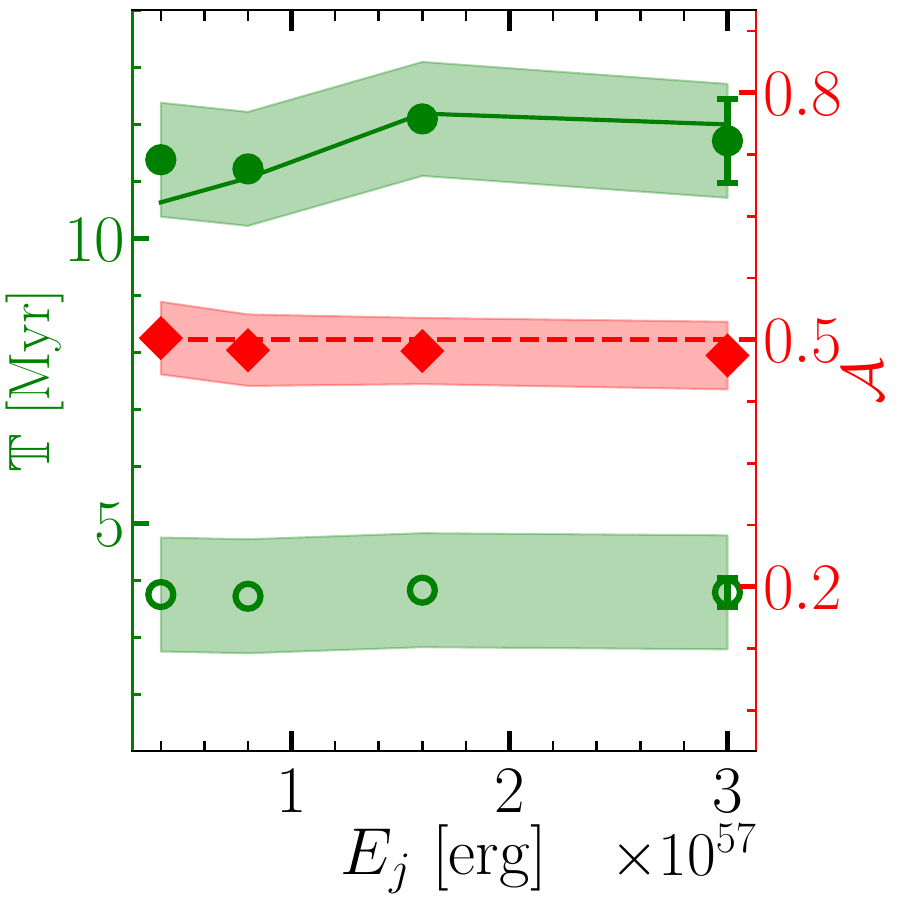}}
    \end{minipage}
    \begin{minipage}{\pw}
    \hspace*{0.2cm}
    \myfbox{\includegraphics[width=\linewidth]{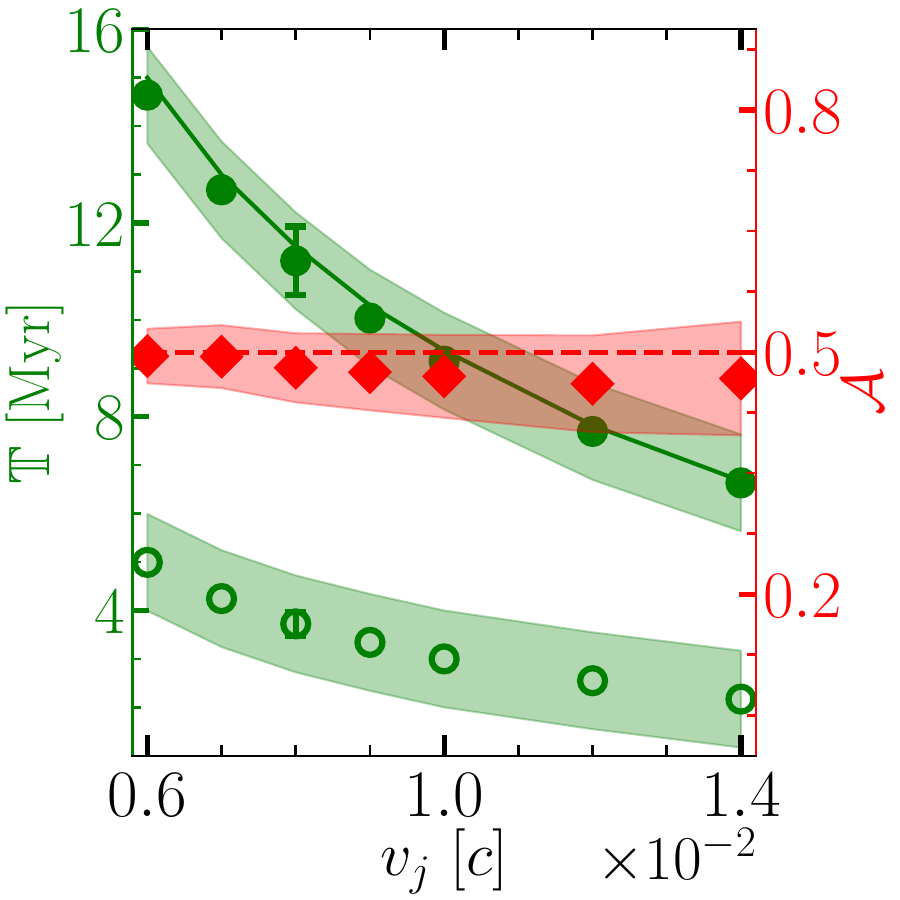}}
    \end{minipage}
    \caption{
    RB age (disks with left axis) and aspect ratio (diamonds with right axis) in simulations similar to $\nomBall$ (error bars, showing convergence with respect to numerical resolution) but with one parameter (abscissa) varied (symbols): outburst energy (top left), velocity (top right), opening angle (bottom left), and duration (bottom right).
    Circles show the age of the bubbles when crossing the FB latitude.
    Shaded regions provide numerical error estimates (see \autoref{appendix:transition}). Curves indicate the aspect ratio (dashed) and RB age (solid) in the ballistic model.
    \label{fig:nomAgeVaried}}
\end{figure}

Figure \ref{fig:nomAgeVaried} explores how the RB age and aspect ratio vary
in the parameter space around $\nomBall$, by changing each injection parameter separately. Also shown is the bubble age when crossing
latitude $|b|=52^\circ$, taken as representative of the head of the FBs. As expected, the ballistic model, shown in solid lines, provides a good description for small parametric variations around $\nomBall$. As the energy is lowered below $\sim10^{57}\erg$, the evolution starts deviating from ballistic in the aging RBs, as indicated by the bottom left panel, albeit not in the younger FBs.

\begin{figure}
    \def\pw{0.49\linewidth}
    \def\ph{0.49\linewidth}
    \fboxvisiblefalse
    \centering
    \begin{minipage}{\pw}
    \myfbox{\includegraphics[width=\linewidth]{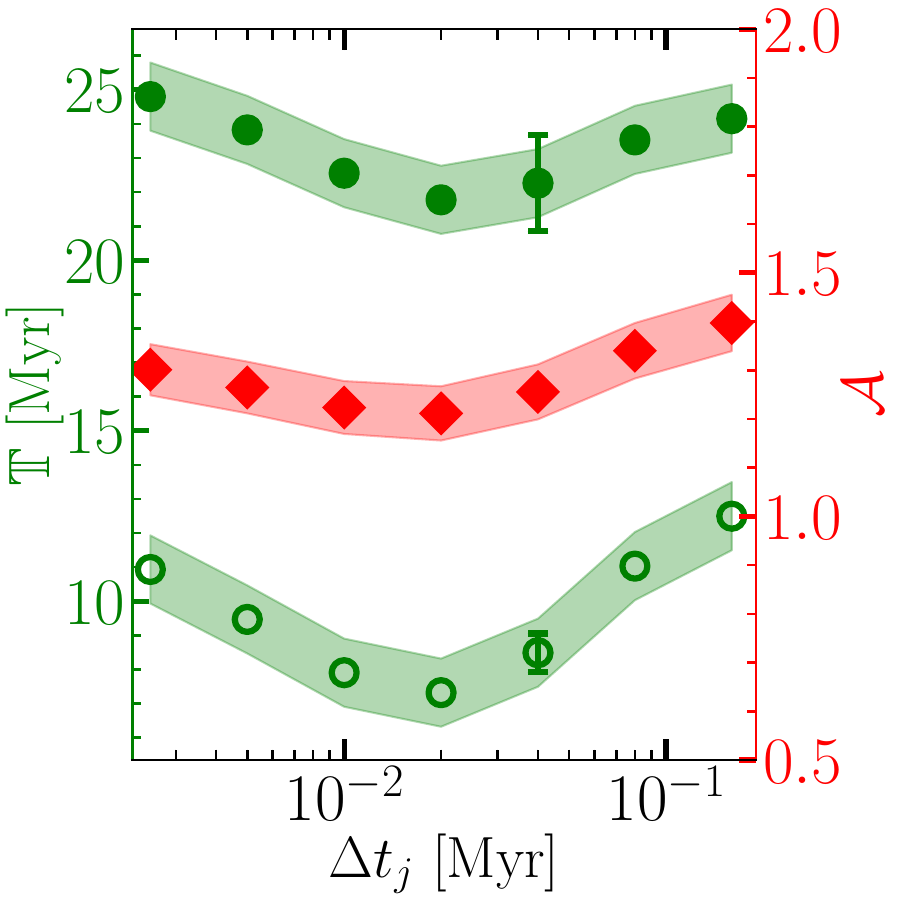}}
    \end{minipage}
    \begin{minipage}{\pw}
    \hspace*{-0.1cm}
    \myfbox{\includegraphics[width=\linewidth]{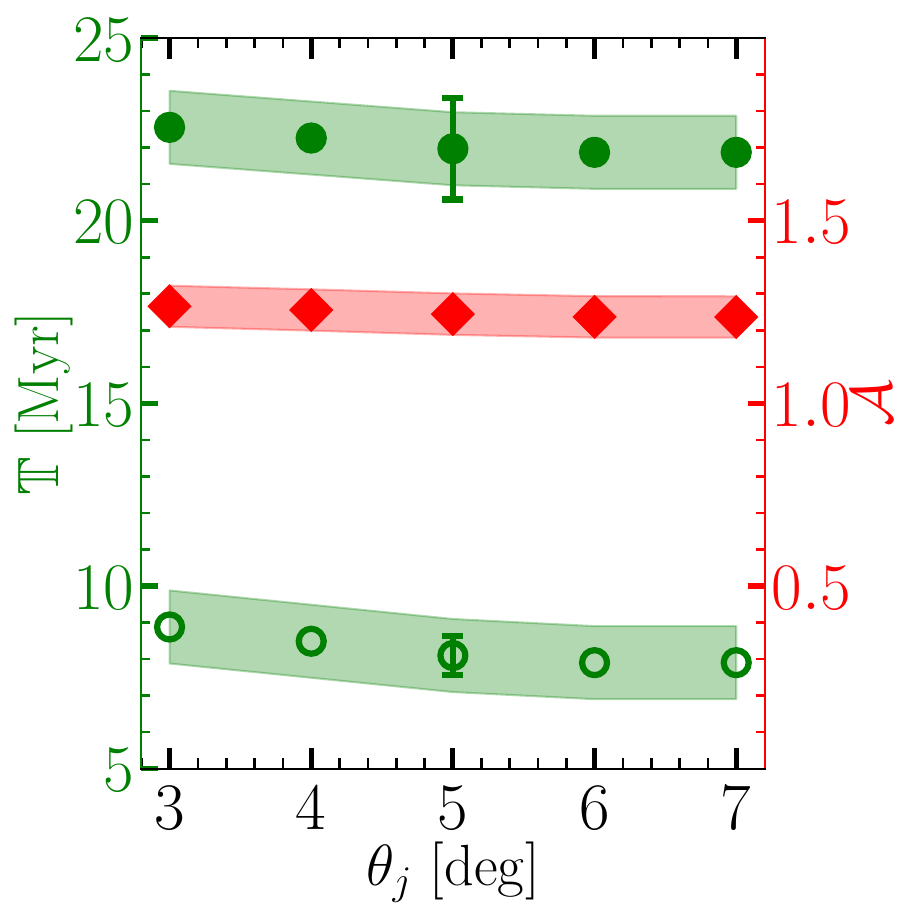}}
    \end{minipage}

    \begin{minipage}{\pw}
    \myfbox{\includegraphics[width=\linewidth]{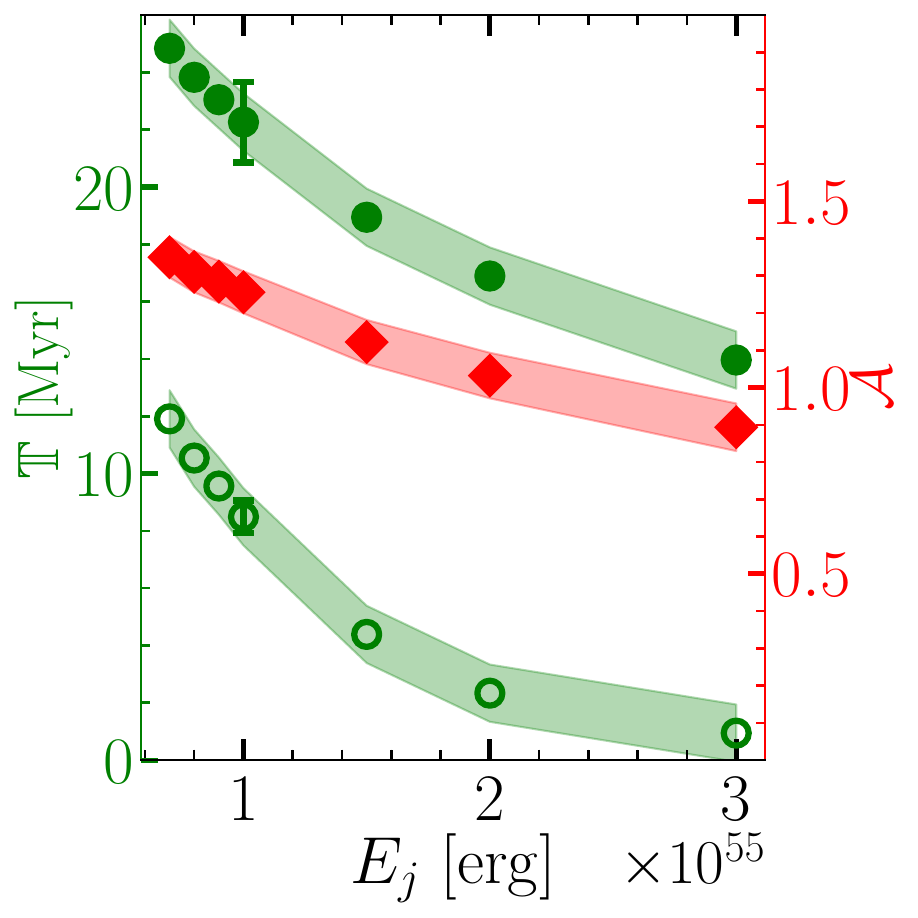}}
    \end{minipage}
    \begin{minipage}{\pw}
    \hspace*{-0.1cm}
    \myfbox{\includegraphics[width=\linewidth]{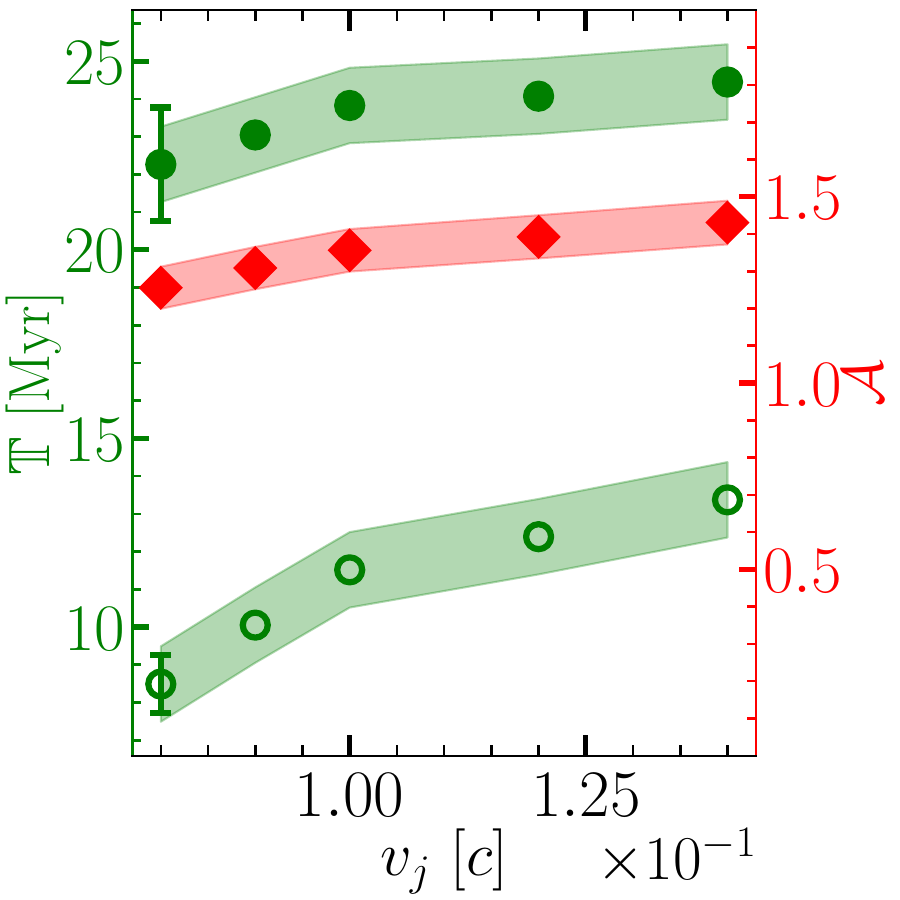}}
    \end{minipage}
    \caption{Same as \autoref{fig:nomAgeVaried}, but for $\nomSlow$ instead of $\nomBall$.
    \label{fig:nomAgeVariedSlow}}
\end{figure}

Analogously, \autoref{fig:nomAgeVariedSlow} explores the parameter space around $\nomSlow$. In this case, a short ballistic phase followed by an extended slowdown evolution implies a greater bubble age, with the $\nomSlow$ bubbles nearly twice as old as their $\nomBall$ counterparts. While $\tage$ depends only on $v_j$ for ballistic bubbles, it is sensitive also to the injected energy in the slowdown case, with modest $E_j$ changes in the bottom-left panel producing bubbles aged $15 \lesssim \tRB/\Myr \lesssim 25$. Our simulations of weak, $E_j \lesssim \text{few}\times10^{54}$ erg outbursts are not presented because the corresponding shocks are too weak and morphologically wide to match the RBs.

\begin{figure}
    \centering{
        \includegraphics[width=0.48\textwidth, trim=0cm 0cm 0.6cm 0cm, clip]{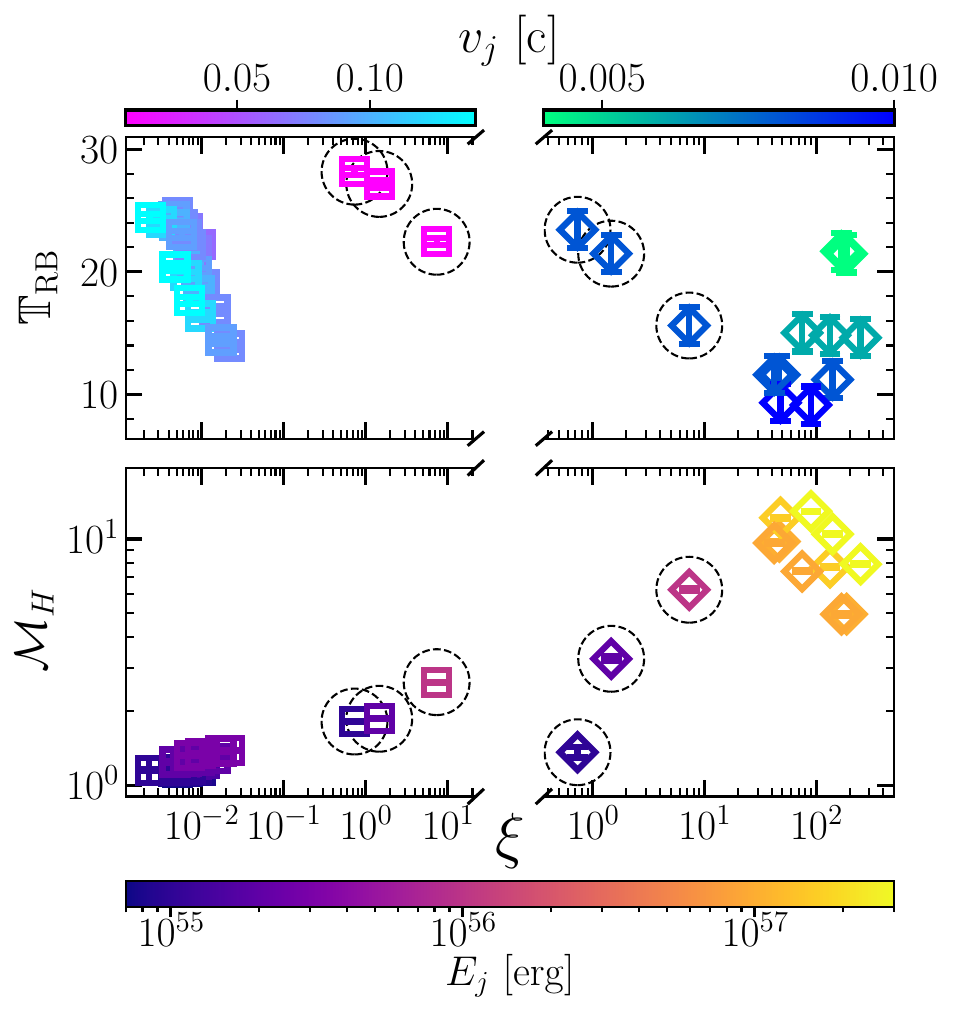}
    }
    \caption{
    RB age (upper panel) and head Mach number (bottom panel) for $\nomBall$ (diamonds) and $\nomSlow$ (squares) simulations of different $v_j$ and $E_j$ values (colorbars) as a function of the ballistic-to-slow parameter $\xi$ (uncorrected for $\xi\ll1$ initial collimation). Error bars reflect a possible range of $3\degree\leq\theta_j\leq7\degree$ (see top-right panels of Figs. \ref{fig:nomAgeVaried} and \ref{fig:nomAgeVariedSlow}). Dashed circles highlight three $\xi\sim1$  simulations not fully converged at our highest resolution.
    \label{fig:enVelScanXi}}
\end{figure}

We perform a two-dimensional parametric scan of the $E_j$--$v_j$ phase space, as shown in \autoref{fig:enVelScanXi}. Converged simulations are obtained in the $\xi \gg 1$ ($\xi \ll 1$) regime, with other parameters fixed to those of $\nomBall$ ($\nomSlow$), carried out without viscosity (with viscosity, but without correcting $\xi$ for initial collimation). The resulting ballistic RBs are aged $10\text{--}20 \Myr$, with a strong, $\Mach_H \simeq 5\text{--}10$ shock, whereas slowdown RBs are somewhat older, $\tRB \simeq 15-25 \Myr$, with weak, $\Mach_H \lesssim 1.3$
shocks. In the intermediate, $\xi\simeq 1$ regime, we perform three simulations (highlighted by dashed circles, both with and without viscosity) with the same $\beta_j=0.008$, $\theta_j=4\degree$, and $\Delta t_j=0.04\Myr$, but different $E_{56}=0.1$, $0.2$, and $1$.
Although these simulations are not fully converged at our highest resolution, and the results depend on viscosity, they suggest that the observed $3\lesssim\Mach_H\lesssim5$ RBs can be produced within $\tRB \simeq 15-25 \Myr$ of an $10^{55}\erg\lesssim E_j\lesssim 10^{56}\erg$ outburst.

\subsection{Joint RB--FB simulations}
\label{subsec:doubleJets}

We now turn our attention to a nested bubble scenario, where two separate outbursts are launched from the GC, temporally separated by a free parameter $\tdel[01]$. The older bubbles then correspond to RBs propagating into an unperturbed CGM, as already discussed in \autoref{subsec:singleJets}. We thus focus on the younger bubbles, representing FBs that evolve into a CGM already shocked by the RBs. For simplicity, we adopt identical injection parameters for the two outbursts. We find that observations can be reproduced even under such a simplifying assumption.

\begin{figure}[h]
    \def\pw{0.49\linewidth}
    \def\ph{0.49\linewidth}
    \fboxvisiblefalse
    \centering
    \myfbox{\includegraphics[width=0.45\textwidth, trim=0cm 0cm 0cm 0cm]{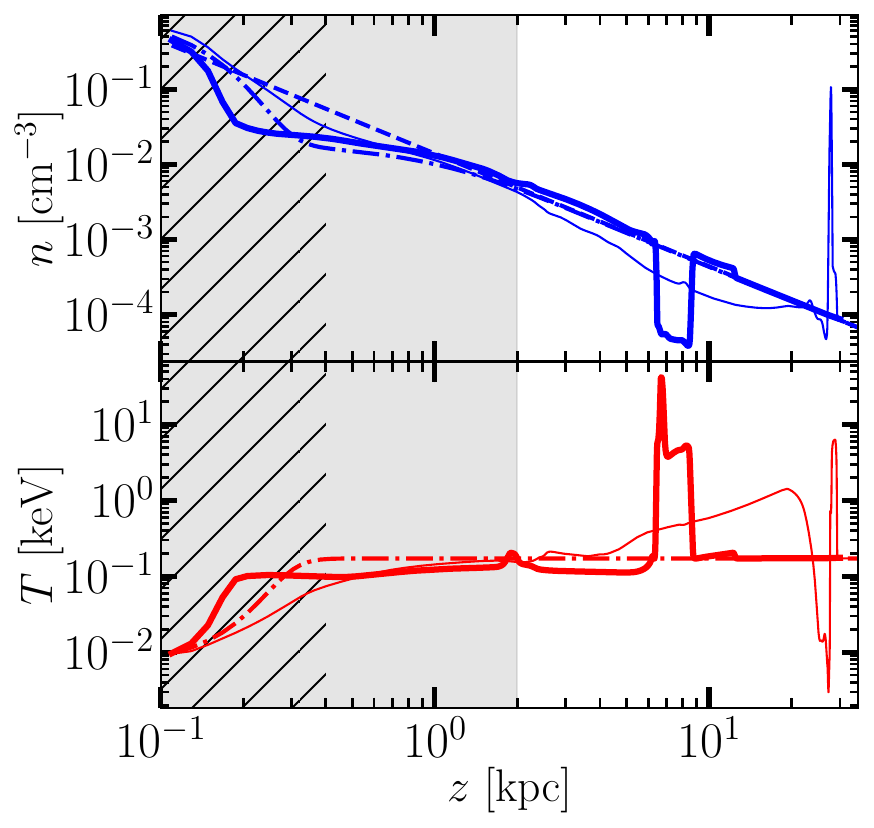}}
    \caption{Density (top) and temperature (bottom) profiles of the RB downstream, averaged over $\theta \lesssim \theta_j$, shown for $\nomBall$ (thin) and $\nomSlow$ (thick). Initial, unperturbed (dashdotted) density and temperature profiles are shown for comparison; the former approaches the halo power-law \citep[dashed]{MillerBregman2015} at $z\gtrsim 2\kpc$, below which the Galactic disk (hatched) and the Bulge (shaded) contributions dominate.
    \label{fig:denProf}}
\end{figure}

The evolution of the FBs is sensitive to their ambient medium, which was recently shocked by the RBs; in particular, a rarefied medium ahead of the FBs prolongs their ballistic phase (\autoref{subsec:secondBurstModel}). To study the modified FB upstream, \autoref{fig:denProf} shows the thermal structure of the present ($\tRB$ after the first outburst) RB downstream, in the $\nomBall$ and $\nomSlow$ simulations. The $z\gtrsim10\kpc$ part of the displayed profiles thus describes the present ($\tFB=\tRB-\tdel[01]$ after second outburst) FB upstream.

In both simulations, the density profile near the $\sim10\kpc$ FB height is diminished by a factor $F>1$, but still crudely follows the same power-law behavior of the initial, unperturbed medium. The $F\simeq 20$ rarefication of $\nomSlow$ is stronger but more localized than $F\simeq 3$ in $\nomBall$.

\begin{figure}[h]
    \centering{
        \hspace{0cm}
        \raisebox{0.3mm}[0pt][0pt]{\includegraphics[width=0.263\textwidth]{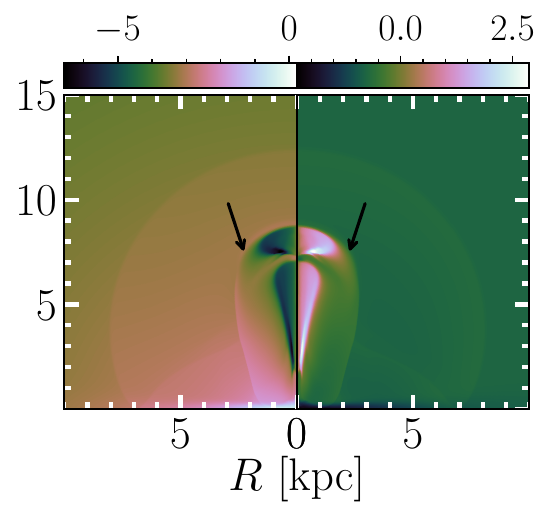}}
        \hspace{-9.25cm}\includegraphics[width=0.277\textwidth, trim=0cm 0cm 0cm 0cm]{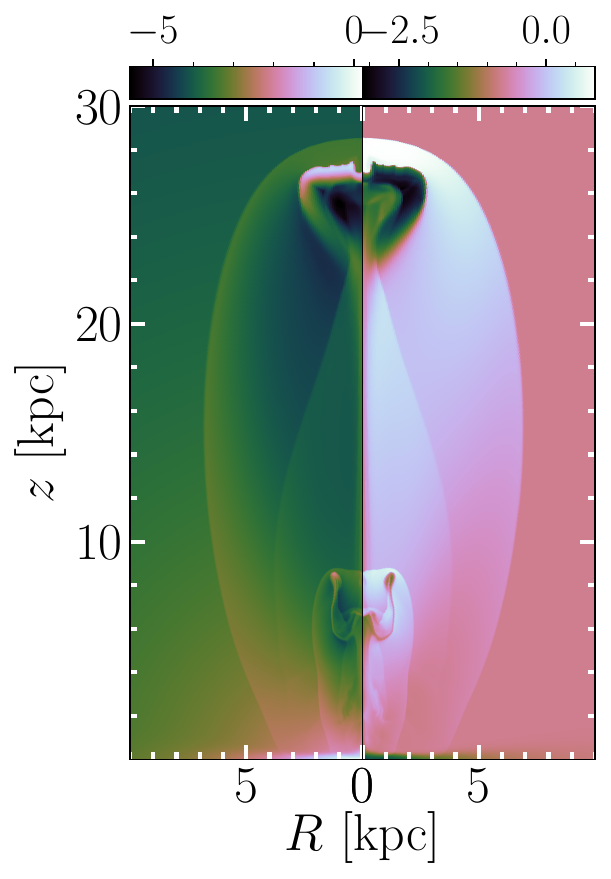}\\
        \hspace{0cm}\includegraphics[width=0.27\textwidth, trim=0cm 0cm 0cm 0cm, clip=False]{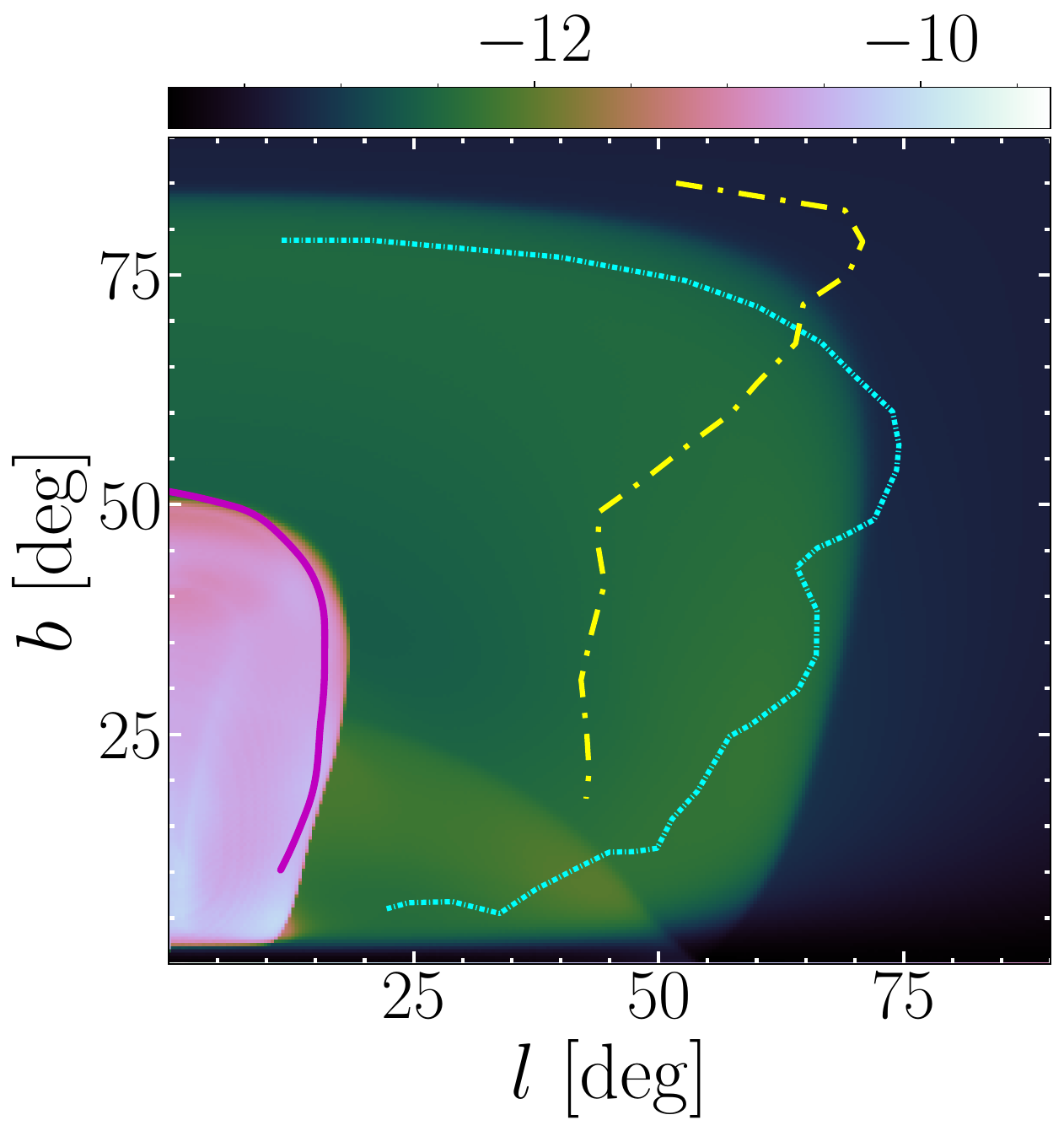}
        \hspace{-9.05cm}\includegraphics[width=0.27\textwidth, trim=0cm 0cm 0cm 0cm, clip=False]{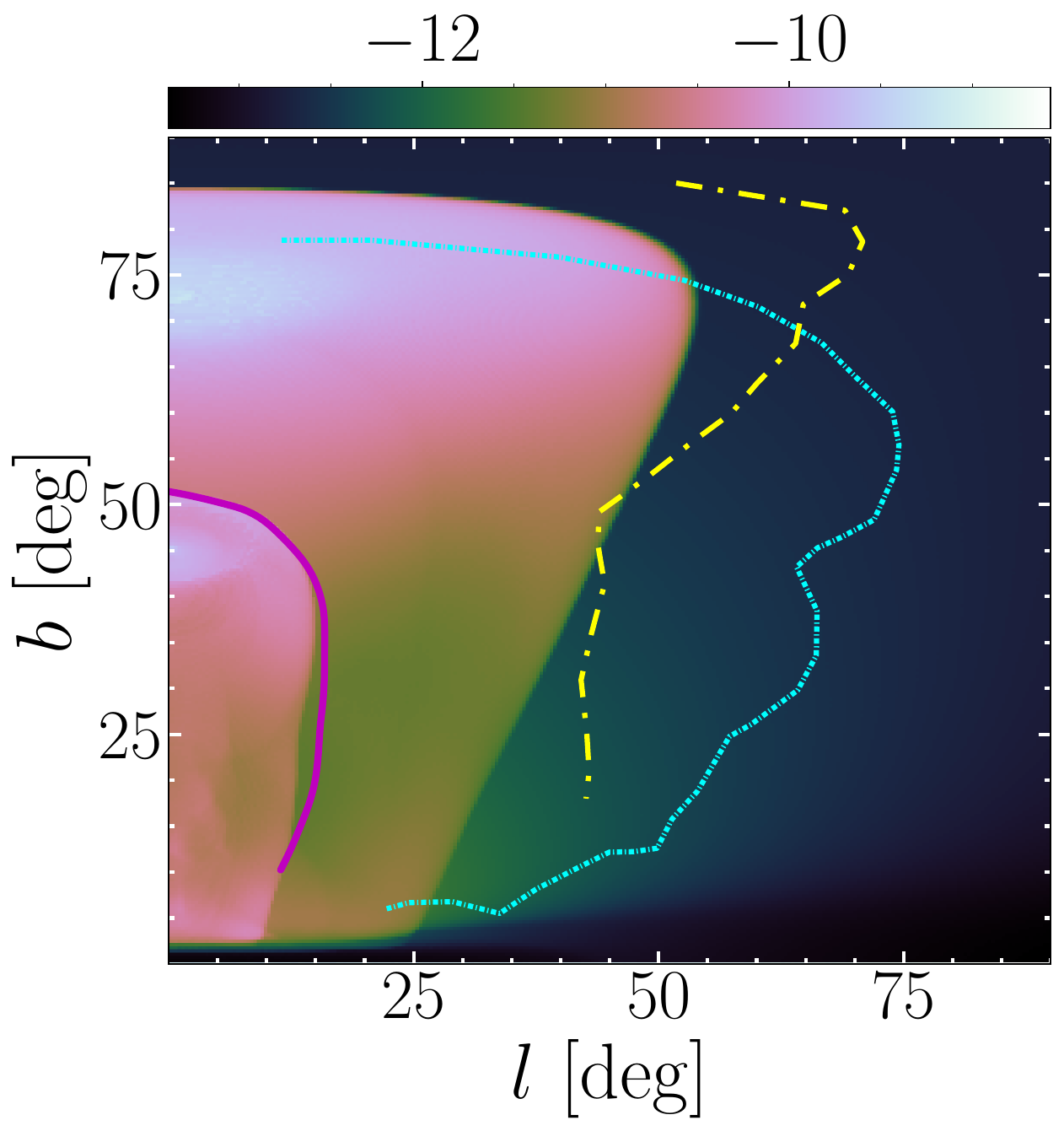}\\
    }
    \caption{RBs and FBs in nominal double-burst simulations
    $\nomBall$ (left column) and $\nomSlow$ (right column). Notation and conventions are identical to \autoref{fig:gen0Nominal}. Observed FB edges (northeast sector; magenta solid) are overlaid on the projected map. In $\nomSlow$, the FB shock (pointed by black arrows) engulfs the relic RB discontinuity structure.
    \label{fig:nominalGB}}
\end{figure}

\Autoref{fig:nominalGB} shows the thermal and projected maps of the simulated, nested bubbles. The FBs rise up to $z\sim 9 \kpc$, with a faint (strong) plume-like (two-component) discontinuity structure trailing the leading shock in $\nomBall$ ($\nomSlow$). The FB interior is $\sim\keV$-hot, enclosed by a strong, $\Mach_H \simeq 4$, forward shock, somewhat weakened as its upstream was heated by the RBs. In projection, both simulations agree well with the observed FB edges.

\begin{figure}
    \def\pw{0.3\linewidth}
    \def\ph{0.6\linewidth}
    \fboxvisiblefalse
    \begin{center}
        \begin{tikzpicture}
            \draw (0, 0) node[inner sep=0]
            {
                \myfbox{\includegraphics[height=\ph, width=0.35\linewidth]{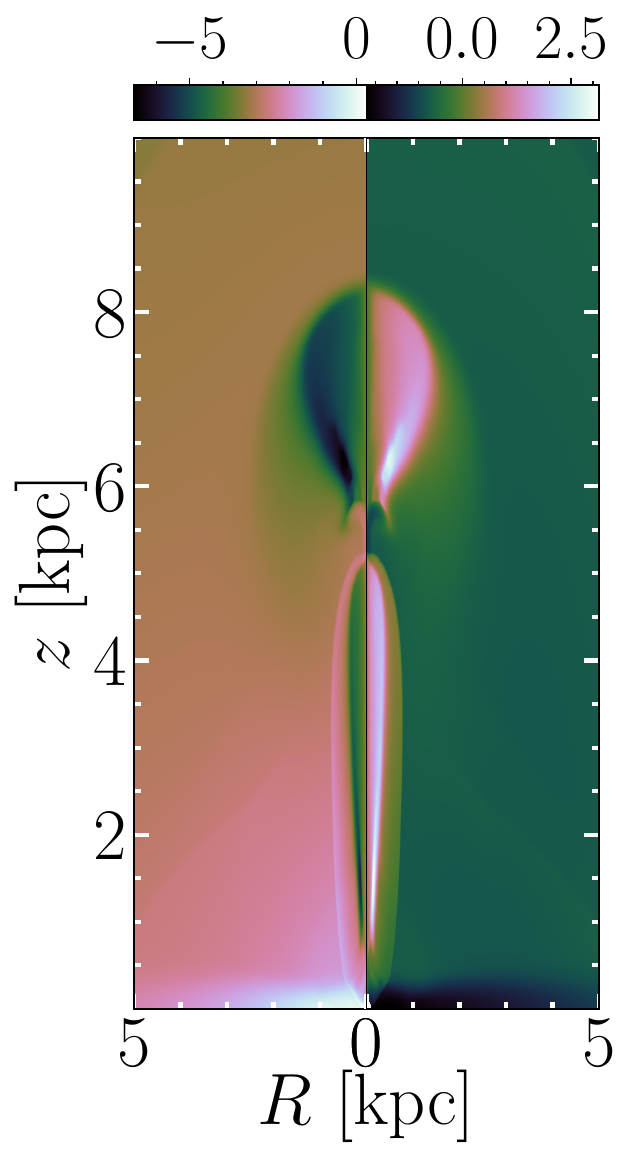}}
                \hspace*{-0.2cm}\myfbox{\includegraphics[height=\ph, width=\pw]{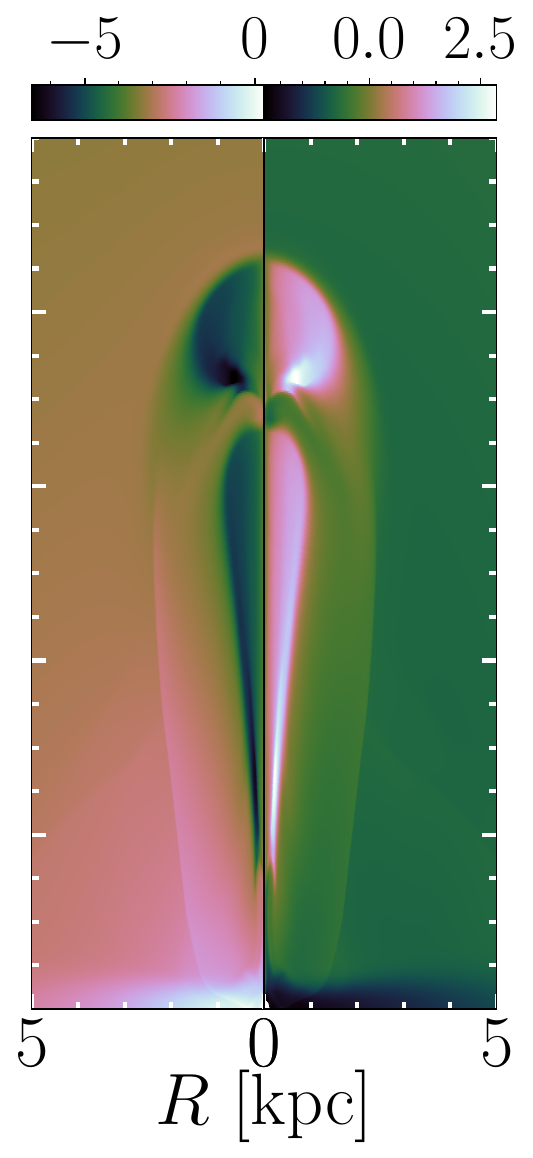}}
                \hspace*{-0.2cm}\myfbox{\includegraphics[height=\ph, width=\pw]{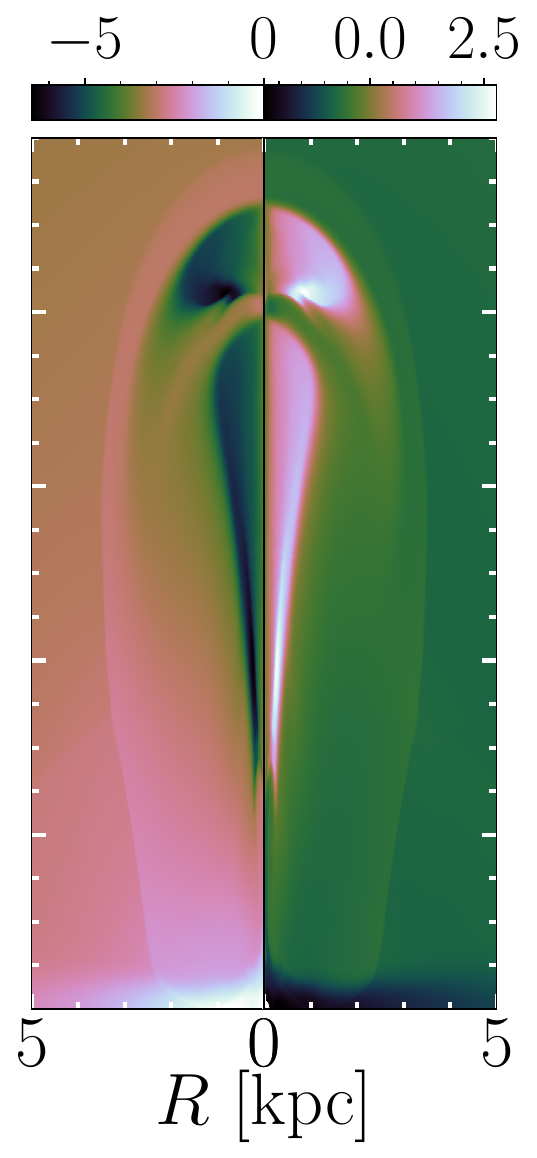}}
            };
            \draw[->,line width=1pt] (-2.7,0) -- (-2.4,-0.1);
            \draw[->,line width=1pt] (-0.6,0.6) -- (-0.3,0.5);
            \draw[->,line width=1pt] (1.9,1.7) -- (2.2,1.6);
    \end{tikzpicture}
    \end{center}
    \caption{Same as the bottom-right panel of \autoref{fig:nominalGB}, zooming into the $\nomSlow$ FB shock (indicated by black arrows) before (left panel; $19\Myr\simeq t\simeq\tdel[01] +0.6\Myr$), during (middle, $t\simeq 21.5 \Myr$), and after (right, $t\simeq 24.4 \Myr$) interacting with the discontinuity structure left behind by the RBs.
    \label{fig:slowRepeater}}
\end{figure}

We find that in order to simultaneously account for the observed morphologies of both RBs and FBs, $\nomBall$ ($\nomSlow$) requires an outburst separation $\tdel[01] \simeq 7.4 \Myr$ ($\tdel[01] \simeq 18.4 \Myr$), with FBs aged $\tFB \simeq 3.8 \Myr$ ($\tFB \simeq 3.9 \Myr$). In $\nomSlow$, a quasi-spherical blob of hot and tenuous gas, enclosed by a radius $\sim 2 \kpc$ discontinuity, trails $\sim 4~\kpc$ behind the RB forward shock (see \autoref{fig:gen0Nominal}). Consequently, the FB shock temporarily weakens and becomes difficult to trace as it plows through this $>$ keV hot medium to reach its present latitude, as shown in  \autoref{fig:slowRepeater}.

\begin{table*}
    \scriptsize
    \centering
        \scalebox{1.1}{
            \begin{tabular}{ccccccccc}
            \hline
            \thead{\textbf{Name}}&\thead{\bm{$E_j$}\\\bm{$[\mathrm{erg}]$}}&\thead{\bm{$\Delta t_j$}\\\bm{$[\mathrm{Myr}]$}}&\thead{\bm{$\beta_j\, (\beta'_j)$}}&\thead{\bm{$\theta_j\,(\theta'_j)$}\\\bm{$[\mathrm{deg}]$}}&\thead{\bm{$\xi\,(\xi')$}}&\thead{\bm{$\tRB$}\\\bm{$[\mathrm{Myr}]$}}&\thead{\bm{$\tFB$}\\\bm{$[\mathrm{Myr}]$}}&\thead{\bm{$\tdel[]$}\\\bm{$[\mathrm{Myr}]$}}\\
            \hline
            $\mathcal{B}_0$&$3\times 10^{57}$&$0.04$&$0.008$&$5$&$140.6$&1$1.2$&$3.8$&$7.4$\\ $\mathcal{B}_1$&$10^{57}$&$0.04$&$0.008$&$5$&$46.9$&$11.5$&$5.1$&$6.4$\\
            $\mathcal{B}_2$&$8\times 10^{56}$&$0.08$&$0.006$&$5$&$66.7$&$15$&$5.6$&$9.4$\\
            $\mathcal{B}_3$&$5\times 10^{56}$&$0.1$&$0.006$&$5$&$41.7$&$15.2$&$5.8$&$9.4$\\
            $\mathcal{S}_0$&$ 10^{55}$&$0.04$&$0.08\,(\sim0.052)$&$4\,(\sim1.0)$&$0.0073\,(\sim0.26)$&$22.3$&$3.9$&$18.4$\\
            $\mathcal{S}_1$&$9\times 10^{54}$&$0.04$&$0.07$&$4$&$0.086$&$21.8$&$2.8$&$19$\\
            $\mathcal{S}_2$&$8\times 10^{54}$&$0.04$&$0.0625$&$4$&$0.096$&$22.3$&$2.9$&$19.4$\\
            \hline
            \end{tabular}}
        \caption{
        Parameters and timescales of the suite of simulations modeling the nested bubbles as two identically parametrized outbursts. The simulations $\nomBall$ and $\nomSlow$ correspond to our nominal simulations; for a description of simulation parameters, see \autoref{subsec:singleJets}. Parentheses designate effective, viscosity-modified parameters inferred from early ballistic evolution (\autoref{appendix:transition}).
        \label{table:twinParams}}
\end{table*}

While the first outburst of $\nomSlow$ yields slowing down RBs, the FBs arising from its identical second outburst are ballistic, like in $\nomBall$. Indeed, the $\nomSlow$ FBs are $\tFB \simeq 8 \Myr$ old (see \autoref{fig:nomAgeVariedSlow}) with an increasing, $\mathcal{A} \simeq1$ aspect ratio (see \autoref{fig:G0RBEvol}) when propagating into an undisturbed CGM, but only $\tFB\simeq 4$ Myr old with a ballistic $\mathcal{A} \simeq0.7$ when closely trailing the RB shock. This result, supplemented by additional simulations below, provides numerical evidence for the \S\ref{subsec:secondBurstModel} conclusion that the phase space of slowing-down FBs shrinks or vanishes due to rarefaction by their preceding RBs.

\begin{figure}
    \centering{
        \includegraphics[width=0.48\textwidth, trim=0cm 0cm 0cm 0cm]{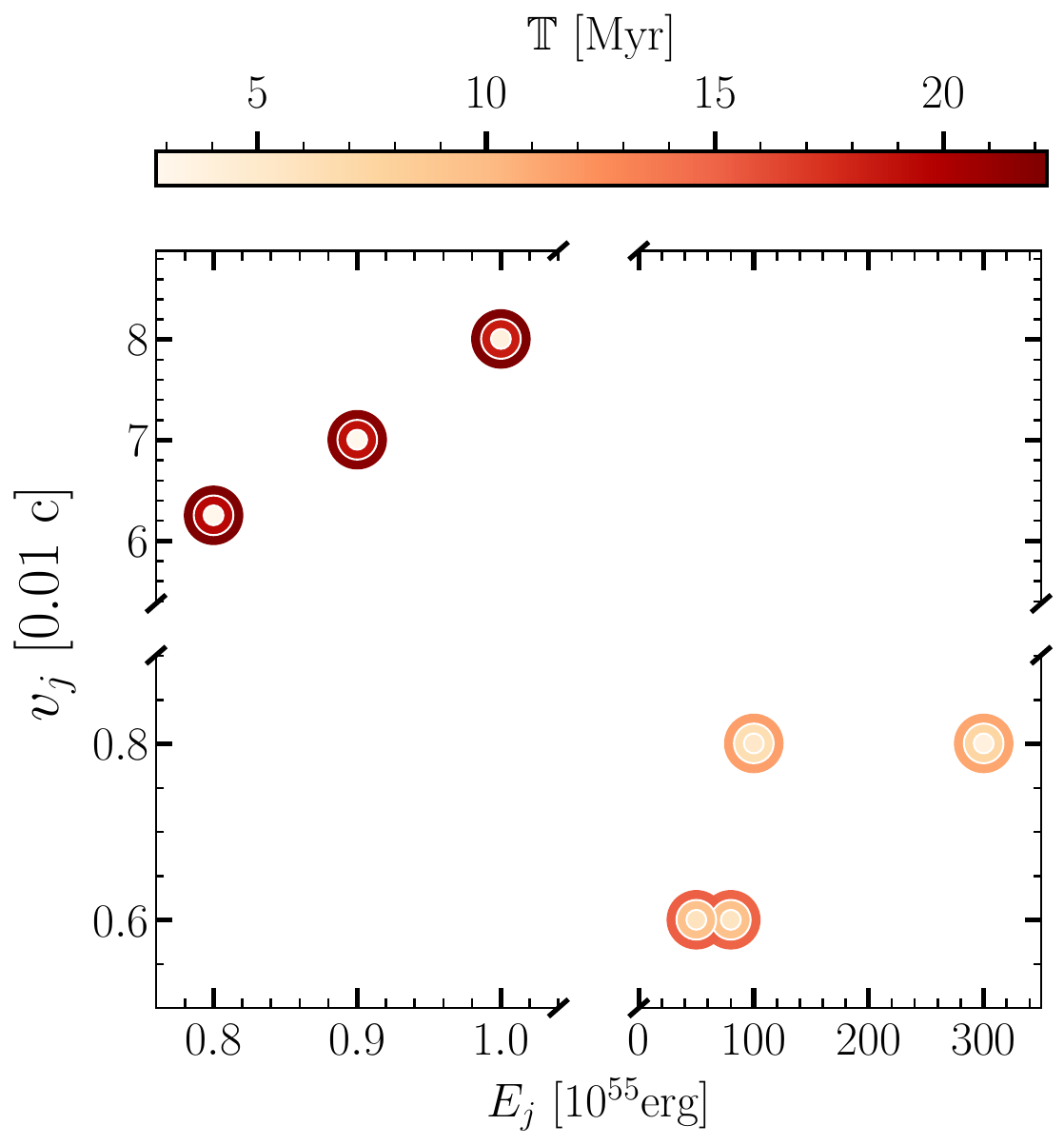}
    }
    \caption{Dependence of timescales (units of Myr, colorbar), associated with the nested bubbles, on the injection parameters $E_j$ and $v_j$, for our suite of simulations (see \autoref{table:twinParams}). Shown are the RB age $\tRB$ (outer ring), delay $\tdel[01]$ between injections (middle ring) and the FB age $\tFB$ (inner circle). \label{fig:doubleJetTime}}
\end{figure}

A suite of simulations similar to $\nomBall$ and $\nomSlow$, each with two consecutive outbursts of identical injection parameters, is listed in \autoref{table:twinParams}, with resulting thermal and projected maps provided in \autoref{appendix:nonNominal}. The age distribution of the nested bubbles, shown in \autoref{fig:doubleJetTime}, indicates $\tFB \simeq 3\text{--}6 \Myr$ for $\xi \gg 1$ FBs, and similarly $\tFB \simeq 3\text{--}4 \Myr$ for $\xi \ll 1$, consistent with spectral constraints.
The X-ray projections of both FBs and RBs are approximately reproduced by each simulation, indicating $\mathcal{A}<1$ ballistic FBs even for $\xi\ll1$, due to RB rarefaction.

\begin{figure}[h]
    \def\pw{0.49\linewidth}
    \def\ph{0.49\linewidth}
    \fboxvisiblefalse
    \centering
    \myfbox{\includegraphics[width=0.48\textwidth, trim=0cm 0cm 0cm 0cm]{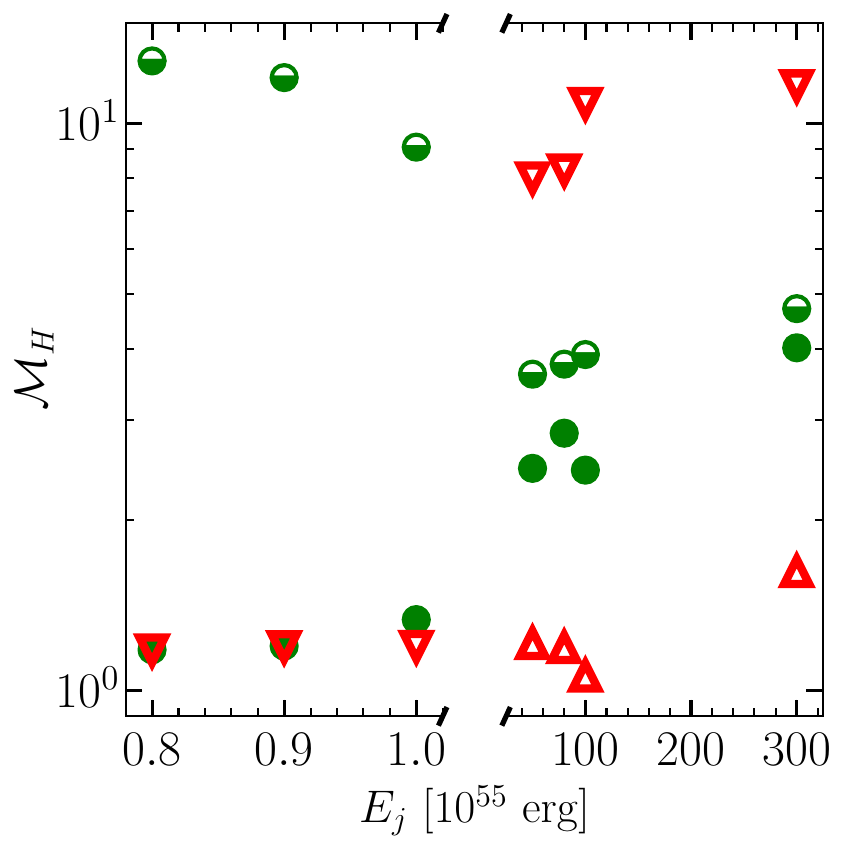}}

    \caption{
    Head Mach numbers of RBs (red triangles) propagating into a relaxed (down triangles) vs. preheated (up triangles) medium, and of their nested FBs (green disks), for our suite of double-outburst simulations (see \autoref{table:twinParams}). FB shocks are initially strong, as seen at intermediate-height (half-filled disks; $z \simeq5 \kpc$),
    but weaken while traversing the hot relic of $\xi\ll1$ RBs.
    \label{fig:machEdges}}
\end{figure}

However, the Mach numbers, presented in \autoref{fig:machEdges}, are sensitive to $\xi$. The RB shocks are too strong ($\Mach_H \gtrsim 8$) for $\xi \gg 1$ and too weak ($\Mach_H \simeq 1.2$) for $\xi \ll 1$ outbursts.
The figure shows FB Mach numbers at both full ($z \simeq 9 \kpc$) and intermediate ($z \simeq 5 \kpc$, below the $\xi\ll1$ RB relic discontinuity structure) height. For $\xi \gg 1$ outbursts, $\Mach_H \simeq 4\text{--}5$ at intermediate-height decreases to $\Mach_H \simeq 2.5\text{--}4$ at full height, only marginally consistent with the strong shocks inferred from observations. The $\xi \ll1$ shocks are initially very strong, $\Mach_H \simeq 10$ at intermediate-height, but weaken to $\Mach_H \simeq 1.2$ at full height as they still traverse the hot RB relic.
The results thus favor the intermediate, $\xi\simeq 1$ regime discussed in \autoref{subsec:estimates} and \autoref{subsec:singleJets}.

\section{Summary and Discussion}
\label{sec:discussion}

As the edges of both RBs and FBs were identified as strong forward shocks, and the necessary outflows were shown to be energetic and collimated nearly perpendicular to the Galactic plane (see {\MK} for the FBs and \autoref{sec:Intro}), we model the bubbles as nested bipolar shells driven by two separate GC outbursts.
We thus generalize a simple, effectively one-dimensional model (\MK), which approximates the evolution of a bubble as stratified (\autoref{sec:burstModels}), and study a suite of numerical simulations to validate the model and explore the outburst parameter space (\autoref{sec:sims}). The model approximately reproduces the observed bubble properties, including their projected morphology (\autoref{fig:edges1}), as verified by the simulations.

Most of the seemingly plausible phase space of outburst parameters, including regimes studied previously and a concurrently reported \citep{Zhang+2025} double-outburst simulation (see discussion in \autoref{appendix:critic}), does not reproduce recent observations. In particular, the shock morphology and strength of the RBs are sensitive to their outburst parameters, and the inner structure of the RBs must be taken into account when modelling their trailing FBs.

Once a collimated outburst sweeps up sufficient mass, it transitions from ballistic propagation to a slowdown phase, which is strongly regulated by momentum conservation (\autoref{subsec:FBModel} and {\MK}), eventually transitioning to isotropic expansion as transverse motion becomes substantial (\autoref{subsec:RBModel}).
While both ballistic ($\xi\gg1$, as defined in \autoref{eq:xiDef}) and marginally slowing-down ($\xi\lesssim 1$) FBs are viable in the absence of the RBs (\MK), prior CGM rarefication by the RBs practically rules out slowdown FB solutions (\autoref{subsec:secondBurstModel}), keeping the FBs ballistic even for the small $\xi\simeq 0.5$ value favored by observations (\autoref{subsec:estimates}).
The RBs can be neither strongly ballistic (morphologically) nor significantly slowed down (as evident from their high Mach number \citep{KeshetGhosh26}, approximate north--south symmetry, and limited downstream rarefaction near the FBs), so must be at the onset of slowdown with $\xi\sim 1$.

Interestingly, integrating the observational evidence available independently for the FBs (\autoref{subsubsec:ObsFBs}) and for the RBs (\autoref{subsubsec:ObsRBs}) indicates that they arise from comparable outbursts: $E_j\sim10^{56}\erg$ energy (up to an uncertainty factor $\sim4$), collimated such that at height $100\pc$ above the GC, their velocity reached $v_j\sim 2000\km\se^{-1}$ within a half-opening angle $\theta_j\sim 4\degree$ (both within an uncertainty factor $\sim2$). Indeed, we find a range of outburst parameters consistent with joint FB and RB observations even if the outbursts are assumed identical, provided that such identical outbursts were separated by $7\text{--}20\Myr$, and each injected $\gtrsim 10^{55} \erg$. Figure \ref{fig:nominalGB} shows examples of nested bubble morphologies consistent in projection with observations in both $\xi\ll1$ and $\xi\gg1$ limits.
The observational evidence favors the intermediate, $\xi\sim 1$ regime, in which our simulations are only marginally converged (see \autoref{fig:enVelScanXi} and \mbox{\autoref{subsec:singleJets}}) and indicate $15\mbox{--}25\Myr$ old RBs.

The RBs and their trailing polarized lobes stretch westward at high latitudes in both Galactic hemispheres, an effect previously attributed to a western Galactic wind. However, the RB shocks were found to be strong \citep{KeshetGhosh26}, and the FBs they enclose were found both to be strong and to show a similar westward stretching in both hemispheres \citep{Keshetgurwich17}. A Galactic wind strong enough to substantially divert the strong FB shocks even after it collides with the strong RB shocks is less plausible than an eastern density gradient, which would facilitates a faster bubble expansion in the west and could be largely preserved by the RB shocks. Indeed, such a density gradient is consistent with the enhanced brightness of the FBs and the RBs in their eastern sectors, despite their uniformly (at least across the former \citep{Keshetgurwich17}) strong shocks.

To examine such a possibility, consider for simplicity the limit of two strong planar shocks, one moving east at $x_-(t)<0$ and the other west at $x_+(t)>0$, with some power-law ambient densities $\rho_+(x>0)<\rho(x=0)$ and $\rho_-(x<0)>\rho(x=0)$, with an eastern density gradient for all $x$. Similarity solutions satisfy $x_+(t)/|x_-(t)|\simeq [\rho_-(x_-)/\rho_+(x_+)]^{1/3}>1$, showing a westward stretching of the bubbles. The FB modeling in \autoref{subsec:bubbleProj} along with the approximate $R_{\Max}\propto \theta_j^{1/2}$ of \autoref{eq:RmaxBall} would then indicate that at a height $z\simeq 5\kpc$ above the Galactic plane, the eastern $\rho_-(x= -2.3\kpc)$ density is comparable in the west to $9.5\rho_+(x\simeq +4.9\kpc)$ in the north and $5.1\rho_+(x\simeq +3.9\kpc)$ in the south.
Interestingly, these estimates correspond to a similar linear, $d\rho/dx\simeq -\rho(x=0)/(6\kpc)$ gradient in both hemispheres.
This conclusion is robust under our deprojection, persisting for example if one examines the bubble at $z=5\kpc$ using the more accurate \autoref{eq:BallRb}.
Such an eastern density gradient seems plausible, especially given the strong RB disturbance; see Ghosh (2026, in prep.).

\section*{Acknowledgements}
We are grateful to K. C. Sarkar, K.~C. Hou, and the late G. Ilani, for useful discussions. This research has received funding from the IAEC-UPBC joint research foundation (grant No. 300/18), and from the Israel Science Foundation (ISF grants 2067/19 and 2126/22).

\bibliographystyle{apsrev4-2-author-truncate}
\bibliography{galacticBubbles}

\appendix
\section{Galactic model}
\label{appendix:galModel}

The Galactic model in our simulations closely follows the implementation of \citep{Sarkaretal15a}. This model includes a rigid gravitational potential, with steady-state equilibrium contributions from baryonic and dark matter (DM) components. The gaseous component comprises the galactic disk, the central molecular zone (CMZ), and the CGM halo. Since our simulations are axisymmetric, we adopt a cylindrical form for all elements in the Galactic model. A summary of the various parameters adopted in this work is provided in \autoref{table:galModel}. For details on the implementation, see {\MK}.

\newcommand{\hstrut}{\rule{0pt}{10pt}\rule[-5pt]{0pt}{0pt}}
\renewcommand{\arraystretch}{1}
\begin{table}[h]
    \centering
    \scalebox{1.1}{
    \begin{tabular}{|l|l|l|}
    \hline
    \textbf{Symbol} & \textbf{Definition}\hstrut & \textbf{Value}\hstrut \\
    \hline
    $M_\mathrm{vir}$ & Virial mass & $1.2\times10^{12} \Msun$\\
    $r_\mathrm{vir}$ & Virial radius & $250$ kpc\\
    $d$ & NFW core radius & $6.0$ kpc\\
    $r_s$ & NFW scale radius & $20.8$ kpc \\
    $c_\mathrm{vir}$ & NFW concentration & $12$ kpc\\
    $a_b$ & Bulge scale radius & $2.0$ kpc\\
    $a_D$ & Disk scale length & $3.0$ kpc\\
    $b_D$ & Disk scale height & $0.4$ kpc\\
    $R_\odot$ & Galactocentric solar radius & $8.5\kpc$\\
    $\rho_\mathrm{cmz, 0}$ & CMZ central density & $50.0~m_p\,\ccinv$\\
    $\rho_{D0}$ & Disc central density & $1.0~m_p\,\ccinv$\\
    $\rho_{h0}$ & Halo central density & $0.019~m_p\,\ccinv$\\
    $k_BT_\mathrm{cmz}$ & CMZ temperature & $8.6\times10^{-5}$ keV\\
    $k_BT_D$ & Disk temperature & $3.4\times 10^{-3}$ keV\\
    $k_BT_h$ & Halo temperature & $0.17$ keV\\
    $f_{D}$ & Disk rotation & $0.975$\\
    \hline
    \end{tabular}}
    \caption{
    Fixed parameters of the Galactic model adopted in this work.
    Mass and length scales
    follow from \citep{McMillan2011,McMillan2017}. \label{table:galModel}
    }
\end{table}

The DM contribution is assumed to follow the NFW \citep{Navarroetal96} model; however, the profile is modified such that it approaches a finite density towards the center of the potential well
\begin{equation}
    \Phi_\mathrm{DM}\paran{R,z} = -\frac{GM_\mathrm{vir}}{\Lambda\paran{c_\mathrm{vir}}}\frac{\ln \paran{1 + r_s^{-1}\sqrt{R^2 + d^2}}}{\sqrt{R^2 + d^2}} \label{eq:phiDM} \,,
\end{equation}
where $r_\mathrm{vir}$ is the virial radius, $r_s$ is the scale radius of the NFW model, $d$ is the radius of the finite density core, $c_\mathrm{vir}\equiv r_\mathrm{vir}/r_s$ is the NFW concentration parameter, and $\Lambda\paran{c_\mathrm{vir}} \equiv \ln\paran{1 + c_\mathrm{vir}} - c_\mathrm{vir}/(1 + c_\mathrm{vir})$. The gravitational potential from the baryonic components include contributions from the stellar bulge and the galactic disk. The stellar bulge potential is given by
\begin{equation}
    \Phi_b\paran{R,z} = -\frac{GM_b}{\sqrt{R^2 + a_b^2}} \label{eq:phiBulge}\,,
\end{equation}
where $M_b$ is the bulge mass and $a_b$ is the bulge scale radius. The disk potential is given by the cylindrical form of the Miyamoto-Nagai potential \citep{Miyamotonagai75}
\begin{equation}
    \Phi_D\paran{R,z} = -\frac{GM_D}{\sqrt{R^2 + \paran{a_D + \sqrt{z^2 + b_D^2}}^2}} \label{eq:phiDisk}\,,
\end{equation}
where $M_D$ is the mass of the disk and $a_D, b_D \geq 0$ are the scale length and scale height of the disk, respectively. The total gravitational potential is the sum of all contributions
\begin{equation}
    \Phi_\mathrm{tot} \paran{R,z} = \Phi_\mathrm{DM}\paran{R,z} + \Phi_b\paran{R,z} + \Phi_D\paran{R,z}\,.
\end{equation}

The density profile for each of the gaseous component $X\in \{D, \mathrm{cmz}, h\}$, where $D$, $\mathrm{cmz}$, and $h$ denote disk, CMZ, and halo components, respectively, is derived by assuming it is in hydrostatic equilibrium with $\Phi_\mathrm{tot}$. The stellar rotation velocity is given by $v_{\phi,\star}(R) = \sqb{-\partialIlAt{\Phi}{R}{z=0}}^{1/2}$.
We define $f_X$ as the ratio of a component rotation velocity to $v_{\phi,\star}(R)$; $f_X$ is nonzero only for the disk component. The isothermal sound speed of each gaseous component $X$ is given by $c_{s,X} = \paran{k_BT_X/\bar{\mu}m_p}^{1/2}$,
where $T_X$ denotes the temperature of the component $X$; note that $T_D$ accommodates contributions from non-thermal sources (cosmic rays, magnetic fields and turbulence). Both $f_X$ and $c_{sX}$ are taken to be independent of $R$ and $z$ in order to obtain analytical expressions for the density distributions; for a detailed derivation, see \citep{Sarkaretal15a}. Using $\rho_{X0} \equiv \rho_X\paran{0,0}$ and $\Phi_0 \equiv \Phi\paran{0,0}$, the density distribution for a component $X$ is given by
\begin{equation}
    \frac{\rho_X\paran{R,z}}{\rho_{X0}} = \exp\sqb{-\frac{\Phi\paran{R,z} - \Phi_0 -\curly{\Phi\paran{R,0} - \Phi_0}f_X^2}{c_\mathrm{sX}^2}}\,,
\end{equation}
while the total density in any numerical cell is given by $\rho_\mathrm{tot}\paran{R,z} = \sum_X \rho_X \paran{R,z}$. The effective rotational velocity of the gas inside a numerical cell is computed as
\begin{equation}
    v_{\phi,\mathrm{eff}}\paran{R,z} = v_{\phi,\star}\paran{R}\sqb{\frac{\sum_X f_X^2\rho_X\paran{R,z}}{\rho_\mathrm{tot}\paran{R,z}}}^\half \,.
\end{equation}

\section{Numerical setup and convergence}
\label{appendix:convTests}

Our simulations are carried out in a 2D grid of spherical polar geometry, the numerical resolution of which is set by $N_r$ cells in the radial direction and $N_\theta$ cells in the polar direction, both linearly spaced.
The azimuthal direction is frozen, restricted to $1$ cell. The box size is held constant: the $\nomBall$ setup extends to $40$ kpc radially, while a radial extent of $30$ kpc is sufficient for $\nomSlow$. Nominally, we use $(N_r, N_\theta) = (2048, 512)$ for $\nomBall$ and $(N_r, N_\theta) = (1536, 768)$ for $\nomSlow$ throughout this work.

\begin{figure}[h]
    \def\pw{0.7\linewidth}
    \def\ph{0.7\linewidth}
    \fboxvisiblefalse
    \centering
    \myfbox{\includegraphics[height=\ph, width=\pw]{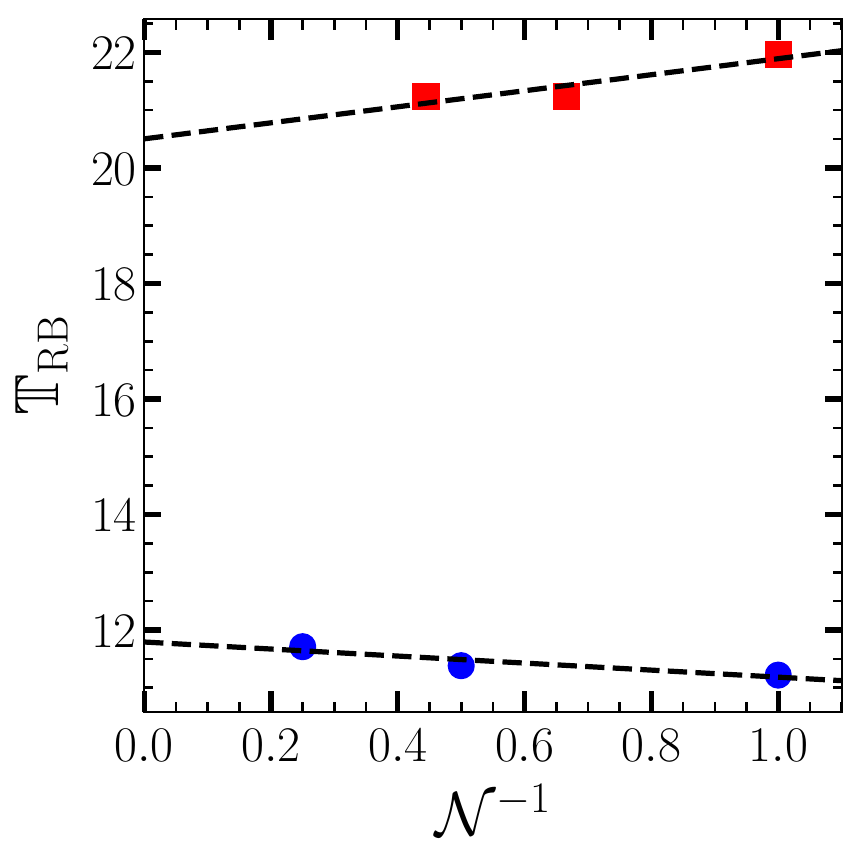}}
    \caption{
        RB age convergence tests for $\nomBall$ (blue circles) and $\nomSlow$ (red squares) as a function of refinement factor $\mathcal{N}$, with corresponding Richardson extrapolations (black dashed).
        \label{fig:convTest}
    }
\end{figure}

All simulations presented in this work were tested to convergence with numerical resolution, unless otherwise stated. Here, we present only the tests for our nominal, single-burst simulations $\nomBall$ and $\nomSlow$ (see \autoref{table:twinParams} for outburst parameters).
This convergence test is performed over a set of numerical grids, obtained by multiplying the number of cells in the nominal resolution by a refinement factor $\mathcal{N}$ in each direction ($\mathcal{N}=1$ denotes the nominal resolution in this test). The set of grids chosen for $\nomBall$ is $\{\mathcal{N}\} = \{1, 2, 4\}$, while for $\nomSlow$ we adopt $\{\mathcal{N}\} = \{1, 1.5, 2.25\}$. For each grid, the bubble age is measured as it reaches $b \simeq 84^\circ$. Results of this test are provided in \autoref{fig:convTest}.

\section{Viscosity-modified outburst parameters\!}
\label{appendix:transition}

In our numerical setup, low-energy, high-velocity outbursts are subject to strong shear during injection, leading to Kelvin-Helmholtz instabilities which we suppress by incorporating viscosity.
Following {\MK}, we adopt the Spitzer isotropic, dynamic viscosity $\mu$ of a non-magnetized plasma \citep{Braginskii1958, Spitzer1962}, capped at $\mu_\mathrm{max} = 10~\g~\cminv\,\sinv$ to prevent prohibitively small simulation timesteps.
Such viscosity collimates the early outflow and slows it down, essentially modifying the outburst parameters $v_j$ and $\theta_j$ with respect to their nominally injected values. The outflow modeling of \S\ref{sec:burstModels} still applies, but only when using the modified parameters (denoted by primes below), which can be determined from the simulation itself.
We verify that no such modification is needed for inviscid simulations.
Here, we demonstrate the determination of $v'_j$ and $\theta'_j$ from the early, $<1$ Myr evolution of the nominal $\xi\ll1$ simulation $\nomSlow$.

\begin{figure}[h]
    \def\pw{0.8\linewidth}
    \def\ph{0.8\linewidth}
    \fboxvisiblefalse
    \centering
    \myfbox{\includegraphics[height=\ph, width=\pw, trim=0cm 0cm 0cm 0cm]{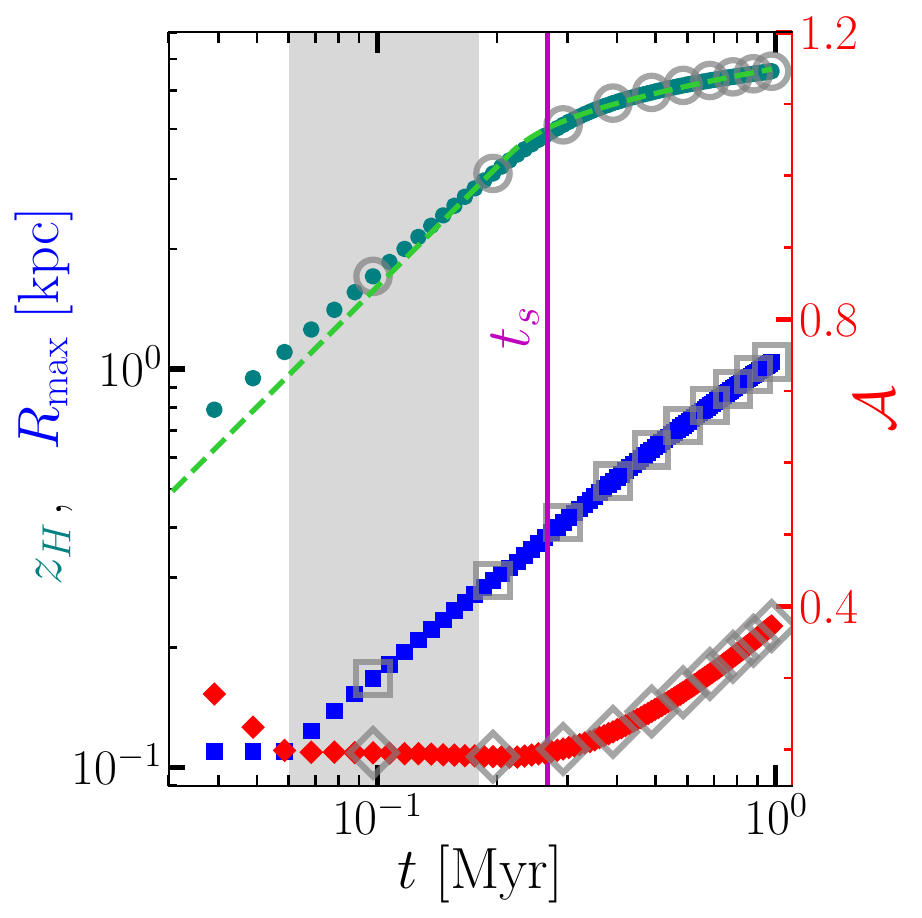}}
    \caption{
        Same as the bottom panel of \autoref{fig:G0RBEvol} (reproduced as large gray empty symbols), but focusing on the early evolution of $\nomSlow$ to estimate the viscosity-modified outburst parameters.
        Highlighted are the $0.06\lesssim t/\mathrm{Myr} \lesssim 0.18$ nearly ballistic epoch (shaded area), the ballistic-to-slowdown transition time $t_s$ (solid; labeled), and a best-fit to the bubble-head trajectory (dashed green) in our model combining Eqs. \eqref{eq:ballCond} and \eqref{eq:zHSlow}.
        \label{fig:S0Ball}
    }
\end{figure}

Figure \ref{fig:S0Ball} shows the $\nomSlow$ evolution, as in \autoref{fig:G0RBEvol}, but at high temporal resolution.
The head monotonically decelerates even at the earliest times accessible, so we approximate the modified ballistic parameters as inferred from the $t\simeq \Delta t_j$ behavior at the end of injection. This corresponds to a head velocity $\beta'_{-2} = 5.2 \pm 0.1$, somewhat slower than the injected $\beta_{-2}=8$.
The concurrent aspect ratio $\mathcal{A} \simeq 0.23$ corresponds to $\theta'_5 = 0.20\pm 0.02$, indicating strong collimation with respect to the injected $\theta_5=0.8$. Combined, the modified outburst parameters imply $\xi' = 0.26 \pm 0.05$ from \autoref{eq:xiDef}, so Eqs. \eqref{eq:Lc} and \eqref{eq:Tc} give $z_s \simeq 4.3 \kpc$ and $t_s\simeq 0.27 \Myr$.
Interestingly, the $z_H(t)$ power-law slope initially increases, saturating at slightly sub-ballistic, $z_H \propto t^{0.9\pm 0.1}$ motion during $0.06 \lesssim t/\mbox{Myr} \lesssim 0.18$ (gray shaded area in the figure).

Alternatively, one can use the ballistic-through-early slowdown evolution to infer the modified three parameters, for example $\beta'$, $\xi'$, and $\tau_z$, of the relevant model phases in Eqs. \eqref{eq:ballCond} and \eqref{eq:zHSlow}. Such a three-parameter fit (dashed green line in the figure) yields $\vJ'=5.2\pm0.1$ and $\xi'=0.19 \pm 0.01$, consistent with the above $t\simeq \Delta t_j$ estimates, and $\tau_z = 0.15\pm0.01$, consistent with our estimate based on the deep slowdown phase (\autoref{subsec:singleJets}).

\section{Nominal nested bubbles in 3D}
\label{appendix:3d}

We run our double-outburst $\nomBall$ setup (see \autoref{subsec:doubleJets})
also in 3D (simulation denoted $\nomBall3D$), to verify the correctness and robustness of the nominal 2D, axisymmetric simulation.
This simulation is performed with the same resolution as $\nomBall$ in the radial and polar coordinates, but adding 128 cells in the azimuthal direction which was frozen in 2D.

\begin{figure}[ht!]
    \centering{
        \includegraphics[width=0.4\textwidth, trim=0cm 0cm 0cm 0cm]{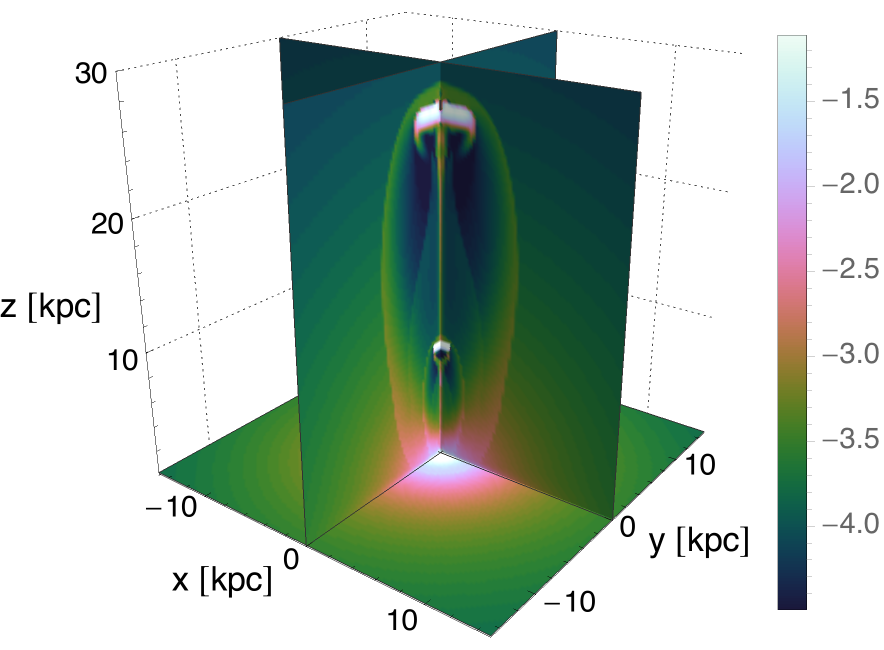}\\
        \includegraphics[width=0.4\textwidth, trim=0cm 0cm 0cm 0cm]{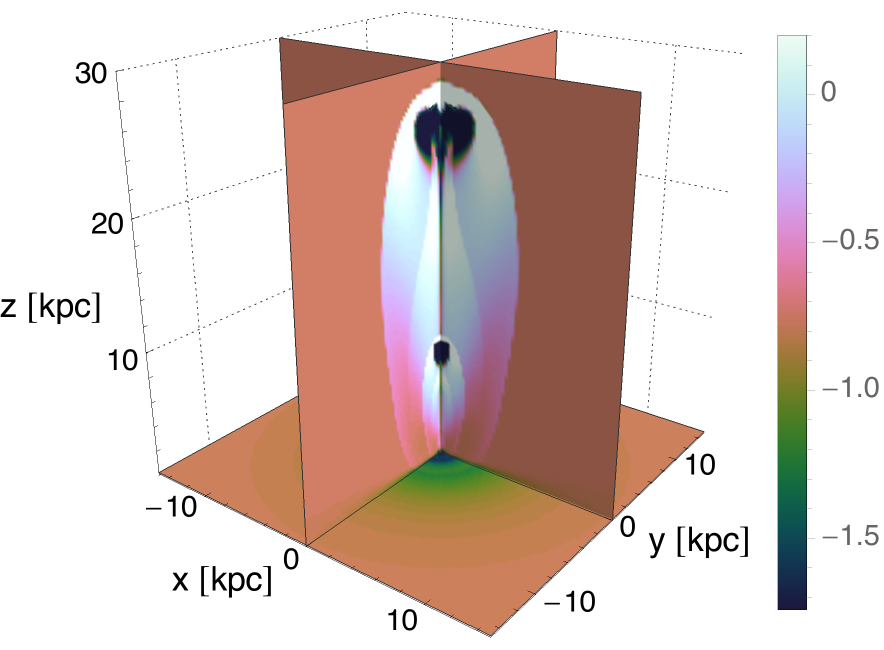}\\
        \includegraphics[width=0.4\textwidth, trim=0cm 0cm 0cm 0cm]{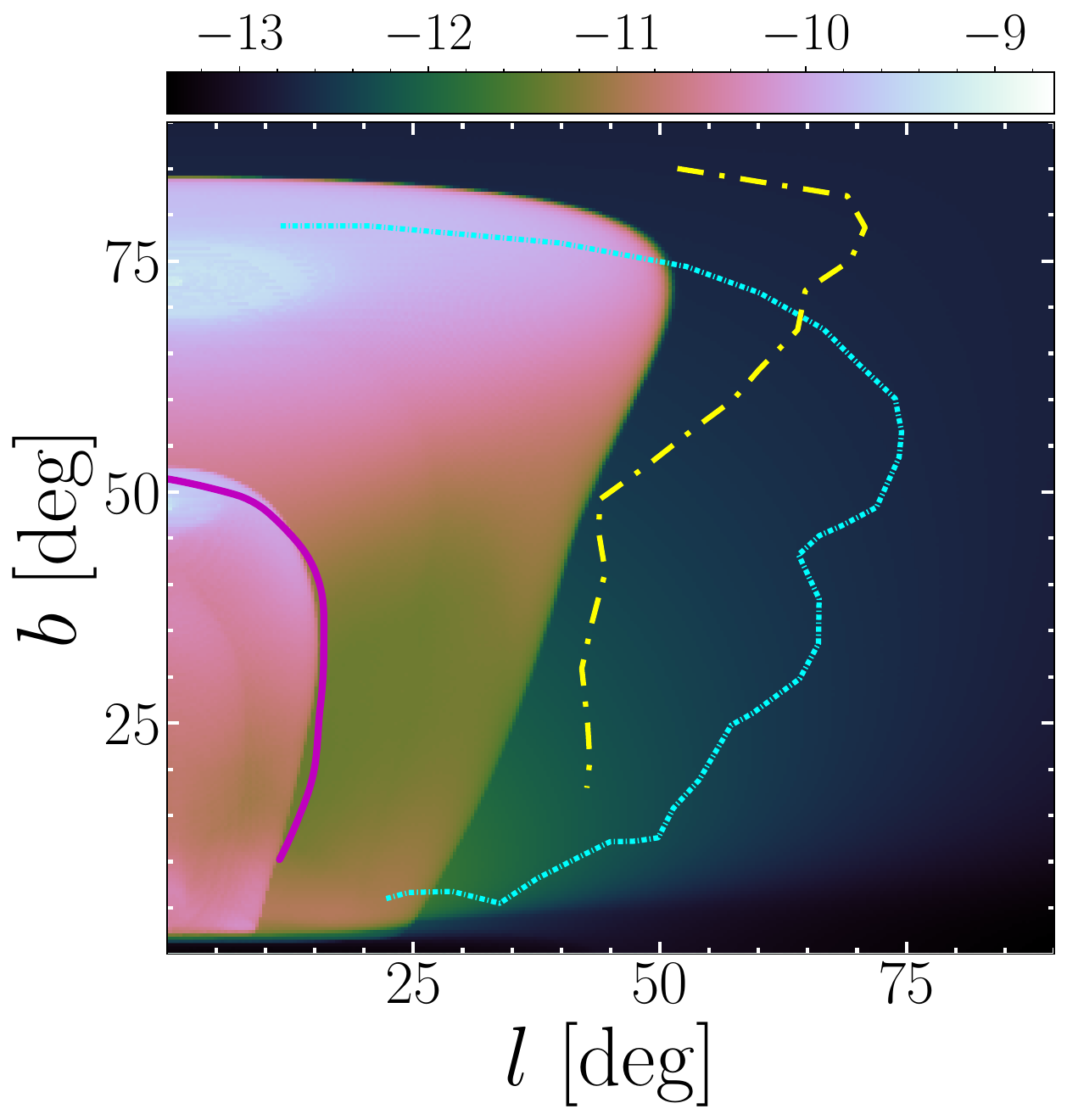}
    }
    \caption{Snapshot at $t\simeq 11.2$ Myr of $\nomBall$3D. Shown are the density (top panel), temperature (middle) and $2\text{--}10\keV$ projection (bottom). Notations are identical to those in \autoref{fig:nominalGB}.
    \label{fig:nominalGB3D}
    }
\end{figure}

Figure \ref{fig:nominalGB3D} shows the density and temperature maps of $\nomBall3D$, which agree well with $\nomBall$. The bubble morphology in projection is in good agreement with the observed edges. While convergence tests could not be performed in 3D due to limited resources, the overall agreement between 2D and 3D simulations captures the robustness of our results. Such a 3D setup facilitates modified outbursts, for example misaligned with the Galactic disk normal, verifying that a small misalignment does not modify our conclusions; see Ghosh (2026, in preparation).

\vspace{0.5cm}
\section{Additional nested-bubble simulations}
\label{appendix:nonNominal}

\begin{figure*}[!htbp]
    \centering{
        \includegraphics[width=0.23\textwidth, trim=0cm 0cm 0cm 0cm]{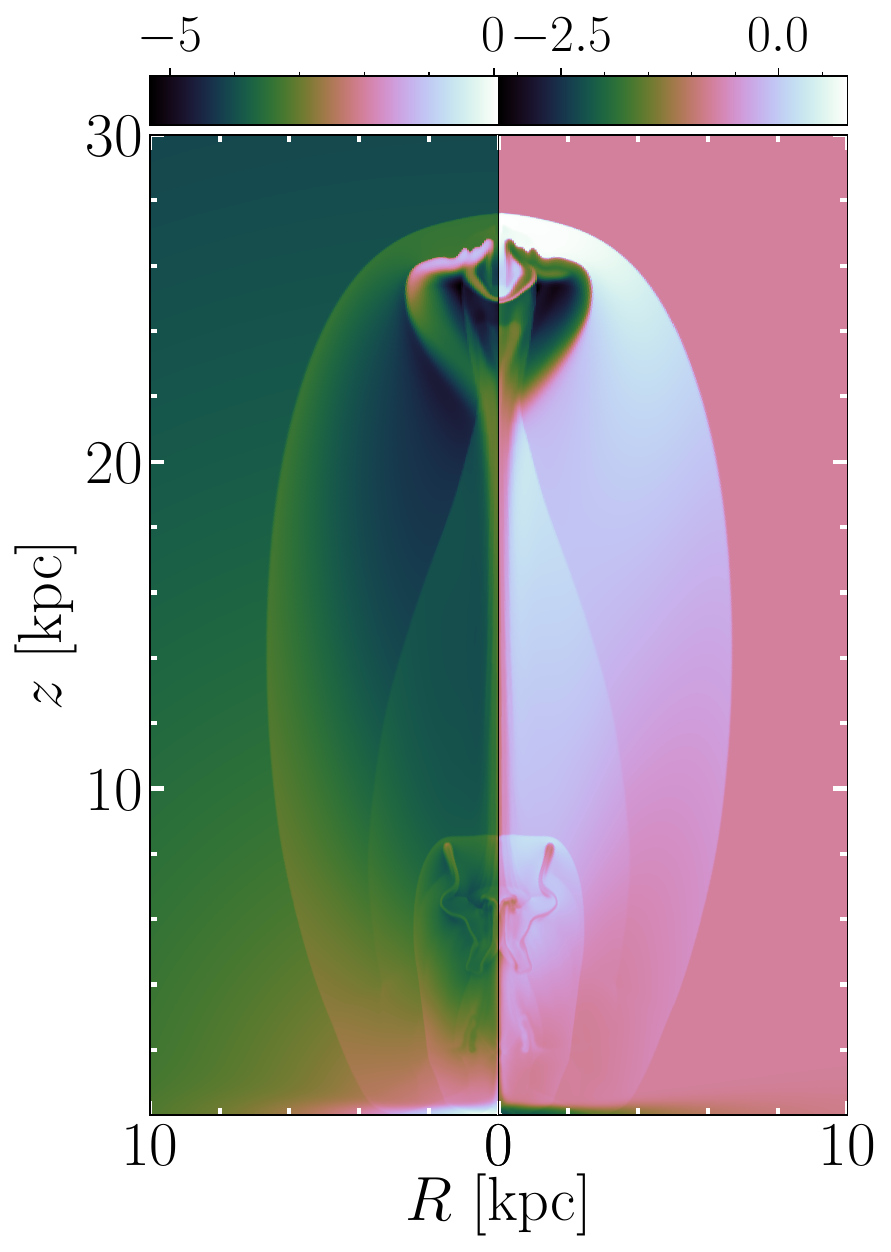}
        \includegraphics[width=0.24\textwidth, trim=0cm 0cm 0cm 0cm]{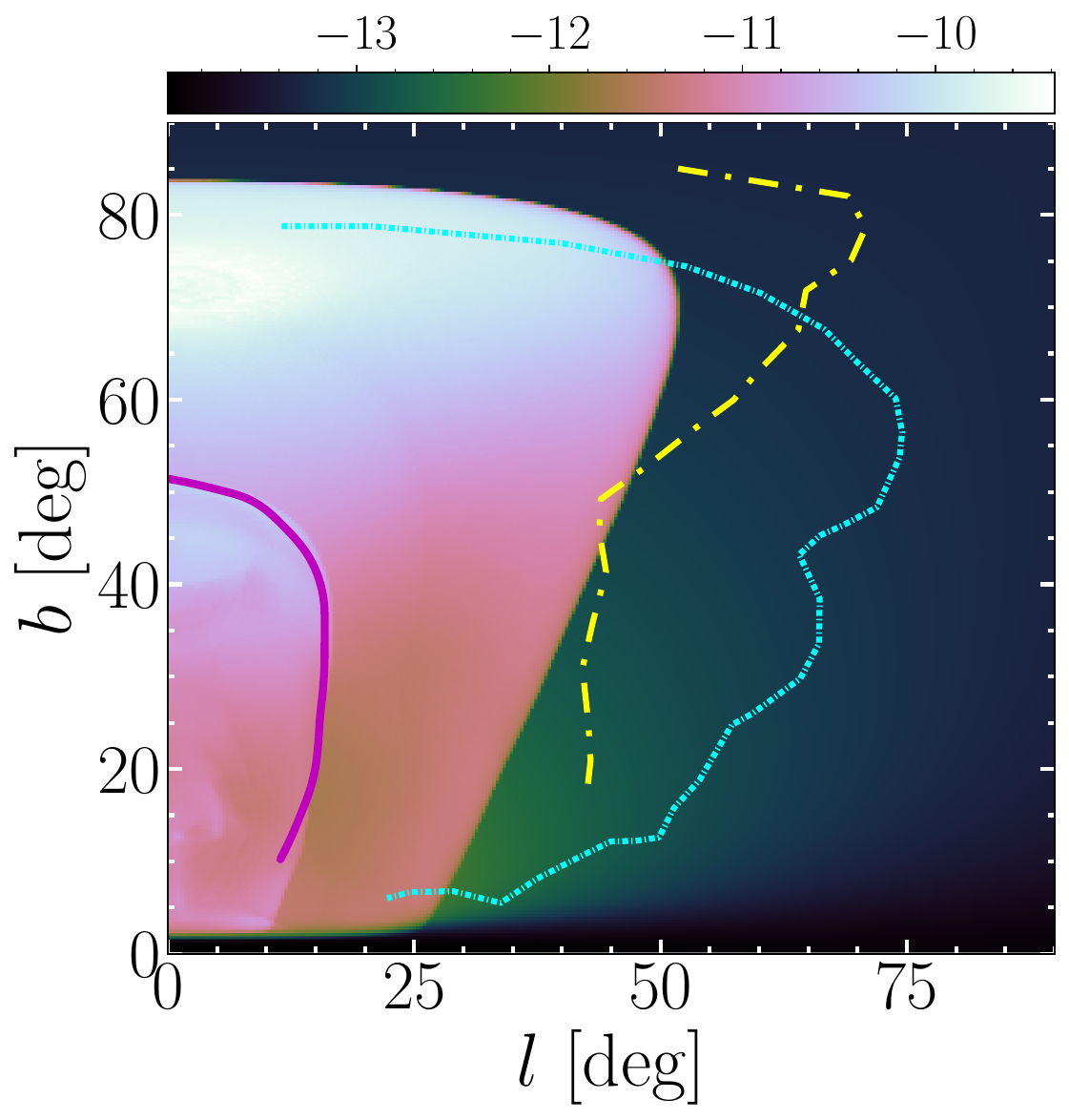}
        \includegraphics[width=0.23\textwidth, trim=0cm 0cm 0cm 0cm]{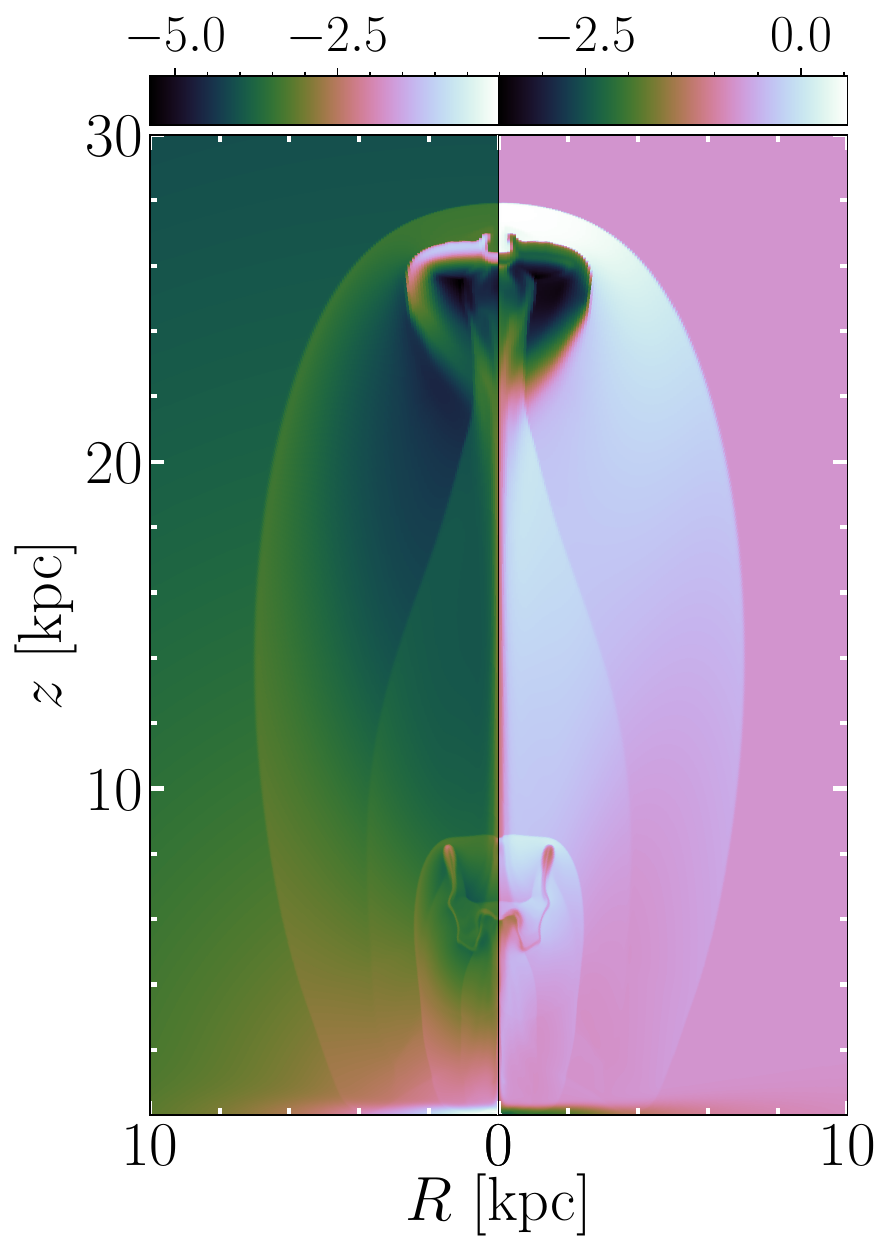}
        \includegraphics[width=0.24\textwidth, trim=0cm 0cm 0cm 0cm]{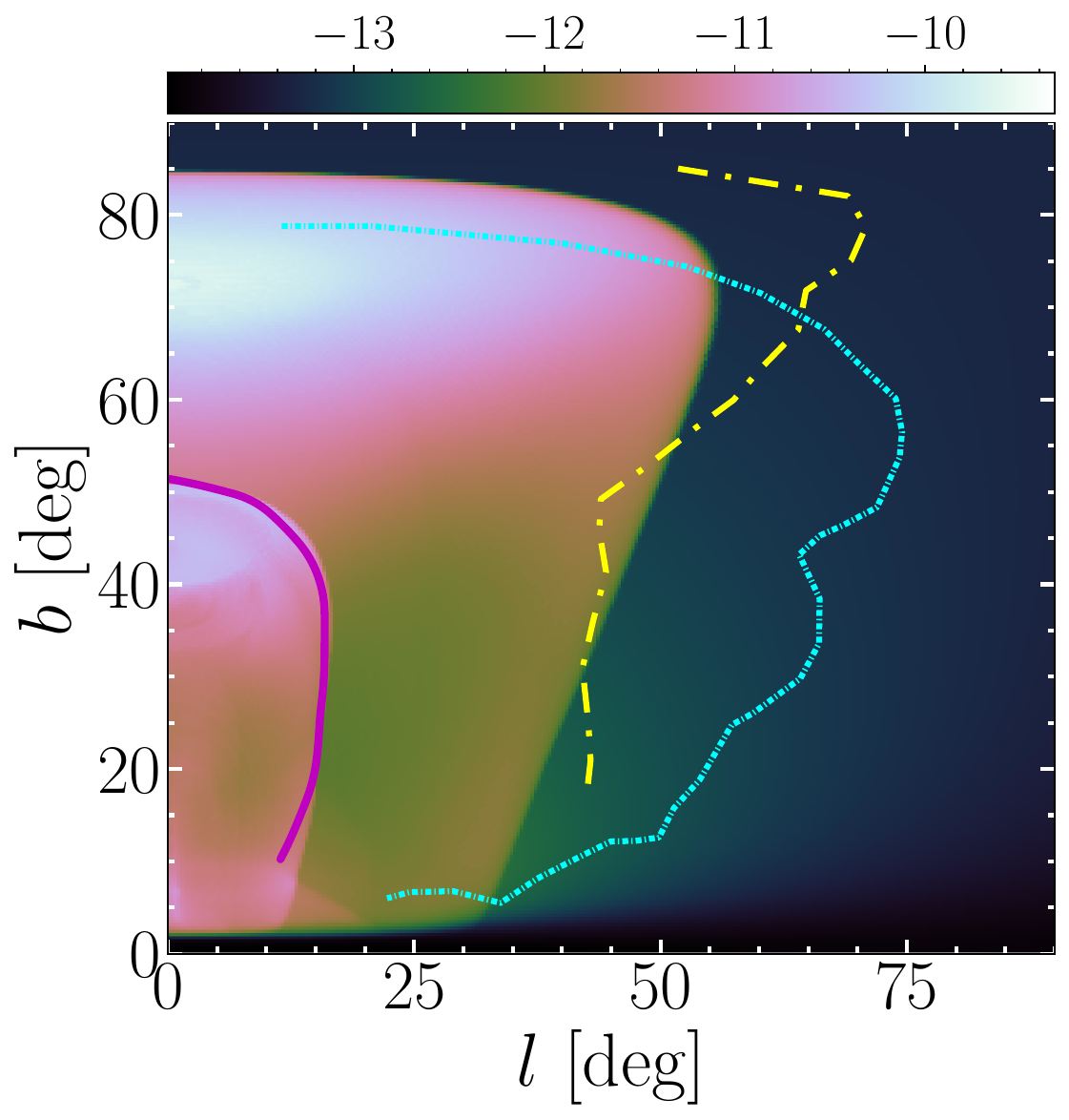}\\
        \includegraphics[width=0.23\textwidth, trim=0cm 0cm 0cm 0cm]{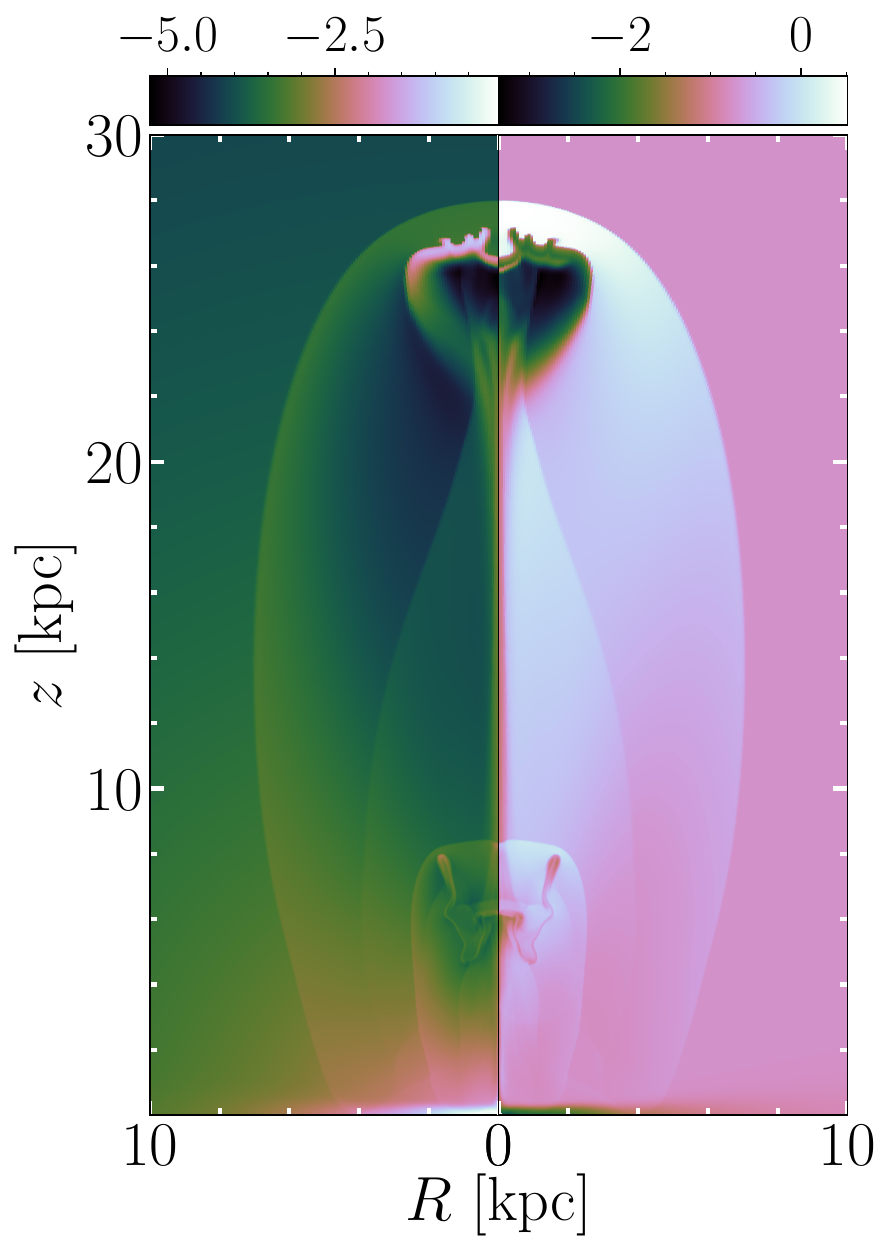}
        \includegraphics[width=0.24\textwidth, trim=0cm 0cm 0cm 0cm]{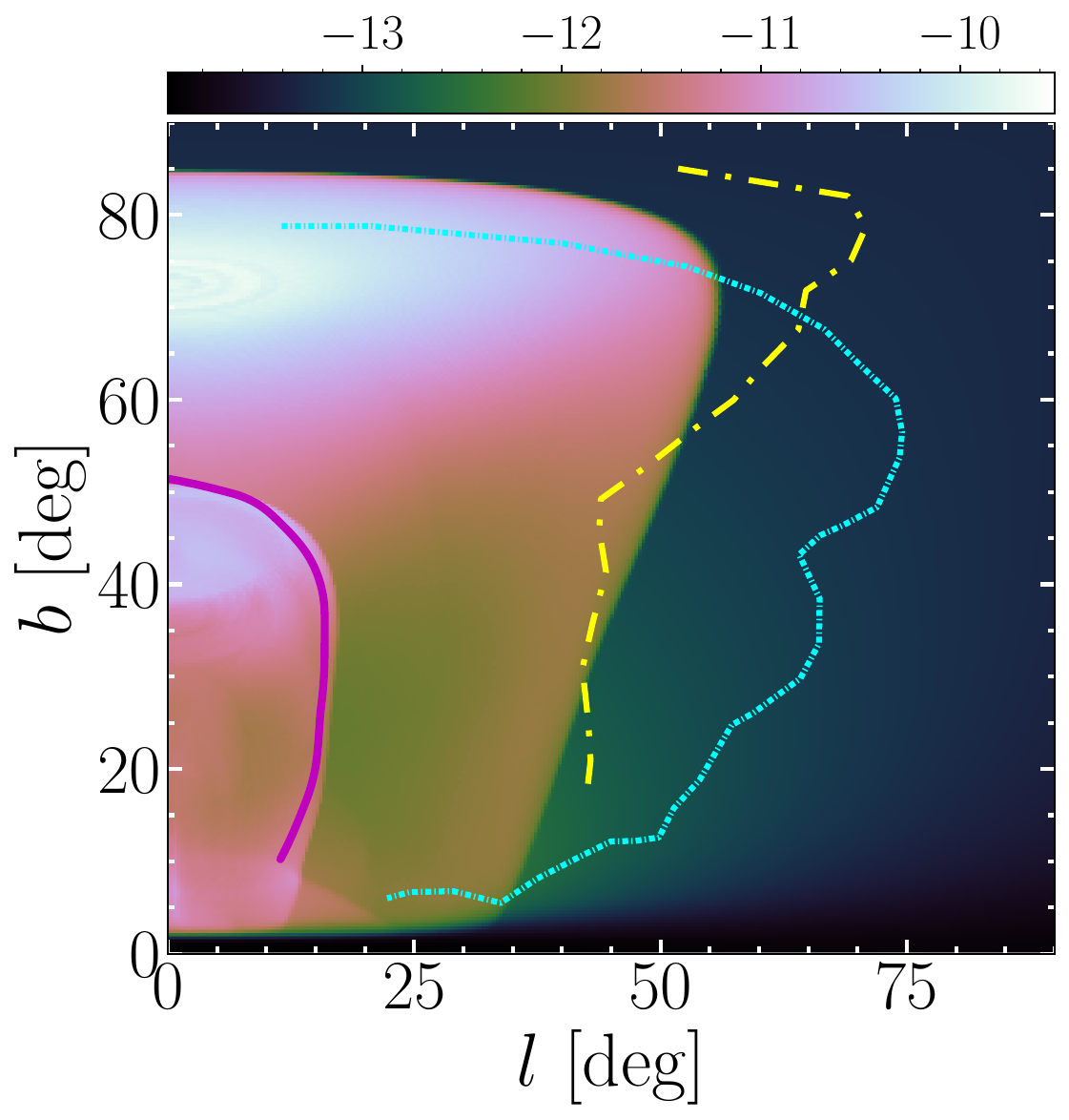}\\
        \includegraphics[width=0.23\textwidth, trim=0cm 0cm 0cm 0cm]{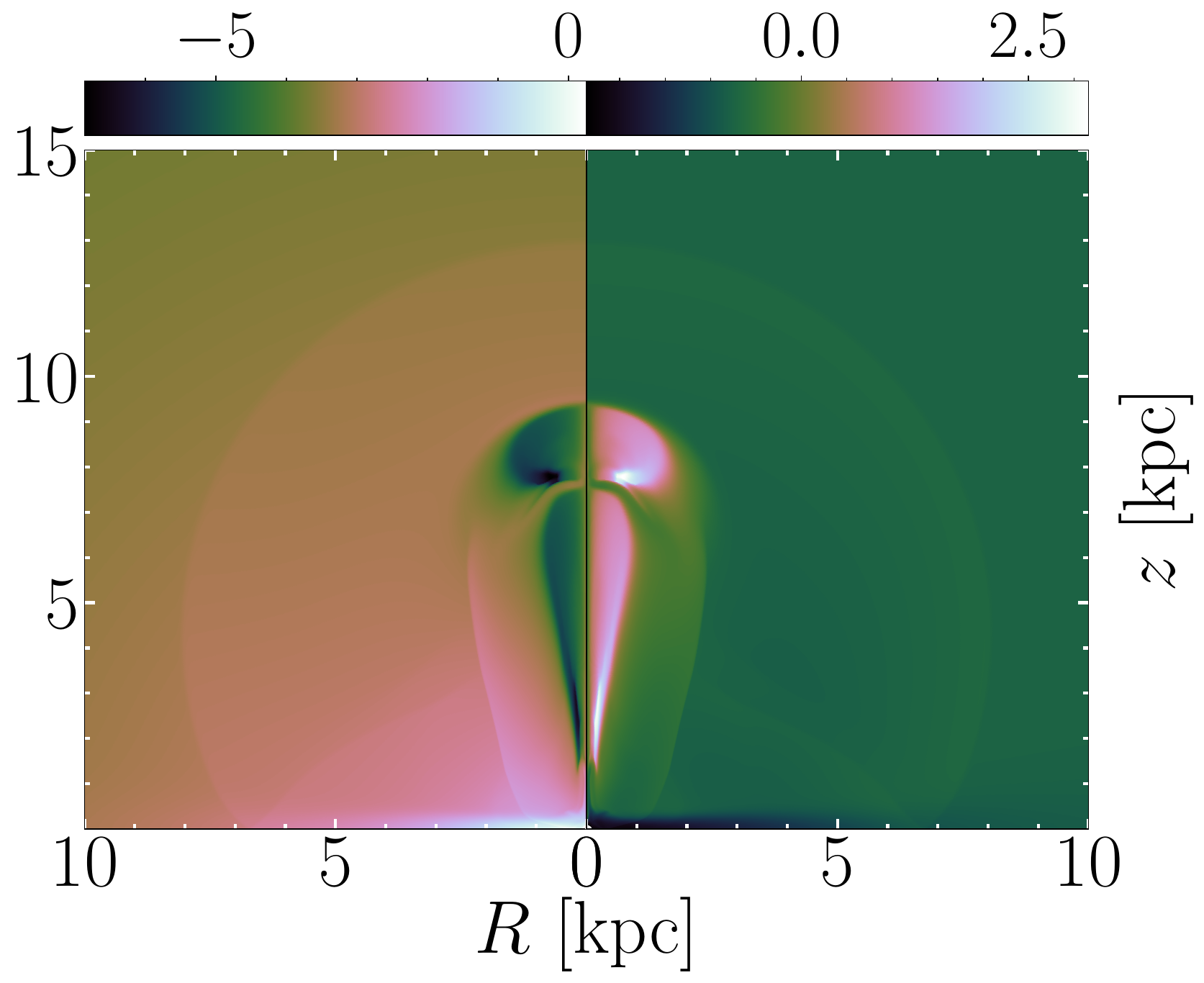}
        \raisebox{-0.15cm}{\includegraphics[width=0.24\textwidth, trim=0cm 0cm 0cm 0cm]{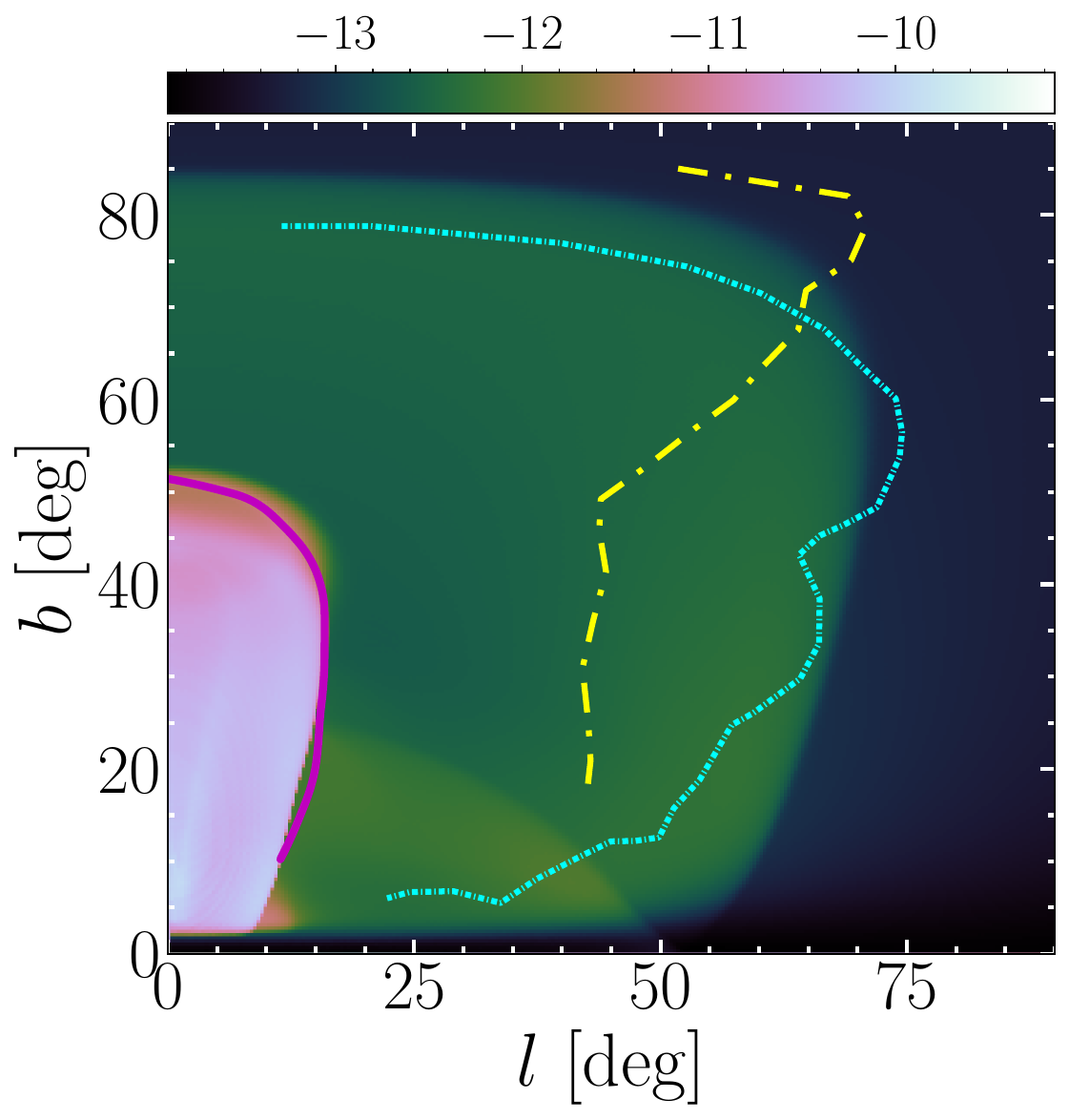}}
        \includegraphics[width=0.23\textwidth, trim=0cm 0cm 0cm 0cm]{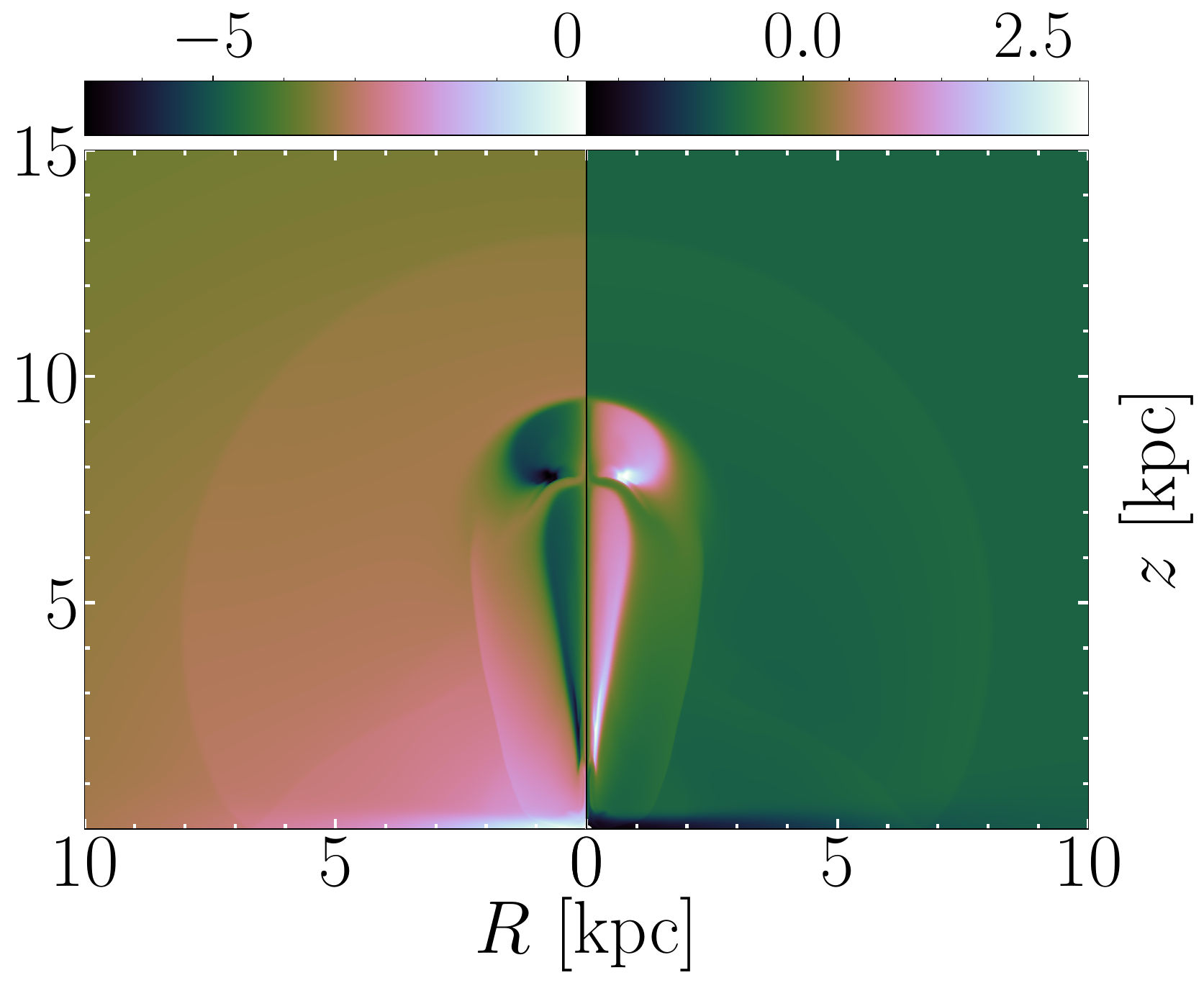}
        \raisebox{-0.15cm}{\includegraphics[width=0.24\textwidth, trim=0cm 0cm 0cm 0cm]{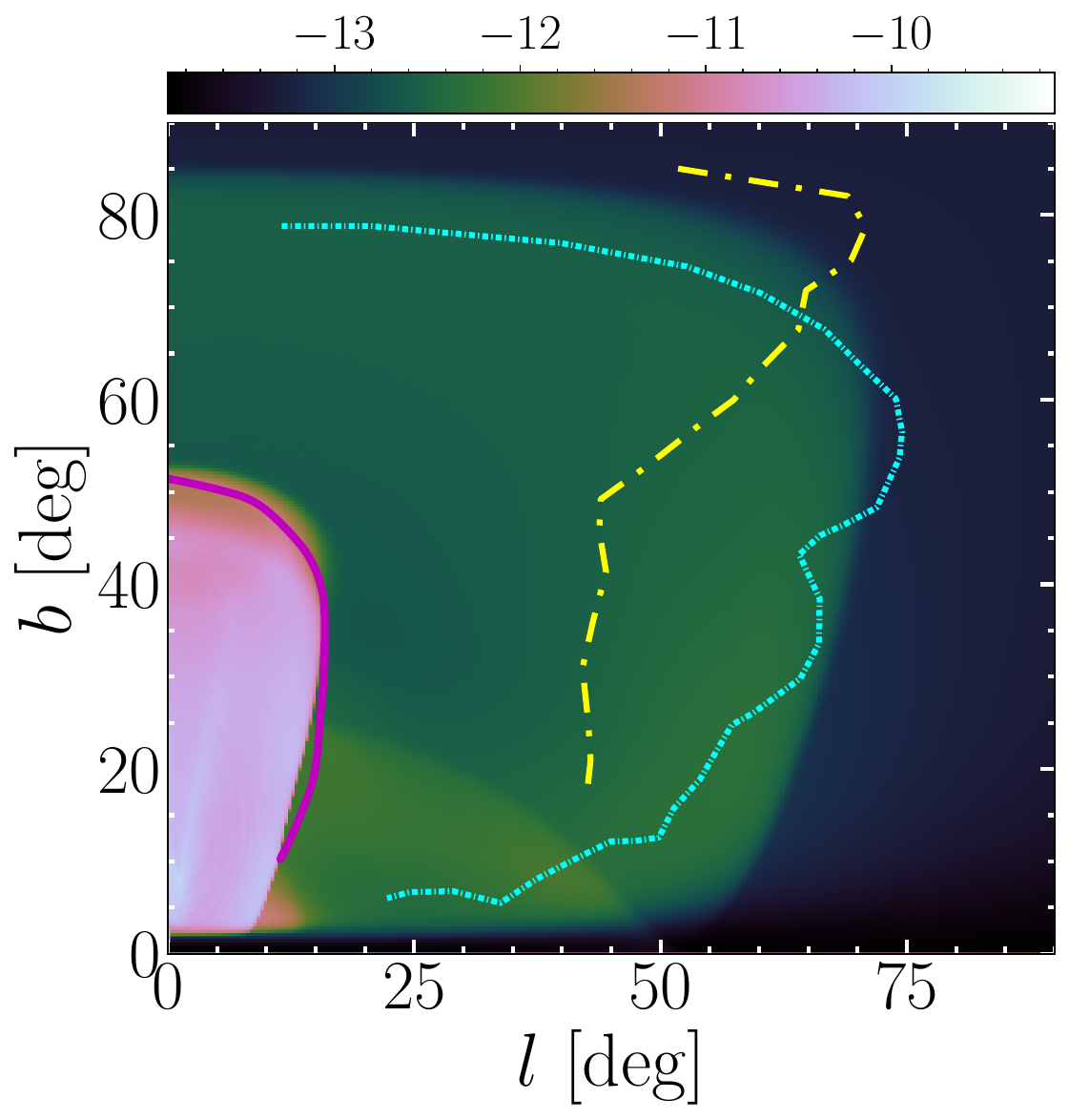}}
    }
    \caption{Nested RBs and FBs in non-nominal simulations (listed in \autoref{table:twinParams}), as the RB approaches $b_{\Max}\simeq 84\degree$.
    A panel pair (thermal on the left and projected X-ray on the right, in \autoref{fig:nominalGB} notations) is shown for each of the simulations $\mathcal{B}_1$, $\mathcal{B}_2$, $\mathcal{B}_3$, $\mathcal{S}_1$, and $\mathcal{S}_2$ (left to right, top to bottom).
    \label{fig:nonNominalGB}}
\end{figure*}

The thermal structure and projected X-ray brightness of the RBs and FBs in our non-nominal double-outburst simulations, listed in \autoref{table:twinParams}, are presented in \autoref{fig:nonNominalGB}.

\section{Critique of concurrent work}
\label{appendix:critic}

Concurrently, a 3D simulation is claimed (\citep{Zhang+2025}, Z25 hereafter) to reproduce the RBs and FBs as forward shocks from two separate GC outbursts, similar to our framework.
However, the Z25 outburst parameters are inconsistent with our constraints, so a few comments are in order.
Overall, we find both the numerical setup and the results of Z25 to be inconsistent with observational constraints, as already noted in part by {\MK}.

In Z25, both RBs and FBs are injected with a velocity $v_j \simeq 1.2\times 10^9 \cm~\sinv$ from a GC cuboid of dimensions corresponding to a height $350\pc$ above the GC and a $3\dgrdot4$ half-opening angle.
The resulting RBs are aged $15$ Myr, arising from a $5.28\times 10^{54}\erg$ outburst of duration $1$ Myr, while the FBs are aged $5$ Myr, originating from a $2.64\times 10^{54}\erg$ outburst which lasted $0.5$ Myr.

The numerical setup of Z25 is not entirely comparable to ours. Injecting the flow at $z\simeq 350$ pc does not incorporate the physical effects of traversing the dense galaxy; the reported injection parameters are not representative of those at small $z$.
The long, $\Delta t_j\simeq 1$ Myr injection duration is not typical of a low luminosity SMBH, and becomes an important dynamical parameter when not negligible with respect to the bubble age. Such issues were already discussed by \MK.

More importantly, while the ambient density in Z25 follows a $\beta=0.5$ beta-model profile similar to ours, Z25 use a very low normalization, with gas density $0.04~m_p~\ccinv$ near the GC. Such a low-density CGM is an order of magnitude more rarefied than inferred from observations \mbox{\citep{MillerBregman2015}}, and is insufficient to account, for example, for the observed thermal \citep{Keshetgurwich18} and non-thermal \citep{Keshet+2023} FB signatures.

Indeed, simulating the Z25 outburst parameters but in our numerical setup, \ie injecting the outflow at $r=0.1\kpc$ into our nominal CGM density, fails to reproduce viable bubbles. Instead, the RB shock is already very weak ($\Mach_H \simeq 1.08$) and nearly spherical ($\mathcal{A} \simeq 1.7$) before reaching even $b=55^\circ$.
A subsequent, lower-energy outburst propagating into a CGM of negligible rarefication would be even weaker and more spherical, failing to account for FB observations, too.

Ultimately, the Z25 results are in tension with key observations, because their:
\begin{enumerate}
    \item FBs are too weak, with $\Mach_H \simeq 1.9$ (as implied by their Fig. 1), in conflict with accumulating evidence for a strong, $\Mach \gtrsim 5$ shock \citep{Keshetgurwich17, Keshetgurwich18, Keshet+2023}.

    \item RBs are too weak, with $\Mach_H \lesssim 1.5$, inconsistent with the inferred $\Mach_H >3$ from the RB edge spectrum \citep{KeshetGhosh26}.

    \item FB morphology is inconsistent with observations. In particular, the simulated bubbles are too tall, reaching $z\simeq 14\kpc$, and $|b|\simeq 60\degree$ in projection. They extend more than $3\kpc$ from the symmetry axis, so are also too wide in projection, at least for the observed eastern edges.
\end{enumerate}
Indeed, we find that an energy a few times higher, and in a slower outflow $>100\pc$ from the GC, are needed to account for the observed RBs and FBs in a nominal-density CGM.

\end{document}